\documentclass[12pt,preprint]{aastex}
\slugcomment{To appear in {\it The Astrophysical Journal, Supplement Series}, Sept.\ 2010}
\shortauthors{Kirkpatrick}
\shorttitle{2MASS Proper Motions}

\begin{document}

\title{Discoveries from a Near-infrared Proper Motion Survey using Multi-epoch 2MASS 
Data\footnote{Some of the spectroscopic data presented herein were obtained at 
the W.M. Keck Observatory, which is operated as a scientific partnership among 
the California Institute of Technology, the University of California and the 
National Aeronautics and Space Administration. The Observatory was made 
possible by the generous financial support of the W.M. Keck Foundation. Other 
spectroscopic data were collected at the Subaru Telescope, which is operated 
by the National Astronomical Observatory of Japan.}}

\author{J.\ Davy Kirkpatrick\altaffilmark{a},
Dagny L.\ Looper\altaffilmark{b},
Adam J.\ Burgasser\altaffilmark{k},
Steven D.\ Schurr\altaffilmark{d},
Roc M.\ Cutri\altaffilmark{a}
Michael C.\ Cushing\altaffilmark{l},
Kelle L.\ Cruz\altaffilmark{e},
Anne C.\ Sweet\altaffilmark{n},
Gillian R.\ Knapp\altaffilmark{f}
Travis S.\ Barman\altaffilmark{g},
John J.\ Bochanski\altaffilmark{c},
Thomas L.\ Roellig\altaffilmark{h},
Ian S.\ McLean\altaffilmark{i},
Mark R.\ McGovern\altaffilmark{j},
Emily L.\ Rice\altaffilmark{m}
}

\altaffiltext{a}{Infrared Processing and Analysis Center, MS 100-22, California 
    Institute of Technology, Pasadena, CA 91125; davy@ipac.caltech.edu}
\altaffiltext{b}{Institute for Astronomy, University of Hawai'i, 2680 Woodlawn 
    Drive, Honolulu, HI 96822}
\altaffiltext{c}{Massachusetts Institute of Technology, 77 Massachusetts Avenue, 
    Building 37, Cambridge, MA 02139}
\altaffiltext{d}{Planck Science Center, MS 220-6, California Institute of 
    Technology, Pasadena, CA 91125}
\altaffiltext{e}{Department of Physics and Astronomy, Hunter College, New York, NY 10065}
\altaffiltext{f}{Department of Astrophysical Sciences, Princeton University, Princeton, NJ 08544}
\altaffiltext{g}{Lowell Observatory, 1400 West Mars Hill Road, Flagstaff, AZ 86001}
\altaffiltext{h}{NASA Ames Research Center, MS 245-6, Moffett Field, CA 94035-1000}
\altaffiltext{i}{Department of Physics and Astronomy, UCLA, Los Angeles, CA 90095-1562}
\altaffiltext{j}{Antelope Valley College, Lancaster, CA 93536}
\altaffiltext{k}{Center for Astrophysics and Space Science, University of California, San Diego, CA 92093}
\altaffiltext{l}{Jet Propulsion Laboratory, Pasadena, CA 91109}
\altaffiltext{m}{American Museum of Natural History, New York, NY, 10024}
\altaffiltext{n}{British Consulate, San Francisco, CA 94104}

\begin{abstract}

We have conducted a 4030-square-deg near-infrared 
proper motion survey using multi-epoch data from the Two Micron All-Sky Survey (2MASS). 
We find 2778 proper motion candidates, 647 of which are not listed in SIMBAD. After 
comparison to DSS images, we find that 107 of our proper motion candidates lack counterparts 
at B-, R-, and I-bands and are thus 2MASS-only detections.  
We present results of spectroscopic follow-up of 188 targets that include the infrared-only sources
along with selected optical-counterpart sources with faint reduced proper motions 
or interesting colors. We also establish a set of near-infrared spectroscopic standards
with which to anchor near-infrared classifications for our objects. Among the
discoveries are six young field brown dwarfs, five ``red L'' dwarfs, three L-type 
subdwarfs, twelve M-type subdwarfs, eight ``blue L'' dwarfs, and several T dwarfs.
We further refine the definitions of these exotic classes to aid future identification of
similar objects. We examine their kinematics and find that both the ``blue L'' and ``red L''
dwarfs appear to be drawn from a relatively old population. This survey 
provides a glimpse of the kinds of research that will be possible through time-domain infrared
projects such as the UKIDSS Large Area Survey, various VISTA surveys, and WISE, and also 
through $z$ or $y$-band enabled, multi-epoch surveys such as Pan-STARRS and LSST.

\end{abstract}

\section{Introduction}

During the first quarter of the twentieth century, cataloging the stellar
constituents of the Solar Neighborhood
became a major focus of astronomical research. Although the closest of the bright stars were
identifiable through dedicated trigonometric parallax programs, such
programs weren't feasible for the wealth of dimmer objects. As a result the search for the
faintest of the nearby stars came to rely heavily on the fact that tangential motion of the 
closest objects would be measurable against the sea of non-moving background objects. As a result, 
these proper motion measurements using multi-epoch photographic plates became the primary means of discovery.

The closest star to the Sun, Proxima
Centauri, was discovered as a distant, common proper motion companion of
$\alpha$ Centauri AB by \cite{innes1915}.  \cite{barnard1916} discovered
the second closest system -- most commonly known as Barnard's Star -- via its
proper motion, which is still the highest known for any star (10$\farcs$37/yr, \citealt{benedict1999}).
\cite{wolf1919} used proper motion to uncover the third closest system, known as Wolf 359, and
\cite{ross1926} did the same to reveal the seventh, eighth, and eleventh nearest systems, known as
Ross 154, Ross 248, and Ross 128, respectively. Taking this work to even fainter limits,
\cite{vanbiesbroeck1944,vanbiesbroeck1961} performed a photographic search for companions 
to nearby stars and
revealed two late-M dwarfs, known as van Biesbroeck 8 (vB 8) and van Biesbroeck 10 (vB 10),
that for many years marked the bottom of the known main sequence.
Deeper, larger scale proper motion surveys by \cite{giclas1971,giclas1978} and by 
\cite{luyten1979a,luyten1979b,luyten1980a,luyten1980b} were 
performed in the mid-twentieth century. These Luyten references tabulate the results of all 
previous motion surveys done through the 1970's. These surveys accounted for a total of 
over 58,000 motion stars with $\mu > 
0{\farcs}18$/yr down to the $B$-band plate limit of the Palomar Observatory 
Sky Survey (POSS) and to brighter limits for stars south of the POSS itself.

A major limitation of these previous surveys was their reliance on optical
data, primarily at $B$ and $R$ bands; a lack of deep, dual-epoch
material in the far southern sky; and an inability to probe close to the Galactic
Plane. Recent large-area surveys have attempted to remove some 
of these biases while still using photographic plate material.
Luyten's surveys, which generally required that objects be detectable on both the $B$- 
and $R$-band plates and be located at relatively high galactic latitude, 
have now been supplemented with the LSPM/LSR proper motion survey of
\cite{lepine2002,lepine2003} and \cite{lepine2005,lepine2008}, which uses two-epoch $R$-band data only 
(when two epochs are available) to push to intrinisically redder and fainter objects. 
Similiar investigations, many concentrating more heavily on the less-complete southern sky, have been 
performed by the Wroblewski \& Torres survey (WT; \citealt{wroblewski2001} and references therein), the 
Calan European Southern Observatory survey (Calan-ESO; \citealt{ruiz2001}), the Automated Plate Measuring 
Proper Motion survey 
(APMPM; \citealt{scholz2000}), the SuperCOSMOS Sky Survey Proper Motion survey
(SSSPM; \citealt{scholz2002}), the Liverpool-Edinburgh High Proper Motion survey
(LEHPM; \citealt{pokorny2004}), and the SuperCOSMOS-RECONS survey (SCR; \citealt{finch2007} and references 
therein). 

CCD-based surveys have also begun to have an impact. The SkyMorph database from the Near
Earth Asteroid Tracking program
uncovered a previously unrecognized late-M dwarf moving at 5$\farcs$05/yr that now ranks 
as one of the closest stars to the Sun (\citealt{teegarden2003}). This discovery highlights 
the fact that deeper visible-light
surveys with shorter epoch differences will be able to detect nearby stars with motions too
large to have been detected with prior searches. Surveys by the Panoramic Survey Telescope 
and Rapid Response System (Pan-STARRS; \citealt{kaiser2004}) in the northern hemisphere 
and Large Synoptic 
Survey Telescope (LSST; \citealt{ivezic2008}) and SkyMapper Telescope (Southern
Sky Survey, aka S3; \citealt{keller2007}) in the southern hemisphere will provide
an ultra-deep CCD-based proper motion catalog for the entire sky.

Limiting the search to optical surveys will, of course, render the census
very incomplete for the faintest, coolest targets that have their peak flux at longer
wavelengths. As photometric searches of
the Two Micron All-Sky Survey (2MASS), the Sloan Digital Sky Survey (SDSS) and the
Deep Near-infrared Survey of the Southern Sky (DENIS) have shown, the Solar Neighborhood 
is replete with, 
and perhaps even dominated by, dwarfs cooler than type M.
Most of these objects have been uncovered
using color selections appropriate for cool dwarfs with solar age and
solar metallicity, but this method will miss cool halo objects with very low 
metallicity and any other objects of unusual color. Therefore, proper motion surveys at
these longer wavelengths are needed to search for classes of very cool, nearby objects that may
have heretofore gone unnoticed.

With this point in mind, researchers have begun to perform proper motion surveys beyond 1 $\mu$m.
The Southern Infrared Proper Motion Survey (SIPS; \citealt{deacon2007}) compares $I$-band SuperCOSMOS
data to 2MASS $J$, $H$, and $K_s$ data, although in this case objects are chosen using 2MASS colors
before being paired up with SuperCOSMOS $I$-band data, hence leading to some of the same
biases and incompletenesses noted above. \cite{sheppard2008} have performed a proper motion
search free of these photometric-selection biases by performing a cross-matching of 2MASS ($J$, $H$, $K_s$)
with SDSS
Data Releases 1 through 5 ($u$, $g$, $r$, $i$, $z$). This work uncovered thirty-six objects with $0{\farcs}2/yr < \mu <
1{\farcs}0/yr$ that were not catalogued previously. Six of these have been followed up 
spectroscopically and the results include one mid-T dwarf, a late-M subdwarf, and an
unusually blue L dwarf. Even more recently, \cite{deacon2009} have
matched up data from the United Kingdom Infrared Deep Sky Survey (UKIDSS;
\citealt{lawrence2007}) Data Release 4 with 2MASS to identify 267 low-mass stars and brown dwarfs
via their proper motions; of these, ten known L dwarfs and one known T dwarf were recovered and
nine new L dwarf candidates identified.

We present here a large-area ($\sim$4030 sq.\ deg., or $\sim$10\% of the sky) proper motion
survey where both epochs of coverage lie longward of 1 $\mu$m. By performing a color-free
proper motion search we can achieve several goals: 1) We can determine the location of unusual
objects in color space to enable future photometric searches for these objects. 2) The kinematic
bias present in a proper motion survey will enable us to find very old brown dwarfs, and such
newfound 
discoveries will help to expand the list of observed low-metallicity, low-temperature 
atmospheres and provide invaluable tests of atmospheric theory.  3) We can begin to explore 
the mass function of substellar halo objects, allowing us for the first time to probe the 
efficiency of brown dwarf formation in low-metallicity environments.

\section{Survey Methodology}

\subsection{The 1X-6X Survey (List 1)\label{list1}}

\begin{figure}
\epsscale{0.9}
\plotone{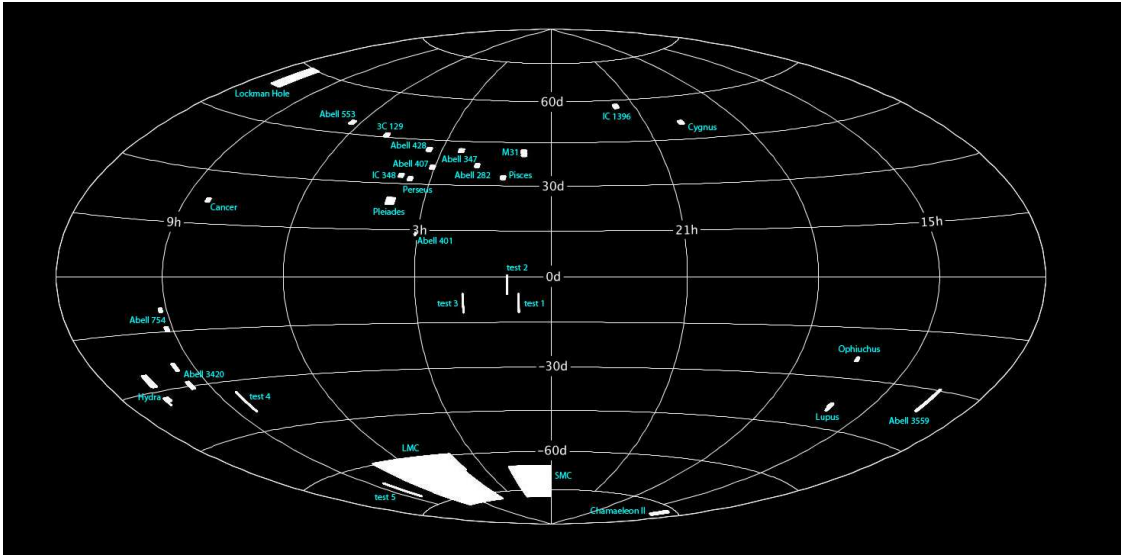}
\caption{2MASS coverage (white area) for the 6X survey shown in a equatorial projection. Each of the targeted fields is labeled with a description in cyan. See the 2MASS Explanatory
Supplement (\citealt{cutri2003}) for more information. \label{6X_coverage}}
\end{figure}

For the first part of our survey we used data from the 2MASS All-Sky 
Point Source Catalog (PSC)
as epoch \#1 and from the 2MASS 6X Point Source Working Database/Catalog as epoch \#2. 
The latter catalog comprises a set of special observations made at the 
end of 2MASS operations and taken with exposures six times longer than the main survey. 
The area covered by the 6X observations and hence
the total area overlapped by both data sets is $\sim$580 sq.\ deg. The epoch
difference is 1-3 years. See Figure~\ref{6X_coverage} for a graphical representation
of the 6X coverage of the sky.

For each of the 137 6X nights, a file was produced during 2MASS 6X data processing
that listed all of the objects in the 
6X scan that failed to match a source within a 0$\farcs$5 search radius in the 2MASS 1X
data. This resulted in 824,383 missing sources. After removing objects that were within
10$\arcsec$ of the 6X scan edges, objects identified as known minor planets, 
objects that were confused with other detections in the scan, and sources of
less than ideal photometric quality (i.e., sources for which none of the 
bands had photometric quality ph\_qual=A), the source list was
reduced to 1,181 sources. All of these were checked for additional problems by looking
at side-by-side images of the field from both 2MASS and the Digital Sky 
Surveys\footnote{See http://archive.stsci.edu/dss/sites.html.}. Problems
identified at this stage were usually false motions induced by source confusion or 
by extended morphology.

After these visual checks, a total of 359 confirmed sources remained, most of which 
are clearly visible
on the DSS plates. 185 (51.5\%) of the sources are not listed in SIMBAD, and 29 
(8.1\%) have no counterpart 
within a 3$\farcs$0 radius in the USNO-B Catalog (\citealt{monet2003}). Of these 
29, eight (2.2\% of the original sample) are 2MASS-only sources
because they have no visible counterpart on the
DSS plates once motion and faint counterparts (such as I-only sources, not included in the USNO-B
Catalog) are accounted for. All eight of these infrared-only discoveries lie in front 
of or in the immediate vicinity of
the Large and Small Magellanic Clouds (LMC, SMC). This is no
surprise since the LMC and SMC were the two largest targets of the 2MASS 6X observing
campaign (see Figure~\ref{6X_coverage}).

\subsection{The 1X-1X Survey}

\subsubsection{Pure Kinematic Selection (List 2)\label{list2}}

The second part of our survey used 1X data only.
Although the 2MASS 1X survey was fundamentally a single-epoch coverage
of the sky, approximately
30\% of the sky was observed more than once in photometric conditions
while the twin telescopes were in operation between 1997 and 2001. Figure 2
shows the sky coverage of the dual-epoch 1X data. Slightly
more than half of this area is in the RA and Dec overlap regions between
adjacent survey tiles.  The remaining area is covered by tiles that were
reobserved to improve minimum data quality, to provide additional
survey validation, or to enable other scientific investigations.
The 2MASS Survey Point and Extended Source Working Databases contain
source information extracted during pipeline data reduction of all
survey observations, and may therefore contain multiple, independent
measurements of sources detected in the regions of the sky observed
more than once.  The 2MASS All-Sky Release Catalogs were constructed
from the Working Databases but used only one measurement
of multiply-detected sources to maintain survey uniformity.

To conduct this proper motion survey we used the 2MASS Survey
Point Source Working Database (which is essentially a union of the 2MASS All-Sky Point
Source Catalog and the 2MASS Survey Point Source Reject Table) and the 2MASS Survey Merged Point Source
Information Table.  The Merged Point Source Information Table contains the averaged positional and
photometric measurements for sources with more than one detection in the Point Source Working 
Database and was created by
positionally autocorrelating
the Point Source Survey Working Database to find ``groups'' of independent source
extractions that fall within 1${\farcs}$5 of each other.  
The mean and RMS position, the mean and RMS 
brightness, and various detection statistics were computed for each 
associated
group of detections and entered into the Merged Source Table.  The merged
source entries were also assigned warning flags if they could not be
unambiguously separated from other nearby groups, if any of their 
constituents
were resolved during pipeline data reduction, or if any of their constituents had measurements that were
possibly contaminated by image artifacts.

We selected proper motion candidates in the Merged Point Source
Information Table by using the IRSA/Gator search engine\footnote{See http://irsa.ipac.caltech.edu.} 
to search for groups
that satisfy the following criteria (conditions in parentheses
after each description refer to the parameters listed in the Merged Point
Source Table):

$\bullet$ Are detected on two or more independent survey scans (sdet$\ge$2);

$\bullet$ Have positional differences of at least $0{\farcs}4$ (0.000111$^\circ$) 
between the first and last detection epoch\footnote{One could extend the positional 
difference to lower values, modulo an increase in the number of false motion sources 
because the positional difference would then be comparable to the single-epoch astrometric error.}
(sep\_jdmax$\ge$0.000111)
and, by default (see above), positional differences of less than $1{\farcs}5$;

$\bullet$ Have average J-band magnitudes brighter than 15.8 mag (j\_mavg$\le$15.8)\footnote{For 
objects with $0.000111 \le sep\_jdmax < 0.000139$ degrees, the magnitude range $15.8 \le J \le 16.0$ mag 
was searched only partially.};

$\bullet$ Are not flagged as confused in the merging process (gcnf=0);

$\bullet$ Contain no extended group members (n\_galcontam=0); and

$\bullet$ Contain no group members with contaminated measurements (ce\_flg=0).

Candidate groups falling within 10 degrees of the Galactic Plane were further
culled from the list, as source confusion there can severely limit the astrometric accuracy.
The apparent proper motion of each candidate group was then computed
by dividing the angular positional difference between the first and last
epoch measurement by the time separating the observations.  We imposed
criteria of $\mu \ge 0{\farcs}2$ yr$^{-1}$ and epoch time differences 
$\ge$ 0.2 yr for all of our proper motion candidates\footnote{The maximum epoch 
difference for any of the 2MASS re-observed regions is about 3 yr.
This means that only those sources with $\mu < 7{\farcs}5/yr$ can be detected for 
cases with the smallest epoch difference, or $\mu < 0{\farcs}5/yr$ for cases with 
the maximum difference. Because the epoch difference varies from spot to spot, it
is not possible to place a fixed value on the maximum proper motion.}. 
The constraint
on epochal difference limits the search to a total area of $\sim$3440 sq.\ deg. 
See Figure 6 of \S A2.1 in the 2MASS Explanatory Supplement (\citealt{cutri2003}).  

Finder charts for each candidate were constructed and examined by
eye.  These charts were comprised of five different images: 2MASS J, H,
and K$_s$ bands; DSS R, B, or V band; and XDSS R band\footnote{The DSS and XDSS
images were pulled from the CADC service. See http://cadcwww.dao.nrc.ca/cadcbin/getdss. DSS
refers to the First Generation Digitized Sky Survey and XDSS refers to the Second Generation
Digitized Sky Survey.}.  
This eliminated spurious sources that clearly showed no motion between the
longer time baseline afforded by the DSS to 2MASS comparison. In the final analysis
we identified 2301 proper motion candidates using these selections, 
384 (16.7\%)  of which
are not listed in SIMBAD.  Of these 2301 proper motion candidates,
225 (9.8\%) have no counterpart 
within a 3$\farcs$0 radius in the USNO-B Catalog. A total of 37 (1.6\% of the 
original sample) are
invisible by-eye on the DSS B-, R-, and I-band images and are thus 2MASS-only sources.

\subsubsection{Kinematic Selection with Imposed Color Criteria (List 3)\label{list3}}

As an adjunct to this second survey, we searched for proper motion
candidates having fainter 2MASS J mags and/or smaller time and positional differences 
between the first and last 2MASS epochs; however, 
we restricted the search to portions of 2MASS color-color space where mid- to late-L
and mid- to late- T dwarfs lie. The common constraints for these color-based surveys were --

$\bullet$ Are not flagged as confused in the merging process (gcnf=0);

$\bullet$ Contain no extended group members (n\_galcontam=0); and

$\bullet$ Contain no group members with contaminated measurements (ce\_flg=0).

A search of L dwarf color space (specifically, $j\_mavg-h\_mavg \ge 0.8$ mag and $h\_mavg-k\_mavg \ge 0.5$ mag)
included the additional constraints $0.000042 \le sep\_jdmax < 0.000111$ deg, $sdet \ge 2$, and epoch
difference $\ge 0.1$ yr. T dwarf color space had two searches performed on it, both
using constraints $j\_mavg \le 16.5$ mag, epoch difference $\ge 0.1$ yr, and total
proper motion $\ge 0{\farcs}2$/yr. The first of these searched color space with
$j\_mavg-h\_mavg \le 0.0$ mag and $h\_mavg-k\_mavg \le 0.4$ mag and applied additional constraints
of $sep\_jdmax \ge 0.000028$ deg and $sdet \ge 2$. The second of these searched T dwarf color
space with $j\_mavg-k\_mavg \le 0.5$ mag and applied additional constraints of 
$sep\_jdmax \ge 0.000111$ deg and $sdet \ge 3$.

Using these criteria, we identified 131 proper motion candidates. Of these, 82
(62.6\%) are not listed in SIMBAD 
and 79 (60.3\%) have no counterpart within a 3$\farcs$0 radius in the USNO-B Catalog.
A total of 62 (47.3\%) are invisible on the
DSS B-, R-, and I-band images and hence are 2MASS-only sources.

\section{List of Proper Motion Candidates}

\subsection{Construction of the Table}

Table~\ref{pm_list} gives composite information on all of these proper motion candidates. Columns 1-4
list general information on each target. Column 1 gives the name of the object as listed 
in the 2MASS All-Sky Survey Point Source Catalog; the name includes the
sexagesimal J2000 position encoded as hhmmss[.]ss$\pm$ddmmss[.]s. Column 2 gives the
list from which the proper motion candidate came; either List 1 (\S~\ref{list1}),
List 2 (\S~\ref{list2}), or List 3 (\S~\ref{list3}), corresponding to the three searches above.
For objects not visible on the B- or R-band plates of the Digitized Sky Surveys, column
3 gives the shortest wavelength band where the object is seen -- I if it visible in
the DSS I-band plates or J if it is seen only in 2MASS. Column 4 lists an asterisk for
those objects we have followed up spectroscopically; follow-up observations for these
are detailed in Table~\ref{spec_followup}.

Columns 5-11 give 2MASS-specific data. Column 5 is the 2MASS-measured, absolute proper motion. For sources
in List 1 this is computed by differencing the positional 
information from the 2MASS All-Sky Point Source Catalog (the earlier epoch) and the
positional information for the same object in the 2MASS 6X Point Source Working 
Database (the later epoch) then dividing 
by the epoch difference; the error in the proper motion is the root sum 
square of the positional uncertainties (available for each source record in both catalogs) 
divided by the epoch difference. For
sources in Lists 2 and 3 the calculations are the same except that the positional difference
comes from the earliest and latest epochs contained in the 2MASS Point Source Working Database.
Column 6 gives the position angle for the proper motion vector, calculated in the standard way
as the number of degrees east of due north.  To gauge the significance of the 
proper motion measurement, we give in column 7 the astrometric displacement divided by the
larger of the astrometric uncertainties in the two epochs. This value can be thought of as the number 
of standard deviations
($\sigma$) that the positional difference lies away from zero motion. Columns 8-10 are the 
default $J$, $H$, and $K_s$ magnitudes and total photometric uncertainties; for List 1 sources 
these are 6X measurements taken from the 2MASS 6X Point Source
Working Database/Catalog, and for List 2 and List 3 sources these are 1X measurements taken from 
the 2MASS All-Sky Point Source Catalog for whichever epoch had the smaller overall 
uncertainties.\footnote{Photometry from epoch 1 would be selected if at least two 
of the following conditions were met in the 2MASS All-Sky Point Source Catalog -- $j\_msigcom$ at epoch 1 $< j\_msigcom$
at epoch 2, $h\_msigcom$ at epoch 1 $< h\_msigcom$ at epoch 2, $k\_msigcom$ at epoch 
1 $< k\_msigcom$ at epoch 2. Otherwise, photometry from epoch 2 is reported instead.} 
Column 7 lists the reduced proper motion at J-band, defined as $H_J$ = J + 5 log($\mu$) + 5, 
computed using the proper motion ($\mu$) and $J$ magnitude reported in columns 5 and 8, respectively.

Additional data for these proper motion candidates were gleaned from other sources.
A 3-arcsec position around each 2MASS source was searched for a corresponding entry in the 
USNO-B1.0 Catalog (\citealt{monet2003}). For sources with USNO-B matches, information 
for the nearest 
such match is listed in columns 12-18. The total USNO-B-measured proper motion, error, and
position angle are listed in columns 12-13. Columns 14-18 give the B1, B2, R1, R2, and I 
magnitudes, respectively. Dual
values of the $B$ and $R$ photometry may exist because the photographic 
survey material covers two epochs of the sky at both of these bands. Errors on the
photographic magnitudes are generally $\sim$0.3 mag (\citealt{monet2003}). 
When doing the positional cross matching, we chose not to proper move the
2MASS PSC values to epoch and equinox J2000.0 because the epoch of the 2MASS
measures is near J2000.0 anyway (2MASS surveyed the sky over the years 1997-2001) 
and the positions in the USNO-B Catalog are
proper moved to epoch and equinox J2000.0 already. As a result, the only 2MASS proper motion
candidates that lack a
counterpart within our 3-arcsec radius are those that are too faint to be
detected in the optical, those that have substantial motions ($>$1$\arcsec$/yr) that
will have moved them beyond our search radius in the $\sim$3 years between the 2MASS
images and the given USNO-B J2000.0 positions, or those with problematic USNO-B
measurements (blends, sources in crowded fields, etc.).

Finally, columns 19-22 report SIMBAD\footnote{See http://simbad.u-strasbg.fr/simbad/.} 
info on each object, current as of mid-2008, if a SIMBAD match was found
within 10 arcseconds of the 2MASS PSC position. This larger match radius
was chosen because positional info in SIMBAD comes from a variety of sources
and could represent an epoch far removed (as much as 50 years) from the 2MASS position.
Column 19 contains the SIMBAD-reported value for proper motion and error, and column 20
contains the computed position angle of the motion vector. Column 21 gives the name of
the object as listed in SIMBAD, and column 22 gives the SIMBAD-reported spectral 
type.

The total number of unique proper motion candidates resulting from the searches above is
2778. Thirteen objects -- 2MASS J01153252$-$7519079, 2MASS J01182974$-$7504544, 
2MASS J03431558+2354453, 2MASS J04112809$-$6859167, 2MASS J04212447$-$5954026, 
2MASS J05205071$-$7755133, 2MASS J05254550$-$7425263, 2MASS J06195065$-$5952247,
2MASS J07025026$-$6102482, 2MASS J13271966$-$3110394, 2MASS J13273959$-$3551009,
2MASS J13375084$-$3549174, and 2MASS J13275892$-$3327329 -- were discovered in both 
the 1X-6X and the 1X-1X surveys. Of the 2778 unique objects, 647 lacked SIMBAD
entries, 332 were not found within 3$\farcs$0 of any source in the USNO-B Catalog,
and 107 were 2MASS-only detections.

\subsection{Diagnostic Plots}

\subsubsection{Judging the Robustness of the 2MASS Measures}

We can use the tabulated data in Table~\ref{pm_list} to judge the credibility of our
2MASS-measured motions. In Figure~\ref{SIMBADpm_vs_2MASSpm} we compare the 2MASS
motions to independent measurements from SIMBAD and find an excellent correlation
for those objects whose motions are the most significant ($>5\sigma$, top panel).
The correlation is not as tight for objects having 2MASS motions
of smaller significance (bottom two panels), as one would expect. 

Despite the excellent correlation in the top panel, it is important to understand 
why a few outliers still exist. One of these, 2MASS J19570113$-$5917285 has a 
2MASS-measured motion of $\mu=0{\farcs}79{\pm}0{\farcs}12$/yr, whereas the SIMBAD value is 
$\mu=0{\farcs}23$/yr. Another outlier, 2MASS J02532235+0608003, has a 2MASS-measured motion of 
$\mu=0{\farcs}86{\pm}0{\farcs}15$/yr, whereas the SIMBAD value is $\mu=0{\farcs}36$/yr. 

Measuring the motion of 2MASS J1957$-$5917 using the 20-year baseline between the DSS images and 2MASS, we find
a motion of $\mu{\sim}0{\farcs}20$/yr, in agreement with the SIMBAD value. Hence, the 2MASS-measured motion
appears to be in error. 2MASS J1957$-$5917 is fairly bright (7.0-8.0 mag in $JHK_s$) and
stars in this brightness range are faint
in the 2MASS 1X 51-millisec Read1 exposures but nearly saturated in the 1.3-sec Read2-Read1 exposures. For
such objects, the astrometry can be less reliable than for stars just brighter (which
have higher SNR and are more robustly measured on the Read1 frames) or just 
dimmer (which are farther from saturation and easier to centroid in the Read2-Read1 frames). Thus,
the 2MASS astrometry needs to be carefully scrutinized for objects in this near-infrared brightness range.

In the case of 2MASS J0253+0608, positional measures using the 45-year baseline 
between the earliest DSS images and 2MASS suggest a motion of $\mu{\sim}0{\farcs}37$/yr, 
also in agreement with the SIMBAD value. The two 2MASS imaging epochs are separated by
only nine months, which means that an additional positional difference of ${\sim}0{\farcs}4$
is being introduced between the 2MASS images. In this case, a study of a few bright 2MASS
stars in the vicinity of this source show a ${\sim}0{\farcs}3$ difference between 
the reconstructed positions of the two epochs. 2MASS J0253+0608 is located near the end of a
six-degree-long 2MASS tile, and it appears that the astrometric solution for one of the
two scans (scan 037 on 991217n) wandered off slightly, creating a ${\sim}0{\farcs}3$ systematic error at that epoch.
It is only because this particular object
shows appreciable, real motion between the DSS and 2MASS images that it survived our by-eye
checks.

Figure~\ref{USNOBpm_vs_2MASSpm} shows the proper motion from the USNO-B Catalog
plotted against the 2MASS-measured value. There is more scatter in the correlations
here than in Figure~\ref{SIMBADpm_vs_2MASSpm} but this is to be expected since the USNO-B
motions are measured in bulk and those from SIMBAD are generally well vetted. In general
the discrepancies here are caused by spurious measurements in the USNO-B.
One example of this, from the upper
panel of the figure, is 2MASS J06105984$-$6512200 (LHS 1832) whose 2MASS motion of 
$\mu=0{\farcs}85{\pm}0{\farcs}07$/yr and USNO-B motion of $\mu=0{\farcs}144{\pm}0{\farcs}054$/yr 
are in clear disagreement. SIMBAD lists this object as having $\mu=0{\farcs}78{\pm}0{\farcs}01$/yr,
agreeing within one sigma of the 2MASS measurement. The USNO-B measurement is presumably 
the result of an incorrect
association across epochs because this is a very crowded field in the outskirts of the
Large Magellanic Cloud. 

A similar situation holds true for the star 2MASS J01132818$-$7056507,
whose discrepancy is obvious on the middle panel of Figure~\ref{USNOBpm_vs_2MASSpm}. This is
the brighter component of the common proper motion pair CCDM J01135-7057AB, which lies in
another crowded field, this time in the outskirts of the Small Magellanic Cloud. 2MASS
measures identical motions of $\mu=0{\farcs}28{\pm}0{\farcs}06$/yr for the two objects, whereas USNO-B reports
non-identical and much higher values of $\mu=0{\farcs}962{\pm}0{\farcs}097$/yr and
$\mu=0{\farcs}541{\pm}0{\farcs}092$/yr for the A and B components, respectively. Comparison
of the DSS and 2MASS images, which represent a 23-year baseline, confirm the 2MASS measures.

Figure~\ref{SIMBADpm_vs_2MASSpm_angles} and Figure~\ref{USNOBpm_vs_2MASSpm_angles}
show the correlation between measurements of the position angle. As with
the two previous figures, these show an excellent correlation for those objects
with 2MASS motions $>5{\sigma}$, with the USNO-B agreement again being less tight than that
of SIMBAD. The top panel of Figure~\ref{SIMBADpm_vs_2MASSpm_angles} shows only one
highly discrepant point, that of 2MASS J02192807$-$1938416, a previously known L1 dwarf whose
2MASS measurement of $\theta=121^o$ disagrees with the SIMBAD value of $\theta=352^o$,
which itself appears to originate from the USNO-B Catalog. A comparison of the DSS R- and I-band
images with 2MASS images shows that the 2MASS value appears
to be correct, the USNO-B value having been calculated from optical data with lower
SNR.

These plots show that the 2MASS-measured proper motions agree well with published values.
Nonetheless, like any other catalog derived from bulk-measured motions, some problems
persist so users are advised to use caution and to perform their own quality assurance
of the results before proceeding with analysis or follow-up.

\subsubsection{Selecting the Choicest Sources}

On average, objects closer to the Sun should have larger proper motions than those 
more distant, so proper motion should be proportional to trigonometric parallax. With this in mind, \cite{luyten1925}
substituted proper motion, $\mu$, for parallax, $\pi$, in the equation for distance
modulus and invented a quantity, $H$, which he referred to as the reduced proper motion.
This is defined by $H_m = m + 5log({\mu}) - 5$, where $m$ is a particular photometric band such as 2MASS J.
The quantity $H$ can be used to create a plot much like an HR diagram, which can then
be used to select potentially nearby stars for further follow-up. However, stars of
higher than average kinematics will have artificially faint values of $H$, making them
also stand out on such a diagram. Thus, reduced proper motion is also a valuable quantity
for selecting potentially old objects for further follow-up.
Figure~\ref{red_motion_vs_JK} shows $H_J$, the reduced proper motion at J-band, plotted against
$J-K_s$ color for objects in the 2MASS proper motion sample.

The $H_J$ vs.\ $J-K_s$ diagram has several distinctive features. First, main sequence stars
earlier than late-K have $J-K_s < 0.8$ and small values of $H_J$ whereas dwarfs later than
mid-M have $J-K_s > 0.9$ and larger values of $H_J$. Dwarfs between late-K and mid-M have
nearly identical colors ($0.8 < J-K_s < 0.9$) and fall at a wide range of $H_J$ bridging
the earlier dwarfs in the upper left and the late-M and L dwarfs in the lower right. 
White dwarfs fall in the lower left portion because they tend to have blue near-infrared colors
but are intrinsically faint for their motion (large $H_J$). T dwarfs, which have a wide range
of near-infrared color ($-1 < J-K_s < 2$) but are intrinsically faint (large $H_J$), span the 
entire lower portion of the diagram. Subdwarfs tend to be bluer in near-infrared
color than their higher metallicity counterparts and because they are older tend to have
higher motions as well. This pushes them down and to the left relative to normal field objects.
The reason that the concentration of objects found with $0.6 < J-K_s < 0.9$ tends to show a
{\it blueward} trend with increasing $H_J$ is because M subdwarfs dominate in the blueward extension
stretching to larger $H_J$ values.

We have used Figure~\ref{red_motion_vs_JK} to select
subdwarf candidates and nearby, late-type candidates for spectroscopic study. 
Color-color diagrams can also be used as further diagnostics for candidate
selection. One example is shown in Figure~\ref{JH_vs_HK}, which plots 2MASS  $J-H$ vs.\ $H-K_s$ color, useful
for selecting L dwarfs (upper right quadrant of each panel) or mid- to late-T dwarfs 
and white dwarfs (lower left quadrant of each panel). 
The diagnostic ability of these reduced proper motion and color-color plots will be
discussed further after discussion of the spectroscopic follow-up below.

\section{Spectroscopic Observations}

Spectroscopic observations of 189 targets were obtained using eleven different 
optical/near-infrared intruments at eight different observatories. Details of this
follow-up are given in Table~\ref{spec_followup}. This table gives the object's 2MASS PSC designation
in column 1, along with telescope and instrument used, the UT date of observation, 
weather conditions, and integration time in columns 2-6. The spectral classifications,
which are discussed in detail later, are given in columns 7 or 8.
In addition, supporting calibration observations of a suite of M and L dwarfs and subdwarfs were 
also acquired, as listed in Table~\ref{sd_spectra}. Details of the observing set-ups and reductions 
are given below.

\subsection{Optical}

\subsubsection{Magellan/LDSS3}

We obtained optical spectroscopy for sources in the southern hemisphere using the 
Low Dispersion Survey Spectrograph (LDSS-3) mounted on the Magellan 6.5 m Clay
Telescope. LDSS-3 is an imaging spectrograph, upgraded by M.\ Gladders from the 
original LDSS-2 (\citealt{allington-smith1994}) for improved red sensitivity. 
We employed the VPH-red grism (660 line/mm) with a 0$\farcs$75-wide longslit to 
obtain 6050 -- 10500 \AA\ spectra across the entire chip with an average resolution of
4.5 \AA. The OG590 longpass filter was used to eliminate second-order light 
shortward of 5900 \AA. Two long exposures were obtained for each target, followed 
immediately by a series of helium/neon/argon arc lamp and flat-field quartz lamp
exposures. 

Data were reduced in the IRAF\footnote{IRAF 
is distributed by the National Optical
Astronomy Observatories, which are operated by the Association of
Universities for Research in Astronomy, Inc., under cooperative
agreement with the National Science Foundation.} environment. Raw science images were trimmed and 
subtracted by a median combined set of bias frames. 
The resulting images were divided by the corresponding normalized, median-combined, 
and bias-subtracted set of flat-field frames. Spectra were then extracted using 
the {\it apall} task with background subtraction but without variance weighting (i.e., 
not ``optimal extraction''). The dispersion solution for each target was determined
 using the tasks {\it refspec}, {\it identify}, and {\it dispcor}, and arc lamp spectra extracted 
using the same dispersion trace; solutions were typically accurate to 0.05 pixels, 
or 0.07 \AA. Flux calibration was determined using the tasks {\it standard} and {\it sensfunc} 
with observations of the spectral standards from \cite{hamuy1994}. 

\subsubsection{Magellan/MagE}

The Magellan Echellette (MagE) Spectrograph (\citealt{marshall2008})
on the Magellan Clay Telescope was also used for follow-up. MagE is a cross--dispersed,
medium-resolution ($R \sim 4,100$) optical spectrograph, spanning
3000 to 10500 \AA. All observations employed the
$0.7\arcsec$ slit, corresponding to a dispersion of $\sim
22$ km s$^{-1}$. The instrument was aligned with the parallactic
angle to avoid atmospheric dispersion.  Each observation was followed
by a Thorium-Argon arc lamp to provide wavelength calibration.  

The data were reduced using a preliminary version of the MagE Spectral
Extractor (MASE; \citealt{bochanski2009}).  MASE is an
IDL data reduction pipeline and GUI, which contains the entire
reduction and calibration process.  Briefly, each frame is bias
corrected and flat fielded.  Wavelength calibration is determined from
ThAr observations and followed by sky subtraction and extraction.  The
pipeline employs simultaneous B-spline fitting
(\citealt{kelson2003}) to both the background and object to
measure the spectrum.  Finally, the orders are combined and a relative
flux calibration is applied. 

\subsubsection{Gemini-North/GMOS and Gemini-South/GMOS}

The Gemini Multi-object Spectrometer (GMOS; \citealt{hook2004}) on the Gemini-North 
and Gemini-South Telescopes was used during
queue observations taken as programs GS-2004B-Q-12, GN-2006A-Q-70, 
GS-2006B-Q-59, and GN-2006B-Q-76. Spectra of known late-type subdwarfs were also taken
for comparison under program GN-2004B-Q-15 (see Table~\ref{sd_spectra}). Observations 
were made using the RG610 order-blocking
filter and 400 line/mm grating blazed at 7640 \AA. In both cases data cover
the wavelength range from 6000 -- 10000 \AA. Two consecutive observations, with
the central wavelength offset slightly between exposures, were taken of each
target to provide complete wavelength coverage despite the array gap. For the
2004B programs the array was read out in a traditional mode whereas in subsequent
semesters the array was read out in nod-and-shuffle mode, the latter providing
improved sky subtraction at longer wavelengths where the OH telluric lines are
more plentiful. In all cases a 0$\farcs$75-wide
slit was used and provided a resolution of $\sim$5.5 \AA.

The GMOS package under the IRAF Gemini tools was used to reduce the data. 
Specific details on our reduction procedures, both for traditional and
nod-and-shuffle GMOS data are given 
elsewhere\footnote{See http://KelleCruz.com.},
but in summary the dark frames were subtracted and data flat fielded using the
{\it gsreduce} task, then sky lines were subtracted using {\it gnsskysub}.
All spectra were extracted using {\it gsextract}.
Standard stars (e.g., LTT 1020 and Feige 34) were taken with the same instrumental setup and were
used to flux calibrate using {\it calibrate}. Two objects, 2MASS 
proper motion candidate 2MASS J06453153$-$6646120 along with the known subdwarf 2MASS J05325346+8246465,
were taken concurrently with the G0 dwarfs CPD$-$65 772 and BD+82 165, respectively. This was 
done to correct the data for
telluric absorption by interpolating across the telluric features in the G dwarf,
multiplying the uncorrected G dwarf spectrum by the corrected version, then multiplying 
the result against the target spectrum.

\subsubsection{Subaru/FOCAS}

Other optical spectroscopic observations were carried out on 
20 and 21 Aug 2007 (UT) at the Subaru Telescope on Mauna Kea, 
Hawai'i, using the Faint Object Camera And Spectrograph (FOCAS; 
\citealt{kashikawa2002}). FOCAS was used with the 
300R grating blazed at 7500 \AA~and the S058 filter to block 
second-order light from wavelengths shorter than 5800 \AA. The 
grating tilt was set so that the wavelength region from 5850 to 
10250 \AA~was covered. Use of a long slit of width $0{\farcs}8$ 
resulted in a resolution of 8.5 \AA. Both nights were clear but 
the seeing was variable over the course of both nights. On 20 
Aug 2007 the seeing varied from $0{\farcs}5$ to $2{\farcs}5$. 
On 21 Aug 2007 the seeing was somewhat more stable, ranging only 
from $0{\farcs}6$ to $1{\farcs}5$. Subaru employs an atmospheric 
dispersion corrector, so keeping the slit aligned with the 
parallactic angle was not necessary.

The data were reduced and calibrated using standard IRAF routines. 
The overscan region of the array was used 
for bias subtraction, and a median of five dome flats taken on the 
first night were used to normalize the response of the detector. 
Because of strong, broad spectral signatures in the flat field 
lamps themselves, special care was needed during the flat field 
correction step. The FOCAS chip is designed to be used with multi-object 
spectroscopy, but for our single-slit observations most of the detector 
area is not used or needed for sky subtraction. This allowed us to 
perform a block average (in the spatial dimension) on a much smaller 
21$\times$3260 pixel region of the median dome flat corresponding to 
the area where the primary target spectrum and sky subtraction region 
would fall in subsequent exposures. This block-averaged,
one-dimensional, vertical slice encompasses the gross undulations of 
the dome flat in the spectral dimension, and we replicated this slice 
across the spatial dimension to create a two-dimensional image. This 
image was then divided back into the median flat to produce a map 
that contains only small-scale flat-field variations. This map was 
then normalized using the IRAF routine {\it imsurfit} and divided into all other data frames 
to remove the flat field signature across the detector.

The individual stellar spectra were then extracted using the {\it apall}
package. Wavelength calibration was achieved using arc lamps taken 
after each program object. Spectra were flux calibrated using an 
observation of LTT 9491 (\citealt{hamuy1994}) taken on the first 
night. We checked this fluxing using another standard, Wolf 1346 
(\citealt{massey1990}), taken on the second night and the 
agreement was excellent. The G0 dwarfs SAO 103565, SAO 10140, and
HD 207827 were acquired after the targets 2MASS J18212815+1414010, 
2MASS J21265916+7617440, and 2MASS J21481633+4003594, respectively
to remove the telluric bands of O$_2$ at 6867--7000 (the Fraunhofer B band) 
and 7594--7685 \AA\ (the Fraunhofer A band) as well as H$_2$O at 
7186--7273, 8162--8282, and $\sim$8950--9650 \AA.

\subsubsection{Keck/LRIS}

Additional optical spectra were obtained with the Keck Low Resolution 
Imaging Spectrometer (LRIS, \citealt{oke1995}) on the 10-m W.~M.~Keck 
Observatory atop Mauna Kea, Hawaii. A 400 lines/mm grating
blazed at 8500 \AA\ was used with a 1$\arcsec$ slit and 2048$\times$2048
CCD to produce 10-{\AA}-resolution spectra covering the range
6300 -- 10100 \AA.  The OG570 order-blocking filter was used
to eliminate second-order light.  The data were reduced and calibrated 
using standard IRAF routines. Flat-field exposures of the interior of the 
telescope dome were used to normalize the response of the detector. 

Individual stellar spectra were extracted
using the {\it apextract} routine in IRAF, allowing for the slight curvature of
a point-source spectrum viewed through the LRIS optics and using a template
where necessary. Wavelength calibration was achieved using neon+argon
arc lamp exposures taken after each program object. Finally,
the spectra were flux-calibrated using observations of 
standards from \citet{hamuy1994}. 

\subsubsection{Keck/DEIMOS}

Other comparison spectra (Table~\ref{sd_spectra}) were obtained on 2004 Jul 05 (UT) using the Deep
Imaging Multi-object Spectrograph (DEIMOS; \citealt{faber2003}) at the 10m 
Keck-II telescope. The instrument was used in single-object mode utilizing a 
0$\farcs$7-wide longslit. The 600 line/mm grating blazed at 7500 \AA\ was used along with a GG495
order-blocking filter to provide continuous wavelength coverage from 4950 --
10250 \AA\ at a resolution of 3.5\AA. Quartz lamps were used for flat
fielding the detectors and a neon/argon/krypton lamp was used for wavelength
calibration. The calibrator star Feige 56 (\citealt{hamuy1994}) was used for
flux calibration. Spectra were not corrected for telluric absorption. Because 
DEIMOS was used in single-object mode, only two of the eight arrays in the
mosaic contained science data. Standard procedures identical to those for the 
LRIS data above were used for data reduction.

\subsection{Infrared}

\subsubsection{IRTF/SpeX}

Most of our near-infrared spectra  
were obtained from the 3.0-m NASA Infrared Telescope Facility using SpeX 
(\citealt{rayner2003}) in prism mode.
Prism mode provided
continuous coverage from 0.7$-$2.5 $\mu$m in a single order on the
1024$\times$1024 InSb array.  Use of the $0{\farcs}5$ slit resulted in a 
resolving power of R${\approx}100$ in $J$-band
to $\approx$300 in $K$-band.  For
accurate sky subtraction, targets were observed in four  
nodded pairs generally having 120-second or 180-second integrations per position. Stars
of spectral type A0 were observed immediately prior or after each target and at a
similar airmass to provide flux calibration. Calibration flat fields were also acquired
along with argon arc lamp exposures for wavelength calibration.
Standard reductions were employed using the Spextool package, version
3.2 (\citealt{cushing2004,vacca2003}).  




\subsubsection{Keck/NIRSPEC}

Additional spectra were obtained with the Near-Infrared
Spectrometer (NIRSPEC, \citealt{mclean1998,mclean2000}) on the 10-m 
W.~M.~Keck Observatory.  All objects were observed in the N3 configuration (see \citealt{mclean2003})
that, at low-resolution, covers
the portion of the $J$-band window from 1.14 to 1.36 $\mu$m where the 
most diagnostic NIR features lie. Use of the $0{\farcs}38$
slit results in a resolving power of 
R~$\equiv~{\lambda}/{\Delta}{\lambda}~{\approx}~2500$.  

2MASS J21481633+4003594 was observed not only with the N3 configuration but also
with the N1 and N2 configurations to provide full coverage from 0.95 to 1.36 $\mu$m. Two other targets --
2MASS J17561080+2815238 and 2MASS J21265916+7617440 -- were observed 
in N1, N2, N3, N4, N6a, and N6b
to cover the entire wavelength range from
0.94 to 2.35 $\mu$m. More information on the specifics of these instrument
configurations can be found in \cite{mclean2003}.

Data were obtained in two sets of 
dithered pairs, with typically a 300-second exposure obtained at each position. 
To measure telluric absorption and to calibrate the flux levels, dwarfs of type A0 were 
observed near in time and airmass to the target objects. Other calibrations 
consisted of neon and argon arc lamp spectra, a dark frame, and a
spectrum of a flat-field lamp.  We employed standard reductions using 
the IDL package, REDSPEC, as described in detail in \cite{mclean2003}.

For the three objects with multiple wavelength coverages, spectra were 
stitched together using the overlaps between pieces. The N4 and N6a spectra
almost touch but don't overlap in wavelength so a best guess was made to the normalization
between the pieces. Synthetic photometry was produced for the final, stitched spectra
and this photometry compared to actual values obtained in the 2MASS All-Sky
Point Source Catalog. (Filter profiles for the 2MASS filters can be found
in sction VI.4.a of the 2MASS Explanatory Supplement.) The program {\it sbands}
in IRAF was used to derive the fluxes in each bandpass. A comparison of
the synthetic colors to the 2MASS $J-H$, $H-K_s$, and $J-K_s$ colors
showed excellent agreement. All synthetic colors were typically within the $1\sigma$ 
limits of the 2MASS measures, except for the $J-H$ color of 2MASS J21265916+7617440 
(which was still within $2\sigma$).

\subsubsection{SOAR/OSIRIS}

2MASS J0645$-$6646 was observed with the Ohio State Infrared Imager/Spectrometer
(OSIRIS) mounted at the 4.1m Southern Astrophysical Research Telescope
(SOAR) located at Cerro Pachon, Chile. We used the 1$\farcs$0-wide slit which provides a 
resolving power of $R \approx 1400$ across the 1.18-2.35 $\mu$m wavelength 
range in three spectral orders. A series of nine 180-s integrations
were obtained along the 24$\arcsec$-long slit. The A0 dwarf HIP 33717 was
also observed to correct for absorption due to the Earth's atmosphere and
to flux calibrate the final spectrum. The data were reduced using a
modified version of the SpeXtool data reduction package (see IRTF/SpeX 
discussion, above). Wavelength calibration was accomplished using the 
OH airglow lines in the science frames. The spectra were then corrected
for telluric absorption using the observations of the A0 V standard
star via the method described in \cite{vacca2003}. The three spectral
orders were then merged into a single spectrum covering the entire
wavelength range.

\subsubsection{Spitzer/IRS}

Four targets -- 2MASS J11061191+2754215, 2MASS J18212815+1414010, 2MASS J18284076+1229207, and
2MASS J21481633+4003594 -- were observed using the Short-Low (SL) module of the  
Infrared Spectrograph (IRS; \citealt{houck2004}) onboard the \textit 
{Spitzer Space Telescope}. These were observed as \textit{Spitzer} AORs 16201728, 16176128,
16202240, and 16201984, respectively.  The SL module covers  
5.2$-$15.3 $\mu$m at R$\approx$ 90 in two orders.  
The observations consisted of a series of  
exposures taken at two different positions along each slit.  The  
total exposure time for each target was 3904 sec, except for 2MASS J1828+1229 which 
was observed for only 2562 sec.  The raw data were  
processed with the IRS pipeline (version S13) at the \textit{Spitzer}  
Science Center and were further reduced as described in \cite{cushing2006}.  
Briefly, the spectra were extracted with a fixed-width  
aperture of 6$''$ and observations of $\alpha$ Lac obtained as part  
of the IRS calibration observations were used to remove the  
instrument response function and to flux calibrate.  

\section{Spectroscopic Classification}

The spectroscopic data discussed above have been used to classify 
each object in the optical and/or near-infrared,
depending upon the wavelength coverage. Although on average the optical
and near-infrared types should agree, there will be cases where they
are different. This should not be regarded as a problem with 
either classification. The two wavelength regimes
sample different levels in the atmosphere, so a difference in spectral
types provides important clues about the underlying physics. Differing spectral
types may also indicate the presence of binarity.
Objects with discrepant types are discussed further in \S~\ref{worthy-section}.

\subsection{Optical Types: Comparing to Pre-existing Anchors}

We have typed the LDSS3, MagE, GMOS, FOCAS, LRIS, and DEIMOS spectra as follows.
Each spectrum has been normalized to one at 8250 \AA\ and overplotted on a suite 
of like-normalized LRIS spectra of primary M, L, and T optical spectral standards 
from \cite{kirkpatrick1991},
\cite{kirkpatrick1999}, and \cite{burgasser2003opt}.
These plots have then been examined by eye to determine the best match and to look
for any peculiarities with respect to the standard sequence. Objects falling midway between 
integral classes (such as L5 and L6) were assigned the half class inbetween (in this case,
L5.5). Objects showing notable peculiarities were given a suffix of ``pec''. If
the peculiarities were determined to be only slight, the suffix ``sl. pec'' was
used instead. For a few of these, the underlying physical reason appears to be
low gravity, so those are also labeled with either a ``(low-g)'' or a $\beta$/
$\gamma$/$\delta$ (\citealt{cruz2009}, Kirkpatrick et al., in prep.) suffix as further
discussed in \S~\ref{lowg-section}. Other peculiar spectra
were later determined to be low metallicity objects and were
reclassified against a suite of subdwarf standards. These objects, given a prefix 
of ``sd'', are described further in \S~\ref{subdwarf-section}.

\subsection{Near-infrared Types: Building a Standard Sequence}

Spectral classification in the near-infrared should be accomplished the same
way as in the optical. Surprisingly, however, a set of near-infrared M and
L dwarf anchors has never been established although those for T dwarfs
have been (\citealt{burgasser2006}). Previous near-infrared classifications,
such as those in \cite{reid2001}, \cite{geballe2002}, and \cite{mclean2003},
have relied on comparing a target
object's spectral ratios with those same ratios derived for an object of
known optical type. These optically classified comparison objects, however, might not
necessarily create a smooth near-infared spectral sequence across all of the
J, H, K spectral windows (1.0-2.5 $\mu$m) since that was not a requirement.
Now that a very large number of spectra have been acquired over these wavelengths
(e.g., the suite of SpeX prism data available at http://www.browndwarfs.org/spexprism 
as well as the set of SpeX spectra acquired here), it is now possible to create a
grid of spectra whose near-infrared features change smoothly and monotonically
as a function of type. This spectral morphological sequence was built to satisfy the
following properties:

\noindent (1) The near-infrared spectral standards should have near-infrared types
identical or very close to the types assigned to them in the optical. In this way
it can be assured that a random M or L dwarf will, on average, have identical or
very similar optical and near-infrared classifications. Primary standards from the
optical classification sequence have been reused here if they satisfy the
other criteria below.

\noindent (2) The near-infrared color of each primary standard should 
be as close as possible to the median $J-K_s$ color of other objects with that same
optical classification. Median colors as a function of optical type are given in Table 3 of 
\cite{kirkpatrick1994} for early- to mid-M dwarfs and Figure 14 of 
\cite{kirkpatrick2008} for late-M and L dwarfs. The goal of this criterion
is to assure that each near-infrared standard describes the average color 
(and hence, presumably, overall morphological shape over the J, H, and K bandpasses)
of objects in that class.

\noindent (3) The resulting spectral sequence should be smoothly continuous
in morphology. That is, each primary standard should be analagous to a frame in a
movie, where each frame is only subtly different from the frames (standards) on either side.

In cases where multiple objects at a single class met the above three 
requirements, objects were given preference if they were relatively bright or 
if they fell within 30 degrees of the celestial equator. The resulting M0-L9 near-infrared
primary standards are listed in Table~\ref{nir_stds}, and the full M0-T8
near-infrared spectral sequence is illustrated in 
Figure~\ref{NIR_M_stds}, Figure~\ref{NIR_L_stds}, and Figure~\ref{NIR_T_stds}. It should be 
noted that no suitable near-infrared
L1 standard was found with an optical type of L1, so an object of optical type L1.5
was used instead. Likewise, at near-infrared type L7 an object of optical type L6 was used.
The optical spectral sequence does not include an L9 class -- in fact, there is a
rather large jump in morphology between known optical L8 dwarfs and the optical T0
standard (\citealt{kirkpatrick2008}) -- so an optical L8 that bridged the morphological
gap between the near-infrared L8 standard and the near-infrared T0 standard of
\cite{burgasser2006} was used in its place.

With this suite of standards in hand, near-infrared spectral classifications for
the SpeX, 
NIRSPEC, and OSIRIS data were
performed as follows. Each target spectrum was normalized to one at 1.28 $\mu$m and
compared to the near-infrared spectral standards from Figure~\ref{NIR_M_stds}, Figure~\ref{NIR_L_stds}, 
and Figure~\ref{NIR_T_stds}
normalized the same way. The core near-infrared type was determined {\it only}
from the 0.9-1.4 $\mu$m portion. With that best match in hand, the corresponding goodness of
fit to the same spectral standard from 1.4-2.5 $\mu$m was judged. In most cases the same
spectral standard also provided the best fit in this region. In other cases, the target
spectrum was notably much bluer or redder so the fit across the H and K
windows was very poor despite the excellent fit in the J window. These objects were
given suffixes of ``blue'' or ``red'' (or for ones only slightly discrepant, ``sl. blue'' or
``sl. red'') to denote the slope of the spectrum relative to the standard. Other spectra
simply had no good match even in the 0.9-1.4 $\mu$m window, so those were given a suffix
of ``pec''. Some of these were deemed to be either low-gravity (``low-g'') or low-metallicity
(prefix of ``sd''), as discussed further in the sections that follow.

Interesting objects and comparable, previously published objects of similar type
are discussed below and tabulated in Tables \ref{lowg_table} through \ref{other_table}. For our noteworthy
discoveries having high-quality spectra with $R \ge 1000$, we also derive radial velocities then
derive distance estimates and tangential velocities to compute total space motions in Table
\ref{RV_table}. Radial velocities are derived using standard cross-correlation to objects of
known radial velocity and similar spectral type. Because of the relativey low-resolution
of our spectra, these radial velocity measurements typically have large uncertainties, 
which we conservatively overestimate to be $\sim$50 km/s.

\section{Objects Worthy of Special Note\label{worthy-section}}

\subsection{Young (Low-gravity) Objects\label{lowg-section}}

Some of the peculiar late-M and L dwarf spectra were recognized as being unusual
because they have much lower gravity than normal field dwarfs. Low gravity in these
objects is the observational by-product 
of youth. These objects are less massive than an older dwarf of the same temperature
(i.e., at the same mass, younger objects have had less time to cool) and have more distended 
atmospheres because they have not yet completed their gravitational contraction.
Both effects contribute to the lowering of the gravity relative to a collapsed, more massive
object of similar temperature.

The first late-M dwarfs exhibiting the telltale signs of low-gravity were identified over 
a decade ago in young
clusters such as the Pleiades (\citealt{steele1995}), the $\rho$ Ophiuchi star formation 
region (\citealt{luhman1997}), and the Taurus-Auriga complex (\citealt{luhman1998}).
Many other low-gravity late-M dwarfs are now recognized both in clusters as well as
in loose, young associations.
The first L dwarf recognized as young through its low-gravity features was the field 
L0 dwarf 2MASS J01415823$-$4633574 (\citealt{kirkpatrick2006}). This object, predicted
to be a member of either the 12 Myr old $\beta$ Pictoris moving group or the 30 Myr old
Tucana-Horologium association from its sky location and low-gravity signatures, now has a 
proper motion measurement by \cite{faherty2009} that is statistically consistent with
membership in Tucana-Horologium. This leads to a
predicted distance of 44$\pm$8 pc and radial velocity of $\sim$+8 km/s (Mamajek, priv.\ comm.),
both of which await verification. Many other
low-gravity field M and L dwarfs have now been recognized because of their low-gravity
signatures (\citealt{kirkpatrick2008,cruz2009}), and these are now the targets of dedicated astrometric
and radial velocity programs that will establish the $UVW$ space motions needed to confirm
membership in young moving groups.

It has been shown that 7.6\%$\pm$1.6\% of field L dwarfs 
exhibit signs of lower gravity and are presumably younger than 100 Myr, in agreement with
theoretical expectations (\citealt{kirkpatrick2008}).
However, proper motion studies are biased toward objects with halo kinematics, so one would not expect
a significant contribution by young objects. Indeed, all six of our young 
discoveries below have measured proper motions that are within one or two sigma of zero and were
primarily targetted for follow-up due to their red colors. 

We identify six objects, 
listed in Table~\ref{lowg_table}, that show the hallmarks of lower gravity. Optical
spectra for all of these are shown in Figure~\ref{oddones_lowg_opt1} and 
Figure~\ref{oddones_lowg_opt2}, and near-infrared spectra
of three are shown in Figure~\ref{oddones_lowg_IR}. As has been true in other studies 
(\citealt{kirkpatrick2008,cruz2009}), these young field objects tend to be found in the 
southern sky where, not coincidentally, most of the young associations are concentrated.
For optical spectral types we use the Greek suffix designations of \cite{cruz2009} in 
which $\alpha$ (usually dropped for convenience) is used to denote objects of normal gravity,
$\beta$ for intermediate gravity, and $\gamma$ for low gravity. Objects of extremely low
gravity are further denoted by $\delta$, as originally outlined in \cite{kirkpatrick2005,kirkpatrick2006}.

\noindent {\it 2MASS J00040288$-$6410358:} The optical spectrum of this object (Figure~\ref{oddones_lowg_opt2})
falls intermediate between that of 2MASS J0141$-$4633 and 2MASS J23225299$-$6151275, which are classified as L0$\gamma$
and L2$\gamma$, respectively, on the scheme of \cite{cruz2009}. We therefore assign 2MASS J0004$-$6410 an
optical spectral type of L1$\gamma$.
Assuming its absolute $J$ mag is identical to that of a normal L1 dwarf, then the magnitude estimates from 
Table 3 of \cite{looper2008flip} can be used to predict a distance of $\sim$57 pc for this object.
The spectrophotometric distance estimate, sky location, and similarity to 2MASS J0141$-$4633 suggest that
2MASS J0004$-$6410 may be a member of the $\sim$50 pc distant, 30 Myr old Tucana-Horologium association
(\citealt{torres2008}).

\noindent {\it 2MASS J02340093$-$6442068:} The optical spectrum of this object (Figure~\ref{oddones_lowg_opt2})
is a good match to that of 2MASS J0141$-$4633. On the classification
scheme of \cite{cruz2009} this gives 2MASS J0234$-$6442 an optical spectral type of L0$\gamma$. 
Assuming its absolute $J$ mag is identical to that of a normal L0 dwarf, then the magnitude estimates from 
Table 3 of \cite{looper2008flip} can be used to predict a distance of $\sim$45 pc for this object.
The spectrophotometric distance estimate, sky location, and similarity to 2MASS J0141$-$4633 suggest that
2MASS J0234$-$6442 may also be a member of the Tucana-Horologium association.

\noindent {\it 2MASS J03032042$-$7312300:} The optical spectrum of this object (Figure~\ref{oddones_lowg_opt2})
falls intermediate between 2MASS J00452143+1634446 (L2$\beta$) and 2MASS J2322$-$6151 (L2$\gamma$). Because 
2MASS J0303$-$7312 has very weak \ion{Na}{1} at 8183-8195 \AA, we believe that a classification of L2$\gamma$
is the more appropriate one. Assuming its absolute $J$ mag is identical to that of a normal L2 dwarf, 
the magnitude estimates from 
Table 3 of \cite{looper2008flip} can be used to predict a distance of $\sim$59 pc for this object. As with
the two objects noted above, the spectrophotometric distance estimate, sky location, and $\gamma$ suffix suggest that
2MASS J0303$-$7312 may also be a member of the Tucana-Horologium association.

\noindent {\it 2MASS J04062677$-$3812102:} The optical spectrum of this object (Figure~\ref{oddones_lowg_opt1})
is a good match to that of 2MASS J0141$-$4633, giving 2MASS J0406$-$3812 an optical type of L0$\gamma$.
Its near-infrared spectrum (Figure~\ref{oddones_lowg_IR})
roughly matches a type of L1 but has some peculiarities. The spectrum is redder than the L1 standard, the H-band
peak is much more triangular, and there are much stronger VO bands at 1.06 and 1.18 $\mu$m than normally seen in an L1. These
discrepancies mimic those seen in the near-infrared spectrum of 2MASS J0141$-$4633 (\citealt{kirkpatrick2006}), which 
provides an excellent match to the overall shape of 2MASS J0406$-$3812. Its L1 near-infrared spectral type and J-band 
magnitude suggest a distance of 90 pc, and the match to 2MASS J0141$-$4633 suggests an age of very roughly 30 Myr (assuming
2MASS J0141$-$4633 itself is a member of the Tucana-Horologium association). Given these clues as well as its sky
location, 2MASS J0406$-$3812 may therefore be a 
member of the 90 pc distant, 30 Myr old Columba association (\citealt{torres2008}).

\noindent {\it 2MASS J05341594$-$0631397:} The optical spectrum of this object (Figure~\ref{oddones_lowg_opt1})
is a good match to the M8$\gamma$ dwarf 2MASS J12073346$-$3932539A, which is a member of the 8 Myr old TW Hydrae Association. (The gravity 
classification suffix $\gamma$ is given to M dwarfs with log(age(yr)) $\approx$ 7; Kirkpatrick et al., in prep.)
The near-infrared spectrum  (Figure~\ref{oddones_lowg_IR}) shows
peculiarities when compared to a standard late-M dwarf but is again almost identical to the low-gravity M dwarf 2MASS J1207$-$3932A.
Using the trigonometric parallax
($\pi$=0{\farcs}01851$\pm$0{\farcs}00103; \citealt{gizis2007}) and J-band magnitude (12.995 mag) of 
2MASS J1207$-$3932A as a guide, we estimate that 2MASS J0534$-$0631 is roughly 223 pc distant. This distance estimate places it
further away than any of the young associations studied by \cite{zuckerman2004} or \cite{torres2008}.

\noindent {\it 2MASS J15575011$-$2952431:} The optical spectrum of this object shows the telltale signs of a low-gravity late-M
dwarf (Figure~\ref{oddones_lowg_opt1}), namely weaker TiO bands and alkali lines together with stronger VO bands. The overall 
spectral morphology is very similar to that of the 1 Myr old Taurus member KPNO-Tau 12 (aka 2MASS J04190126+2802487). A
tentative core type of M9 is therefore assigned, and a suffix of $\delta$ is given because that is the suffix given
to M dwarfs like KPNO-Tau 12 with log(age(yr)) $\approx$ 6 (Kirkpatrick et al., in prep.) The
near-infrared spectrum also shows the hallmarks of lower gravity (Figure~\ref{oddones_lowg_IR}) -- a triangular-shaped H-band 
peak and a strong VO band at 1.08 $\mu$m -- which are again similar to those features seen in KPNO-Tau 12.
Given that 2MASS J1557$-$2952 and KPNO-Tau 12 have similar optical classifications and J-band magnitudes (2MASS measures of
16.316 and 16.305 mag, respectively), we can assume they fall at roughly the same distance, which for KPNO-Tau 12 is assumed to be 
that of the Taurus Molecular Cloud (140 pc; \citealt{luhman2006}). Given the sky location, estimated distance, and young 
implied age of 2MASS J1557$-$2952, it may be a member of the broad Scorpius-Centaurus complex.

Radial velocities, tangential velocities, and total space motions for these six low-gravity objects
are given in Table \ref{RV_table}. The measured proper motions for these objects, with the exception of 2MASS J1557$-$2952,
are within 2$\sigma$ of zero motion, and an accurate derivation of the tangential velocity is further
complicated by the fact that these objects are more distant than most other objects listed in
Table \ref{RV_table}. The radial velocity measurements, which are independent of distance, are also within 2$\sigma$ of zero 
motion. Despite the large uncertainties, the lack of high radial velocities
in this list is nonethless consistent with our interpretation of these objects as young.

\subsection{Unusually Red L Dwarfs}

The first unusually red L dwarf, 2MASS J22443167+2043433 (\citealt{dahn2002}), was discovered 
during the search for red AGN with 2MASS. Its 2MASS color of J-K$_s$=2.45$\pm$0.16 is far redder 
than the mean J-K$_s$ color of other L6.5 dwarfs (see Figure 14 of \citealt{kirkpatrick2008}), and 
its near-infrared spectrum is markedly odd compared to most other late-L dwarfs (\citealt{mclean2003}). 
We now believe that the redness of 2MASS J2244+2043 is caused by low gravity (\citealt{kirkpatrick2008}) and 
low-gravity L dwarfs can become even redder than this, as evinced by the TW Hydrae association member 
2MASS J12073346$-$3932539B, a late-L dwarf with J-K $\sim$ 3.05 (\citealt{gizis2007}).

As mentioned in the section above, a cause has been proposed (\citealt{kirkpatrick2006}) and a classification system developed 
(\citealt{cruz2009}) for these low-gravity, red L dwarfs.
Our proper motion survey has uncovered a few other red L dwarfs, however, whose peculiarities cannot as easily be attributed to 
low gravity, and for these we assign a suffix of ``pec'' to indicate their peculiar morphology in the near-infrared. 
The oddest of these, 
2MASS J21481633+4003594, has been discussed at length by \cite{looper2008} along with
another bright example, 2MASS J18212815+1414010, both of which were discovered during our proper motion survey. 
We discuss three other red L dwarf discoveries below:

\noindent {\it 2MASS J13313310+3407583:} In the optical (Figure~\ref{oddones_redL_opt}) this object most closely resembles
an L0 dwarf. In the near-infrared (Figure~\ref{oddones_redL_IR}) it most closely resembles an L1 at J-band. If the flux
of the 2MASS J1331+3407 spectrum is normalized to the J-band of the L1 near-IR standard,
the H- and K-band portions of the 2MASS J1331+3407 spectrum have too much flux relative to the standard (hence the ``red''
designation). This redness is also obvious in the shallow depth of the H$_2$O bands between the J- and H-band peaks and
again between the H and K peaks.  The 2MASS color of J-K$_s$ = 1.448$\pm$0.035 mag is $\sim$0.2 mag redder than 
the median color of optically defined L0 dwarfs in Figure 14 of \cite{kirkpatrick2008}.

\noindent {\it 2MASS J23174712$-$4838501:} This object most closely resembles an L4 in the optical 
(Figure~\ref{oddones_redL_opt}), the only major deviation being that its 8432 \AA\ TiO band is weaker than the L4 standard 
and more like that of an L5 or L6 dwarf. In the near-infrared, the spectrum continues to match that of an L6 up to 
$\sim$1.1 $\mu$m. At longer wavelengths -- even in the longer half of the J bandpass -- there is excess flux 
relative to both the L6 and L7 near-infrared standards. 
The 2MASS color of J-K$_s$ = 1.969$\pm$0.048 mag is $\sim$0.2 mag redder than 
the median color of optically defined L4 dwarfs in Figure 14 of \cite{kirkpatrick2008}. 

\noindent {\it 2MASS J23512200+3010540:} Overall this object matches the L5 (or L6) optical standard well 
(Figure~\ref{oddones_redL_opt}).  In the near-infrared (Figure~\ref{oddones_redL_IR}) its J-band spectrum is a
good match to the L5 near-infrared standard, but there is excess flux at H- and K-bands compared to the standard.
Curiously, the 2MASS color of J-K$_s$ = 1.831$\pm$0.118 mag 
places it near the peak color of optically defined L5 or L6 dwarfs in Figure 14 of \cite{kirkpatrick2008}, but the error
in this color measurement is large.

These three red L dwarfs along with the other two discoveries
published by \cite{looper2008} have substantial proper motions -- 0.44$\pm$0.06, 
0.24$\pm$0.11, 1.39$\pm$0.26, 0.37$\pm$0.17, and 0.28$\pm$0.07 arcsec/yr for 2MASS J1331+3407, 2MASS J1821+1414, 2MASS J2148+4003, 
2MASS J2317$-$4838, and 2MASS J2351+3010, respectively. For estimated distances of roughly 13.8, 9.8, 9.9, 24.4, and 23.1 pc, these
correspond to transverse velocities of roughly 28.9, 11.2, 65.5, 43.0, and 30.8 km/s, respectively. The two most extreme red L dwarfs, 
2MASS J2148+4003 and 2MASS J2317$-$4838, have transverse velocities well above the median transverse velocity of 24.5 km/s for 
field L dwarfs (\citealt{vrba2004}), but still well below the median transverse velocity of 250 km/s for L subdwarfs
(a sample of three from \citealt{schilbach2009}). Table \ref{RV_table} lists our measured radial velocities and estimates of total space motion
for these five objects. Using the median space velocity of the three 2MASS J2148+4003 measures, we find that the average
space velocity of this sample is 165 km/s. It therefore appears that these red L dwarfs are drawn from a considerably older population than the 
young, low-gravity M and L dwarfs discussed in the previous section. Analysis of the spectral energy distribution of 2MASS
J2148+4003 by \cite{looper2008} points to either high-metallicity or low-gravity as a cause for the spectral peculiarities,
but neither of these would be a likely explanation for an older object. It should be noted that our three radial velocity measurements
for 2MASS J2148+4003 are discrepant, ranging over almost 200 km/s, which is a significant variation despite the large
uncertainties. Follow-up observations at higher resolution, which may shed additional light on the red L dwarf phenomenon, are desired 
and easily accomplished since this object is relatively bright (2MASS J=14.147 and K$_s$=11.765).

\subsection{M and L Subdwarfs\label{subdwarf-section}}

The Galaxy's oldest constituents are rare in the Solar Neighbohood and sometimes difficult to discern with limited 
photometric information. Proper motion surveys are particularly well attuned to identifying such objects because
of their high kinematics, which result from a long lifetime of gravitational encounters with objects of greater
mass (giant molecular clouds, star clusters, etc.). Of
particular interest are the low-mass stellar and substellar members of this
population, as they contain a fossilized history of star formation near the brown dwarf limit at the Milky
Way's earliest epochs. As a group they also contain empirical information about the brown dwarf cooling rate.

Old objects also tend to have low metal content since they are composed of matter devoid of
substantial metal enrichment by earlier generations of stars. For the purposes of this paper, we use the term
``subdwarfs'' to describe those objects whose spectral morphology clearly indicates
a metal-poor makeup. The classification of
subdwarfs has seen great advancement over the past decade, including the discovery of subdwarfs at and below
the hydrogen-burning limit (\citealt{gizis1997,lepine2007,burgasser2009}).
Our proper motion survey has identified several new, low temperature subdwarfs, ranging in type from late-K
through late-L. Optical spectra with types of M9.5 or earlier have been classified on the system 
of \cite{lepine2007}. Because this new system has not yet been extended into the near-infrared
and because so few of the extant subdwarfs have been classified on it so far,
our near-infrared data are typed using an extension to those wavelengths of the older \cite{gizis1997} system.

To check for differences between the new \cite{lepine2007} system and the \cite{gizis1997} system,
we have taken a suite of late-M dwarf and subdwarf optical spectra (Table \ref{sd_spectra}), along with 
optical spectra of the \cite{lepine2007} standards (kindly provided by Sebastien L{\'e}pine), and
typed them according to the \cite{lepine2007} prescription. Of the thirty new standards, we recover
the \cite{lepine2007} prefix class (sd, esd, or usd) for all thirty. We recover the spectral 
subclass (K7, M2.5, M7, etc.) for twenty-four out of thirty; for the remaining six, our computed
type is only a half subclass different from the \cite{lepine2007} subclass (e.g, we obtain a
classifcation of usdM5.5 as opposed to the published type of usdM5). This
sample of spectra, however, contains no normal dwarfs.

When the \cite{lepine2007} prescription is run on the spectroscopic sample of Table \ref{sd_spectra},
we find the following. As expected, all of the normal M dwarfs were identified as such, although
the \cite{lepine2007} system does not specify a way to obtain the subclass of dM stars.\footnote{This may be
because the weakening of TiO due to dust formation in the later types  makes the classification 
with respect to TiO and CaH bands ambiguous beyond M6 (\citealt{reid1995}).}
Of seven objects previously classified as extreme subdwarfs (``esd''), six
retain their ``esd'' classifications. One is reclassified as an ultrasubdwarf (``usd''), a
distinction which is new to the \cite{lepine2007} system. Subtypes sometimes vary by a full subclass
or more between the old and new esdM (or usdM) classifications, but this is not suprising given that
earlier published types are themselves not always homogeneously classified. More surprising, however,
is the fact that of the ten previously typed sdM's (not including LSR J1610$-$0040, which is a true oddball; 
\citealt{cushing-schizo}), only four warrant a classifcation of sdM on the new system.

The six objects that fail to earn a subdwarf classification on the \cite{lepine2007} system are shown
in Figure \ref{sd_or_not}. In two cases -- LHS 3189 and LP 97-817 -- a comparison of their spectra
to those of normal dwarfs confirms that types of dM, not sdM, are justified. In the other four
cases -- LHS 377, LHS 1035, LHS 1135, and LHS 2067 -- there appears to be sufficient evidence to
classify these objects as something other than normal M dwarfs, since they appear to have evidence
of low-metallicity as demonstrated by their stronger bandstrengths of CaH relative to TiO. For these
four objects, we suggest types of ``d/sdM'' to denote a morphology intermediate between
normal dwarfs and sdM's. These are used as the adopted types in Table \ref{sd_spectra}.

Subdwarf discoveries from our proper motion survey are listed in Table \ref{sd_table}
and discussed below. Optical spectra are shown in
Figures \ref{oddones_sd_opt3} - \ref{oddones_sd_opt2}. Near-infrared spectra are shown in Figures 
\ref{oddones_sd_IR2} - \ref{oddones_sd_IR6}.

\subsubsection{Borderline Subdwarfs}

Three objects have near-infrared spectra which support a type intermediate between normal M7
and sdM7 (i.e., d/sdM7)\footnote{Note that the near-infrared sdM7 being plotted is LHS 377, an object classified in the
optical (see discussion above) as d/sdM5.}. We also have optical spectra for two of these objects, and in both cases a 
normal dwarf spectral type, not a subdwarf type, appears warranted:

\noindent {\it 2MASS J00554279+1301043:} Whereas the optical spectrum of this object (Figure \ref{oddones_sd_opt3})
supports a type of M6, the near-infrared spectrum supports a type intermediate between M7 and sdM7
(Figure \ref{oddones_sd_IR2}).

\noindent {\it 2MASS J18355309$-$3217129:} Whereas the optical spectrum of this object (Figure \ref{oddones_sd_opt3})
supports a type of M6.5, the near-infrared spectrum supports a type intermediate between M7 and sdM7
(Figure \ref{oddones_sd_IR2}).

\noindent {\it 2MASS J23470713+0219127:} The near-infrared spectrum supports a type intermediate between M7 and sdM7
(Figure \ref{oddones_sd_IR2}). We have no optical spectrum for this object.

\subsubsection{Late-M Subdwarfs}

\noindent {\it 2MASS J00054517+0723423:} In the optical the spectrum of this object matches a normal M4 dwarf 
(Figure \ref{oddones_sd_opt3}). The near-infrared spectrum of this object (Figure \ref{oddones_sd_IR1}), however,
best fits an sdM6. Note that the near-infrared spectrum of the M4 standard provides a poor match, as it
shows pronounced differences below 1 $\mu$m and its infrared water bands are too shallow.

\noindent {\it 2MASS J04470652$-$1946392:} This is our latest M-type subdwarf discovery. Figure \ref{oddones_sd_opt1}
shows the optical spectrum compared to that of the sdM8 standard from \cite{lepine2007}. The overall bandstrengths
and continua are very similar. In addition to the strong hydride bands of CaH near 6350 and 6950 \AA\ and FeH
band at 9896 \AA, this spectrum also shows the late subdwarf hallmarks of strong metal lines -- \ion{Ca}{1} at 6517 \AA, 
\ion{Ti}{1} (7209/7213, 8433, and 9600-9700 \AA), and \ion{Ca}{2} (8498, 8542, and 8662 \AA).
Using the \cite{lepine2007} $\zeta_{TiO/CaH}$ metallicity index along with the relation between the 
CaH2 and CaH3 indices and spectral type, we classify this object in the optical as an sdM7.5. In the near-infrared
(Figure \ref{oddones_sd_IR1}) this object best matches the spectrum of LSR J2036+5100, which has also been classified as
an sdM7.5. Using the trigonometric parallax of LSR J2036+5100 from \cite{schilbach2009} and the 2MASS J-band magnitudes
of LSR J2036+5100 and 2MASS J0447$-$1946, we estimate a distance to 2MASS J0447$-$1946 of 102 pc.

\noindent {\it 2MASS J10462067+2354307:} The near-infrared spectrum of this object (Figure \ref{oddones_sd_IR1})
best fits an sdM6. We have no optical spectrum.

\noindent {\it 2MASS J16130315+6502051:} The near-infrared spectrum of this object (Figure \ref{oddones_sd_IR1})
best fits an sdM6. Again, we have no optical spectrum.

\subsubsection{L Subdwarfs}

Our discovery of three new L subdwarfs substantially increases the number of known members of this class
(Table \ref{sdLs}). Because there are few known examples, there are no classification anchors on
which to pin spectral types, so the closest match to the normal L dwarf sequence is used instead. 

\noindent {\it 2MASS J06453153$-$6646120:} This is the latest subdwarf discovery from our survey. The optical spectrum,
shown in Figure \ref{oddones_sd_opt2}, most closely matches an L8 dwarf but there are peculiarities. Because 
the FeH and CrH bands are noticeably weaker in 2MASS J0645$-$6646 than in the L8 standard,
we label this as an sdL8. In the
near-infrared (Figure \ref{oddones_sd_IR3}) 2MASS J0645$-$6646 shows a markedly blue continuum compared to a
normal L8. The curious absorption trough just shortward of 1.6 $\mu$m (Figure \ref{oddones_sd_IR6})
appears to be enhanced FeH, echoing
the strong hydride bands seen in the optical. The 2MASS color of J-K$_s$ = 1.259$\pm$0.050 mag is bluer than 
all of the optically defined L8 dwarfs in Figure 14 of \cite{kirkpatrick2008} and $\sim$0.5 mag bluer than
the median L8 color. At $\mu$ = 1.57$\pm$0.08$\arcsec$/yr, 2MASS J0645$-$6646 is the highest proper motion object 
found in the 1X-6X survey. Using its 2MASS J-band
magnitude and assuming its absolute J-band magnitude is comparable to that of a normal L8 (\citealt{looper2008flip}),
we estimate a distance to 2MASS J0645$-$6646 of $\sim$16 pc.

\noindent {\it 2MASS J11582077+0435014:} The optical spectrum of this object, shown in Figure \ref{oddones_sd_opt2},
most closely matches a normal L7 but bands of TiO, FeH, and CrH are much stronger in 2MASS J1158+0435. For this
reason, we classify the object as an sdL7. In addition to the strong alkali lines seen in late L dwarfs,
\ion{Ti}{1} at 8433 \AA\ is also seen and this is another oddity sometimes seen in late-type subdwarfs. 
In the near-infrared (SpeX data in Figure \ref{oddones_sd_IR3}; NIRSPEC data in Figure \ref{oddones_sd_IR5}) the best overall
match to the suite of standard L dwarfs from Figure \ref{NIR_L_stds} is the L7 but there are marked differences. At J-band the 
FeH bands are much stronger than in the normal L7 and the overall continumm at H- and K- bands is markedly 
suppressed. The 2MASS J-K$_s$ color of this object, 1.172$\pm$0.084 mag, is bluer than 
all of the optically defined L7 dwarfs in Figure 14 of \cite{kirkpatrick2008} and $\sim$0.6 mag bluer than
the median L7 color. A detail of the NIRSPEC N3 spectrum of this object is shown in Figure \ref{oddones_sd_IR5}.
Both the FeH bands and the \ion{K}{1} doublet lines are stronger than in the L7 standard. In lieu of a trigonometric
parallax, we estimate a distance of $\sim$16 pc to 2MASS J1158+0435 using its 2MASS J-band
magnitude and the absolute J-band magnitude of a normal L7 from \cite{looper2008flip}.

\noindent {\it 2MASS J17561080+2815238:} In the optical this object most resembles an
L1 dwarf but shows several discrepancies. Figure \ref{oddones_sd_opt1} illustrates that 2MASS J1756+2815 has stronger
TiO, FeH, and CrH bands than the standard L1, leading us to assign a classification of sdL1. Note
also that 2MASS J1756+2815 has stronger \ion{K}{1} line cores and, particularly, line wings than the L1 standard.
Lines of \ion{Rb}{1} and \ion{Cs}{1} are also stronger.
In the near-infrared (Figure \ref{oddones_sd_IR4}) the most striking feature is the suppression of the H- and K-bands
relative to a normal L1 dwarf. Also noticeable are its stronger lines of 2.20 $\mu$m \ion{Na}{1} and stronger
bands of 2.30 $\mu$m CO at $K$-band. Its 2MASS J-K$_s$ color of 0.899$\pm$0.053 mag is bluer than 
all of the optically defined L1 dwarfs in Figure 14 of \cite{kirkpatrick2008} and $\sim$0.4 mag bluer than
the median L1 color. A detail of the NIRSPEC N3 spectrum of this object is shown in Figure \ref{oddones_sd_IR4b}.
Note that the \ion{K}{1} doublet lines are stronger and broader than in the L1 standard although, curiously, the
strengths of the FeH bands here are not markedly different. Because of this, we are not able to classify this object
in the near-infrared as an L subdwarf, but denote it as a blue L1. (Blue L dwarfs are discussed further below.)
This is the only object so far identified that has this strongly dual nature, so it is worthy of concentrated
follow-up work. Using the 2MASS J-band magnitude and the \cite{looper2008flip} 
relation noted above, we estimate a distance of $\sim$35 pc. 

\subsubsection{Discussion of L Subdwarfs}

Table \ref{RV_table} lists the radial velocities, tangential velocities, and total space motions measured
for the three new L subdwarfs and one late-M subdwarf. We find an average tangential velocity of 171.5 km/s, or average
space motion of 223 km/s, for this sample of four. Although high, this velocity is not as high as the average 
tangential velocity of 263.5 km/s found for late-M and L subdwarfs in \cite{schilbach2009}. Given that the
L subdwarfs comprising the bulk of the \cite{schilbach2009} sample are much bluer than those found here -- 
mainly because the earliest L subdwarf discoveries were made as by-products of T dwarf searches focused on
very blue objects in J-K$_s$ color -- they may also represent a substantially more metal poor and older population
than those found here.
Nonetheless, our new L subdwarf discoveries are kinematically distinct from the field L dwarf population, whose
median transverse velocity is 24.5 km/s (\citealt{vrba2004}).

There are several reason to believe that these three L subdwarfs, with the possible exception of the
enigmatic 2MASS J1756+2815, are 
objects low in metallicity: (1) Stronger hydride bands are the by-product of an atmosphere where
metal-metal molecules such as TiO or VO are less abundant. With less absorption by the oxides, the
hydride bandstrengths are more prominent. 
(2) In a cool, metal-deficient atmosphere, the relative absorption by collision-induced H$_2$ is stronger than in 
an atmosphere with a rich array of lines and bands from metal-bearing species. Because this broad H$_2$
absorption is strongest at H- and K-bands (\citealt{borysow1997}), this leads to a bluer near-infrared
continuum, and hence bluer J-K$_s$ color, for L subdwarfs when compared to L dwarfs of higher metallicity 
(\citealt{burgasser2003}).
(3) Somewhat counterintuitively, the presence of metal lines (such as \ion{Ti}{1} and \ion{Ca}{1})
and the increased strengths of the alkali lines (\ion{Na}{1}, \ion{K}{1}, \ion{Cs}{1}, and \ion{Rb}{1})
in low-temperature atmospheres may indicate retarded condensate formation in a metal-poor
environment. The rarity of metal atoms in the atmosphere means that condensates are more difficult to
form, and thus these metals remain in the atmosphere to lower temperatures.

Using one of these new discoveries (2MASS J1158+0435) along with a new discovery from the SDSS team
(SDSS J1416+1348; Schmidt et al.\, submitted), we are able to build an empirical sequence in metallicity
for spectral type L7. Figure \ref{sdL7s_opt} and Figure \ref{sdL7s_nir} show the optical and near-infrared
spectra of these two objects compared to the standard L7, DENIS J0205$-$1159 (presumably of solar metallicity),
and the first late-L subdwarf discovered, 2MASS J05325346+8246465. 2MASS J1158+0435 and SDSS J1416+1348 are
very similar in spectral appearance, although the SDSS object shows slightly stronger hydride bands (CaH,
CrH, and FeH), slightly stronger TiO, and a bluer near-infrared continuum (by 0.14 mag in J-K$_s$). The spectrum
of 2MASS J0532+8246 is even more extreme than these, particularly in the strength of its CaH band and in its
overall near-infrared color, which is 0.77 mag bluer than in SDSS J1416+1348. (The slightly weaker TiO, CrH,
and FeH bands in 2MASS J0532+8246 may indicate that this object is a bit later in type than the other
two subdwarf comparisons.) Because of its more extreme nature, we propose that 2MASS J0532+8246 should actually
be labelled as an esdL7 to indicate an even more unusual spectral morphology. Further investigation
will be needed to determine if such morphological traits easily translate into a lower value of [Fe/H]. In one example,
the sdL5 2MASS J06164006$-$6407194 has a spectral morphology that is not nearly as abnormal as that
of 2MASS J0532+8246 yet kinematic investigations place the former in the outer halo 
(\citealt{cushing2009}) and the latter in the inner halo (\citealt{burgasser2008-parallax}). Single objects may not
be representative of their population, however, so larger collections of L subdwarfs will be needed to further
gauge the variation in spectral appearance of these two, very old samples.

\subsection{Unusually Blue L Dwarfs}

Although L subdwarfs have particularly blue $J-K_s$ colors due to increased 
relative collision-induced opacity by H$_2$ at $H$- and $K_s$-bands,
there is a class of L dwarfs that shows unusually blue near-infrared colors not 
obviously caused by low metallicity.
These so called ``blue L dwarfs'' have unusually blue $J-K_s$ colors for their 
optical spectral types and were first pointed out in follow-up
of 2MASS sources by \cite{gizis2000} and \cite{cruz2003}. Subsequent follow-up 
of SDSS sources revealed objects with peculiar near-infrared
spectral morphologies. Specifically, spectral indices measuring the depth of 
H$_2$O bands versus those measuring FeH or \ion{K}{1} give 
discordant spectral types, with the H$_2$O-based index always
suggesting a much later L subclass (\citealt{knapp2004,chiu2006}).
Some of these blue L dwarfs are suspected to be the composite spectra of an
L dwarf + Tdwarf binary (e.g., \citealt{burgasser2007-0805}).
One intriguing blue L dwarf, 2MASS J11263991$-$5003550, for which binarity seems not to be the
cause, has been extensively studied by \cite{folkes2007} and \cite{burgasser2008}.
The latter authors argue that the spectral features are the result of thin and/or
large-grained clouds, the cause for which appears not to be easily attributable to
either surface gravity or low metallicity.
Another example -- an object known as 2MASS J17114559+4028578 
(G 203-50B) -- is particularly interesting because it shares common proper motion 
with an M dwarf whose optical spectrum suggests a metallicity very
similar to that of the Sun (\citealt{radigan2008}). 

\cite{faherty2009} performed a kinematic study of ten unusually blue L dwarfs
and found a median transverse velocity of 99$\pm$47 km/s, substantially larger than
that of the field L dwarf population (24.5 km/s; \citealt{vrba2004}) but still
smaller than the tranverse velocity for M subdwarfs (196$\pm$91 km/s). This
suggests that many of the blue L dwarfs may be old, though not as old as the
halo population. It therefore remains a possibility that these objects
represent a slightly lower-metallicity population than normal field dwarfs
but not so low in metal content that they can be confidently labeled as
subdwarfs. For very cool atmospheres such as these, a small change in 
metal content can perhaps have more prominent effects on the formation of
condensates (i.e., cloud properties) than on the strength of gas phase
molecular transitions.

As a result of our proper motion study, we report another eight blue L dwarfs (Table \ref{blueL_table}),
spectra for which are shown in Figures \ref{oddones_blueL_opt1} - \ref{oddones_blueL_IR3},
For the purposes of this paper, we define a ``blue L dwarf'' to be an object 
whose 1.5-2.5 $\mu$m continuum is significantly bluer than expected, given the continuum shape
of the near-infrared spectroscopic standard that best fits its spectrum at $J$-band, but whose
cause is not obviously low metallicity as judged primarily by the strength of hydride features. 
We discuss each of these eight objects below:

\noindent {\it 2MASS J00150206+2959323:} The optical spectrum of this object (Figure 
\ref{oddones_blueL_opt2}) is very similar to the
L7 optical standard except that the FeH and CrH bandstrengths are somewhat stronger in 2MASS J0015+2959.
Our near-infrared spectrum (Figure \ref{oddones_blueL_IR3}) is best fit by
either the L7 or the L8 near-infrared standard. The depth of the water band between the J- and H-bands
is well matched by the L8 standard, but the flux peaks at H- and K-bands in 2MASS J0015+2959 
are significantly suppressed relative to this same standard. This continuum suppression is even more
extreme at K-band if the L7 standard is used as the comparison. The 2MASS J-K$_s$ value of 1.676$\pm$0.109
mag is only $\sim$0.15 mag bluer than the median color of other optically defined L7
dwarfs (\citealt{kirkpatrick2008}), although the uncertainty in this measurement is large.

\noindent {\it 2MASS J03001631+2130205:} Our near-infrared spectrum (Figure \ref{oddones_blueL_IR2})
is an excellent match to the L6 near-infrared standard in the J-band. At H-and K-bands, however,
the continuum is markedly bluer than that of the standard, earning 2MASS J0300+2130 a ``blue L''
classification. Its 2MASS J-K$_s$ color of 1.566$\pm$0.098 mag is $\sim$0.35 mag bluer than
the median color for other optically defined L6 dwarfs.

\noindent {\it 2MASS J10461875+4441149:} We have only a near-infrared spectrum of this object
(Figure \ref{oddones_blueL_IR2}) but it shows a bluer continuum at wavelengths longward of
1.5 $\mu$m compared to the L5 near-infrared standard that it best matches shortward of this.
Except for this, the overall feature strengths through the spectrum look very similar to the
standard itself. 

\noindent {\it 2MASS J11181292$-$0856106:} Figure \ref{oddones_blueL_opt2} shows the optical spectrum,
which is a very good match to the optical L6 standard. In the near-infrared (Figure \ref{oddones_blueL_IR2})
the best J-band fit is again to the L6 standard, but the overall bandstrengths do not match well.
At H- and K-bands the flux in 2MASS J1118$-$0856 is considerably quenched relative to the L6
standard. Its 2MASS J-K$_s$ color of 1.560$\pm$0.102 mag is $\sim$0.35 mag bluer than
the median color for other optically defined L6 dwarfs.

\noindent {\it 2MASS J13023811+5650212:} The optical spectrum, shown in Figure \ref{oddones_blueL_opt1},
is very similar to the L2 optical standard. Our near-infrared spectrum (Figure \ref{oddones_blueL_IR1})
is noisy but shows a suppressed continuum at H- and K-band relative to the near-infrared L3 standard
that it best matches at J-band. Its 2MASS J-K$_s$ color of 1.388$\pm$0.173 mag is only $\sim$0.1 mag bluer than
the median color for other optically defined L2 dwarfs, but the uncertainty on this color measure is large.

\noindent {\it 2MASS J14403186$-$1303263:} The optical spectrum, shown in Figure \ref{oddones_blueL_opt1},
is very similar to the L1 optical standard. The near-infrared spectrum (Figure \ref{oddones_blueL_IR1})
is most like the L1 near-infrared standard at J-band but shows a suppressed continuum at longer wavelengths.
Most striking is the much stronger H$_2$O absorprtion trough between J- and H-bands. Its 2MASS J-K$_s$ 
color of 1.128$\pm$0.082 mag is $\sim$0.2 mag bluer than
the median color for other optically defined L1 dwarfs.

\noindent {\it 2MASS J19495702+6222440:} The optical spectrum of this object (Figure \ref{oddones_blueL_opt1})
best fits the L2 optical standard over the wavelength region from 8000 - 9000 \AA\ but is otherwise much too
flat (blue) for an early-L dwarf. Longward of 9000 \AA, the
spectrum of 2MASS J1949+6222 is suppressed relative to the standard, although the FeH and CrH bandstrengths
are very similar. Shortward of 8000 \AA\ the spectrum of 2MASS J1949+6222 shows excess flux relative to the standard,
and the 7400-\AA\ band of VO is much stronger than in the standard L2. Our near-infrared spectrum of
2MASS J1949+6222 (Figure \ref{oddones_blueL_IR1}) is noisy but nonetheless shows a somewhat suppressed continuum
at H- and K-bands relative to the near-infrared L2 standard that it best matches at J-band.
Surprisingly, its 2MASS-measured J-K$_s$ color of 1.609$\pm$0.167 mag is $\sim$0.1 mag {\it redder} than
the median color for other optically defined L2 dwarfs, but the photomeric error is substantially larger
than this. Given the odd nature of its optical spectrum, we feel that 2MASS J1949+6222 is
not in the same class of object as the other blue L dwarfs found during our survey and is worthy of additional
follow-up.

\noindent {\it 2MASS J21513979+3402444:} The near-infrared spectrum of this object (Figure \ref{oddones_blueL_IR3})
best fits the L7 near-infrared standard. At longer wavelengths, the overall continuum is suppressed relative
to that same standard, earning this object a ``blue L'' designation.

A full list of known blue L dwarfs is given in Table~\ref{blue_Ls}. For the new discoveries, we
estimate a distance using the 2MASS J-band magnitude and the predicted absolute J-band magnitude given by the optical spectral type,
as given by the 
relation of \cite{looper2008}. (The near-infrared spectral type is used when an optical type is not available.)
With this larger sample of blue L dwarfs we find a mean transverse velocity of $\sim$66 km/s, which is somewhat lower than the number
computed by \cite{faherty2009} but still within their 1$\sigma$ errors. As stated earlier, the blue L dwarfs in this list probably 
represent a mixture of different types. Follow-up of suspected binaries such as SDSS J0805+4812 (\citealt{burgasser2007}) 
should be done to eliminate objects that are extrinsically blue, leaving a purer sample of intrinsically blue L dwarfs for
further study.

In Table \ref{RV_table} we compute radial velocities and total space motions for our blue L dwarfs. Discounting the two blue L dwarfs
with large uncertainties in V$_{tan}$, we find an average space motion of 141 km/s, similar to the value found for the
red L dwarfs but smaller than the value found for late-M and L subdwarfs. This raises the intriguing possibility that
the red L dwarf and blue L dwarf phenomena are related, or at least occur in L dwarfs of similar age. It is even
possible that these two phenomena occur in the same objects
and that viewing angle determines the spectral appearance. This might be the case if clouds are not
homogenously distributed in latitude or if properties such as grain size and cloud thickness vary in latitude. 
In this scenario the red dwarf phenomenon would coincide with the less frequent case of
pole-on viewing angle, and the blue L dwarf phenomenon would relate to the more frequent case of equator-on viewing angles.
This hypothesis can be tested by measuring the rotational velocities of objects in the two samples to see if the red
L dwarfs in general show smaller $V_{rot}$ values. Even then, the reason for the odd distribution of clouds -- thicker or large-grained
clouds at the poles and thiner or smaller-grained clouds near the equator -- and its predilection for
objects in this range of ages would have to be explained theoretically. Whether or not this hypothesis has any basis in
reality, it is still possible to gather more observational clues even on the red L and blue L samples currently identified.

\subsection{T Dwarfs}

T dwarfs share the same locus as hotter stars and L dwarfs in the standard J-H vs.\ 
H-K$_s$ diagram. In particular, early-T dwarfs are extremely difficult to identify
in 2MASS because they have near-infrared colors similar to M dwarfs,
which are detected in far greater numbers since they are visible to much greater distances.
With our proper motion survey, however, we have been able to identify T dwarfs regardless 
of color selection by using kinematic criteria. 

As expected, our survey recovered previously identified, later T dwarfs such as the T8 dwarf
2MASS J09393548$-$2448279, one of the coolest dwarfs currently known
(\citealt{burgasser2008-2mass0939}). Our survey also uncovered some previously unidentified 
early-T dwarfs (Table \ref{T_table}). \cite{looper2007} has already discussed three of these
discoveries -- 2MASS J11061191+2754215 
(T2.5), 2MASS J13243553+6358281 (T2 pec), and 2MASS J14044948$-$3159330 (T3). A fourth object,
2MASS J15111466+0607431 (T2 pec), was independently identified by \cite{chiu2006} using data
from the Sloan Digital Sky Survey. A fifth object is presented here
for the first time:

\noindent {\it 2MASS J21265916+7617440:}
The optical spectrum is shown in Figure \ref{oddones_T_opt}. The overall 
continuum shape matches that of the L7 optical standard although there are difference 
in bandstrengths of some molecular features. The main disagreement between the two
spectra is in the strengths of the FeH and CrH bands, which are weaker in 2MASS J2126+7617.
(The spectrum of the optical standard,
unlike that of 2MASS J2126+7617, has not been corrected for telluric absorption and thus does
not match well in the 9300 - 9850 \AA\ water band.) Despite the late-L dwarf classification in the optical, 
the near-infrared spectrum (Figure \ref{oddones_T_IR}) shows methane absorption features at both 
H and K bands and warrants a T spectral type. A T0 standard spectrum best matches the 
near-infrared features, but there are
clear discrepancies. In 2MASS J2126+7617 the FeH band at 9796 \AA\ is too strong, there is
excess flux near the 1.28 $\mu$m peak at J-band, the H-band methane feature is too strong, and
the K-band portion of the spectrum is slightly suppressed.

These features can be reproduced if this object is assumed to be an L dwarf + T dwarf binary.
Fitting to the suite of single spectra and synthetic composite spectra in \cite{burgasser2010}
and using a modified ${\chi}^2$ statistic (called G$_k$, from \citealt{cushing2008}), 
we find that the best binary fit is a 
statistically significant better match than the best fit to a single source (see Equation 4 in
\citealt{burgasser2010}). Fits to spectroscopic composites whose components are scaled using 
the \cite{looper2008} absolute magnitude vs.\ spectral type relation suggest that the binary
is comprised of an L7 and a T3.5 dwarf (see Figure \ref{2MASS2126_fit}), with an error of just 
under one spectral subclass for each
component. Given that the secondary of 2MASS J2126+7617 is expected to fall very near the spectral
type where the $J$-band bump is located (i.e., the local maximum in the M$_J$ vs.\ spectral type
relation for T dwarfs), this potential binary could be as important a probe of atmospheric flux
distribution over the L/T transition as the binary 2MASS J1401$-$3159 (\citealt{looper2008flip}),
also discovered during our proper motion survey.

\subsection{Other Interesting Objects}

Other noteworthy objects not meeting the categories above are discussed here and listed
in Table \ref{other_table}.

\noindent {\it 2MASS J04205430$-$7327392:} This object is detected in the 2MASS 1X data at J=13.159$\pm$0.024 mag 
and K$_s$=10.203$\pm$0.021 mag. Curiously, this object is
invisible in the 2MASS 6X data, which of course probe to much deeper limits (SNR=10 limits of J$\approx$17.3 mag
and K${_s}\approx$15.7 mag; \citealt{cutri2003}). Given the extremely red color
(J-K$_s$=2.956$\pm$0.032 mag), presumed photometric (as opposed to astrometric) variability, and sky location, 
we believe this object may be a nova
in the outskirts of the Large Magellanic Cloud.

\noindent {\it 2MASS J09073735$-$1457036:}  Like the previous object, this source is clearly
visible in the 1X data, in this case with J=15.322$\pm$0045 mag and K$_s$=14.847$\pm$0.149,
but completely disappears in the 6X data. Because of its blue color (J-K$_s$=0.475$\pm$0.156),
it was observed as a candidate T dwarf in the survey of \cite{burgasser2002} and was not
found in deeper infrared images obtained on 2000 May 20 (UT). The 1X observation may have
been of a previously unknown asteroid, uncataloged because this observation lies a
full 30 degrees off the ecliptic. 

\noindent {\it 2MASS J16002647$-$2456424:} The near-infrared spectrum of this object 
(Figure \ref{oddones_other_IR}) shows a morphology
most like an M7, except that the 0.6-1.0 $\mu$m slope is steeper and
the depth of the H$_2$O band near 1.5 $\mu$ is deeper than in a normal M7, 
the latter giving the H-band a triangular shape. This H-band morphology is also seen in low-gravity
objects, but this spectrum does not clearly exhibit other low-gravity features. We have no optical
spectrum of this object.

\noindent {\it 2MASS J18284076+1229207:} The near-infrared spectrum of this object (Figure \ref{oddones_other_IR}) 
shows the same characteristics as the near-infrared spectrum of 2MASS J1600$-$2456 and is also best
typed with a normal M7. In the case of 2MASS J1828+1229, we have an optical spectrum (not shown) which
looks like a normal M8 dwarf.

\noindent {\it 2MASS J21403907+3655563:} The near-infrared spectrum of this object (Figure \ref{oddones_other_IR})
best fits the M8 standard over the J-band. At longer wavelengths the spectrum has excess flux relative to
the M8 standard. This object may be a late-M analog to the ``red L'' dwarfs.

\section{Prospects for Future Brown Dwarf Surveys}

Having now used proper motion to identify interesting objects that were not easily selected via their
2MASS colors alone, we look toward other surveys that will be able to extend this search. An extension in
this case can be either to the 90\% of
the sky the 2MASS Proper Motion Survey was not able to reach or to even fainter magnitudes or longer
wavelengths than probed here.

Two ongoing surveys that have both near-infrared sensitivity and multi-epoch coverage are the 
Large Area Survey of the United Kingdom Infrared Deep Sky Survey (UKIDSS-LAS) and surveys being done by
the Visible and Infrared Survey Telescope for Astronomy (VISTA).
The multi-epoch aspect of the UKIDSS-LAS will cover a similar area as the 2MASS Proper Motion
Survey (4000 sq./ deg.), although in a single filter, but the sensitivity will be considerably deeper ($J \approx 20.0$ mag;
see http://www.ukidss.org/surveys/surveys.html) and the time baseline may be up to seven years. These 
areas also cover regions probed deeply by the Sloan Digital Sky Survey (SDSS) at optical wavelengths;
for objects in common to both surveys, these data will provide an even longer time baseline. VISTA (http://www.vista.ac.uk)
has several surveys planned, the largest of which, the VISTA Hemisphere Survey (VHS), will cover the entire
hemisphere in two near-infrared bands approximately four magnitudes deeper than 2MASS. Three others
(UltraVISTA, the VISTA Magellanic Survey, and VISTA Variables in the Via Lactea), though much 
smaller in extent, will have multi-epoch coverage.

Three other large-area surveys, all optically based, will enable astronomers to study the entire sky repeatedly in the time domain.
The ongoing Panoramic Survey Telescope and Rapid Response System 
(Pan-STARRS; http://www.pan-starrs.ifa.hawaii.edu) plans to reach a cadence of 6000 sq.\ deg.\ per
night, surveying the entire sky (as seen from Hawaii, so $\sim$30,000 sq.\ deg.) over and over again in
five or six bands, the longest wavelength of which will be either the z- or y-band. 
The SkyMapper Telescope (Southern
Sky Survey, aka S3; \citealt{keller2007}) aims to observe all
$\sim$20,000 sq.\ deg.\ south of the celestial equator at u, v, g, r, i, and z bands with cadences
ranging from hours to years.
The Large
Synoptic Survey Telescope (LSST; http://www.lsst.org) will cover over 20,000 sq.\ deg.\ of the 
Southern Hemisphere in six bands, covering the entire area roughly one thousand times during the
course of the ten-year survey. Its longest wavelength filters will be z- and y-bands. Even though
based in the optical, all three of these surveys will probe deeply enough at $z$ and/or $y$ to canvass
the Solar Neighborhood for L and T dwarfs.

Figure \ref{IR_color_color_colored} shows the standard near-infrared color-color diagram with
our most interesting discoveries from the 2MASS Proper Motion Survey color-coded. With a notable
exception or two -- such as the extremely red color for 2MASS J2148+4003 -- most of these objects
fall in the same locus as normal late-M and L
dwarfs. M subdwarfs, however, by virtue of their bluer J-H and H-K$_s$ colors, fall below the locus
of K and M dwarfs on this diagram and could be selected by the
improved, very accurate colors of UKIDSS and
VISTA alone. Our late-M and L subdwarfs, show a wider range of colors. Sometimes these
objects are identifiable from their colors alone but sometimes (as in the case of 2MASS J1756+2815)
they are indistinguishable from much more common M dwarfs.

With the addition of proper motion information, plots like Figure \ref{red_motion_JK_colored} can be
used to find interesting objects. In this plot, the subdwarfs again distinguish themselves by
being bluer than the other late-type dwarfs and also by having very large reduced proper motions
(although this is partly a selection effect since we used this plot to prioritize candidates for
follow up). Low-gravity M and L dwarfs, red L dwarfs, blue L dwarfs, and early-T dwarfs share the
same locus on this diagram and cannot be easily distinguished from one another. Near-infrared
surveys like UKIDSS-LAS and the various VISTA surveys, even with a proper motion component, will have 
trouble distinguishing these objects from the sea of normal L dwarfs.

If we use the 2MASS J band as a proxy for $y$, we can envision what a similar reduced proper motion
diagram might look like for Pan-STARRS or LSST. In Figure \ref{red_motion_IJ_colored} we plot I-J
color as a proxy for i-y. As with Figure \ref{red_motion_JK_colored}, subdwarfs tend to separate
cleanly from the other objects, but other interesting types again fall in the same color 
locus. It is almost certainly the case that in the absence of spectroscopic follow-up, the only
hope of distinguishing these objects is to combine photometry from large-area surveys at 
different wavelengths and look for other optical/infrared colors that might help segregate
the types. Color combinations that come closest to the way these objects are defined may,
of course, have the best degree of success. An optical color such as i-z that can be used 
as a proxy for optical spectral type (\citealt{hawley2002}, \citealt{knapp2004}) can be plotted 
against J-K$_s$, in much the same way
that Figure 14 of \cite{kirkpatrick2008} can separate red L dwarfs and low-gravity objects from blue L
dwarfs and low-metallicity objects.

Finally, there is one new survey that -- by virtue of its all-sky coverage, multi-epoch data (though only
with an epoch difference of six months), and sensitivity at wavelengths where brown dwarfs have their
peak flux -- has the best chance of canvassing the immediate Solar Neighborhood completely for all
M, L, and T dwarf exotica. The Wide-field Infrared Survey Explorer (WISE; http://wise.ssl.berkeley.edu) was launched on 14 Dec 2009 and
promises to give a view of local brown dwarfs like no other previous study. Combining its four bands of
data (centered near 3.4, 4.6, 12, and 22 $\mu$m)
with other existing surveys will aid greatly in our ability to distinguish these exotica from the more
mundane M, L, and T dwarfs.

\section{Acknowledgments}

We would like to thank John Stauffer and Maria Morales-Calderon for giving us some telescope time 
in exchange for instrument expertise on 2005 Dec 09 (UT) at Keck-II. We thank Lee Rottler for 
assistance in installing and troubleshooting various science software packages in support of this 
paper. We thank Francesca Colonnese for help with intermediate versions of the tables for the 1X-6X 
survey and Rob Simcoe for taking some of the Magellan-Clay observations. We are indebted to the
support staff at each of the observatories we used; their expertise was instrumental in making this 
survey a success. We acknowledge use of The Digitized Sky Surveys, which were produced at the Space 
Telescope Science Institute under U.S. Government grant NAG W-2166. The images of these surveys are 
based on photographic data obtained using the Oschin Schmidt Telescope on Palomar Mountain and the 
UK Schmidt Telescope. This publication makes use of data products from the Two Micron All Sky Survey, 
which is a joint project of the University of Massachusetts and the Infrared Processing and Analysis 
Center/California Institute of Technology, funded by the National Aeronautics and Space Administration 
and the National Science Foundation. This research has made use of the NASA/IPAC Infrared Science Archive,
which is operated by the Jet Propulsion Laboratory, California Institute of Technology, under contract
with the National Aeronautics and Space Administration. Our research has benefitted from the M, L, and
T dwarf compendium housed at DwarfArchives.org, whose server was funded by a NASA Small Research Grant, 
administered by the American Astronomical Society. We are also indebted to the SIMBAD database,
operated at CDS, Strasbourg, France . We further wish to recognize and acknowledge the very significant 
cultural role and reverence that the summit of Mauna Kea has always had within the indigenous Hawaiian 
community. We are most fortunate to have the opportunity to conduct observations from this mountain.

\clearpage

\begin{deluxetable}{cccccccccccccccccccccc}
\tabletypesize{\scriptsize}
\rotate
\tablewidth{9in}
\tablenum{1}
\tablecaption{List of 2MASS Proper Motion Candidates\tablenotemark{a}\label{pm_list}}
\tablehead{
\colhead{2MASS Designation} & 
\colhead{List} &
\colhead{Band} &
\colhead{Spec.} &
\colhead{2MASS PM} &
\colhead{2M Theta} &
\colhead{Motion Sig.} &
\colhead{J} &
\colhead{H} &
\colhead{K$_s$} & 
\colhead{H$_J$} &
\colhead{USNO-B PM} &
\colhead{US Theta} &
\colhead{B1} & 
\colhead{B2} & 
\colhead{R1} & 
\colhead{R2} & 
\colhead{I} & 
\colhead{SIMBAD PM} & 
\colhead{SI Theta} & 
\colhead{SIMBAD Name} & 
\colhead{Spec.\ Type} \\

\colhead{J2000 coords} &
\colhead{} &
\colhead{} &
\colhead{} &
\colhead{(arcsec/yr)} &
\colhead{(deg)} &
\colhead{} &
\colhead{(mag)} &
\colhead{(mag)} &
\colhead{(mag)} &
\colhead{(mag)} &
\colhead{(arcsec/yr)} &
\colhead{(deg)} &
\colhead{(mag)} &
\colhead{(mag)} &
\colhead{(mag)} &
\colhead{(mag)} &
\colhead{(mag)} &
\colhead{(arcsec/yr)} &
\colhead{(deg)} &
\colhead{} &
\colhead{} \\

\colhead{(1)} &
\colhead{(2)} &
\colhead{(3)} &
\colhead{(4)} &
\colhead{(5)} &
\colhead{(6)} &
\colhead{(7)} &
\colhead{(8)} &
\colhead{(9)} &
\colhead{(10)} &
\colhead{(11)} &
\colhead{(12)} &
\colhead{(13)} &
\colhead{(14)} &
\colhead{(15)} &
\colhead{(16)} &
\colhead{(17)} &
\colhead{(18)} &
\colhead{(19)} &
\colhead{(20)} &
\colhead{(21)} &
\colhead{(22)} 
}

\startdata
J00020936$-$7715065& 1&  -&  -&   0.26$\pm$0.12&  143&   2.09&    8.942$\pm$0.021&    8.462$\pm$0.026&    8.375$\pm$0.023&  11.00&        ---           &   ---&   11.54&   11.07&   10.19&   10.14&    9.76&        0.21$\pm$0.01&   118&   NLTT 58838&                  ---\\
J00021076$-$4309560& 2&  -&  -&   1.00$\pm$0.14&  137&   6.74&   12.597$\pm$0.026&   12.425$\pm$0.023&   12.445$\pm$0.024&  17.59&       0.656$\pm$0.032&   181&   ---  &   20.77&   ---  &   13.04&   12.40&        1.02$\pm$0.14&   135&   LHS 1005  &                  DAs\\
J00023356+4343145  & 2&  -&  -&   0.32$\pm$0.06&   81&   4.90&   12.629$\pm$0.021&   12.066$\pm$0.021&   11.954$\pm$0.018&  15.18&       0.321$\pm$0.004&    78&   16.10&   15.65&   14.28&   14.31&   13.79&        0.35$\pm$0.02&    77&   G 171-37  &                  ---\\
J00024468$-$2428426& 2&  -&  -&   0.41$\pm$0.07&   99&   5.88&   11.765$\pm$0.021&   11.204$\pm$0.023&   10.957$\pm$0.025&  14.84&       0.404$\pm$0.003&    97&   17.18&   17.54&   14.73&   14.98&   13.22&        0.42$\pm$0.01&    98&   NLTT 10   &                  ---\\
J00024961$-$7636194& 1&  -&  -&   0.21$\pm$0.07&  267&   3.20&    9.884$\pm$0.023&    9.285$\pm$0.024&    9.118$\pm$0.023&  11.49&       0.219$\pm$---  &   237&   13.14&   12.76&   11.98&   11.94&   11.60&        0.22$\pm$0.01&   237&   NLTT 29   &                  ---\\
...\\
...\\
...\\
\enddata
\tablenotetext{a}{The full data set is contained in the online table.}
\end{deluxetable}

\clearpage

\begin{deluxetable}{lcccccll}
\tabletypesize{\scriptsize}
\rotate
\tablewidth{8in}
\tablenum{2}
\tablecaption{Spectroscopic Follow-up of Discoveries\label{spec_followup}}
\tablehead{
\colhead{2MASS Designation} & 
\colhead{Telescope} &
\colhead{Instr.} &
\colhead{Obs.\ Date} &
\colhead{Weather} &
\colhead{Int.} &
\colhead{Opt.\ Sp.\ Ty.} &
\colhead{NIR Sp.\ Ty.} \\
\colhead{J2000 coords} &
\colhead{}&
\colhead{} &
\colhead{UT} &
\colhead{} &
\colhead{sec} &
\colhead{} &
\colhead{} \\
\colhead{(1)} &
\colhead{(2)} &
\colhead{(3)} &
\colhead{(4)} &
\colhead{(5)} &
\colhead{(6)} &
\colhead{(7)} &
\colhead{(8)}
}

\startdata
2MASS J00040288$-$6410358 & Magellan  &  LDSS3  &  2006 Nov 21   & patchy clds  &  2700 &  L1 $\gamma$    &  ---\\    
2MASS J00054327+0836483   & Magellan  &  MagE   &  2008 Nov 27   & clear        &  1500 &  M4.5           & --- \\        
2MASS J00054517+0723423   & IRTF      &  SpeX   &  2007 Oct 12   & clear        &   480 &  ---            &  sdM6 \\
                          & Magellan  &  MagE   &  2008 Nov 27   & clear        &  1500 &  M4             &  --- \\
2MASS J00150206+2959323   & IRTF      &  SpeX   &  2005 Oct 18   & cirrus       &   960 &  ---            &  L7.5 pec (blue) \\ 
                          & Magellan  &  LDSS3  &  2008 Aug 13   & clear        &  1800 &  L7             &  --- \\    
2MASS J00163761+3448368   & IRTF      &  SpeX   &  2005 Sep 09   & cirrus       &   960 &  ---            &  M8.5  \\       
2MASS J00184613$-$6356122 & Magellan  &  LDSS3  &  2006 Nov 21   & patchy clds  &   900 &  L2             &  ---\\
\enddata
\tablecomments{Additional spectroscopic data for 2MASS J1821+1414 and 2MASS J2148+4003 are presented in Looper et al. 2008.}
\tablenotetext{a}{The near-infrared classification here is based only on a NIRSPEC N3 spectrum covering 1.15-1.35 $\mu$m.}
\tablenotetext{b}{2MASS J1043+1213: This discovery has since been independently confirmed and published by Chiu et al. 2006.}
\tablenotetext{c}{2MASS J1520$-$4422AB: Near-infrared spectra types are taken from Burgasser et al. 2007.}
\tablenotetext{d}{2MASS J1924+5506: This shares common proper motion with an object 10$\farcs$0 distant and at $PA = 115^\circ$. 
  This other object has a 960-sec IRTF-SpeX spectrum from 2005 Oct 19 (UT) that shows it to be a white dwarf.}
\tablenotetext{e}{2MASS J2148+4003: Spectral type based on a NIRSPEC N1+N2 spectrum covering 0.95-1.30 $\mu$m.}
\tablenotetext{f}{2MASS J0115$-$7519: This object was found in both the 1X-6X and the 1X-1X survey.}
\tablenotetext{g}{2MASS J0100+1711: This object is also known as LHS 1174.}
\end{deluxetable}

\clearpage

\begin{deluxetable}{lcccccllll}
\tabletypesize{\scriptsize}
\rotate
\tablewidth{9in}
\tablenum{3}
\tablecaption{Supporting Observations of M and L Dwarfs and Subdwarfs\label{sd_spectra}}
\tablehead{
\colhead{2MASS Designation} & 
\colhead{Other Name} &
\colhead{Telescope} &
\colhead{Instr.} &
\colhead{Obs.\ Date} &
\colhead{Weather} &
\colhead{Int.} &
\colhead{Opt.\ Sp.\ Ty.} &
\colhead{New Opt.} &
\colhead{Adopted} \\
\colhead{J2000 coords} &
\colhead{}&
\colhead{}&
\colhead{} &
\colhead{UT} &
\colhead{} &
\colhead{sec} &
\colhead{} &
\colhead{Sp.\ Type\tablenotemark{a}} &
\colhead{Type} \\
\colhead{(1)} &
\colhead{(2)} &
\colhead{(3)} &
\colhead{(4)} &
\colhead{(5)} &
\colhead{(6)} &
\colhead{(7)} &
\colhead{(8)} &
\colhead{(9)} &
\colhead{(10)}
}

\startdata
2MASS J00114208$-$2057129& LHS 1035                   & Gemini-North & GMOS & 2004 Oct 06 & clear &  600 & sdM6 &     ---\tablenotemark{b}    & d/sdM5 \\
2MASS J00433134+2054316  & LHS 1135                   & Gemini-North & GMOS & 2004 Oct 06 & clear &  900 & sdM6.5 &   ---\tablenotemark{b}    & d/sdM5 \\
2MASS J01031971+6221557  & Gl 51                      & Keck-II & DEIMOS    & 2004 Jul 05 & clear &  420 & M5 &       ---                     & M5 \\
2MASS J01572792+0116433  & LP 589-7                   & Gemini-North & GMOS & 2004 Sep 12 & clear & 1200 & esdM5 &    esdM5.5                 & esdM5.5 \\
2MASS J05103896+1924078  & LHS 1742a                  & Gemini-North & GMOS & 2004 Sep 13 & clear & 1200 & esdM5.5 &  esdM6.5                 & esdM6.5 \\
2MASS J05325346+8246465  & ---                        & Gemini-North & GMOS & 2004 Nov 12 & clear & 3600 & (e?)sdL7 & ---                     & (e?)sdL7 \\
2MASS J05585891$-$2903267& APMPM J0559$-$2903         & Gemini-North & GMOS & 2004 Nov 08 & clear & 1200 & esdM7 &    esdM7.5                 & esdM7.5\tablenotemark{c} \\
2MASS J08223369+1700199  & LSR J0822+1700             & Gemini-North & GMOS & 2004 Nov 08 & clear & 2400 & esdM6.5 &  usdM8                   & usdM7.5\tablenotemark{d} \\
\enddata
\tablenotetext{a}{Typed on the subdwarf classification system of \cite{lepine2007} using the spectra obtained here.}
\tablenotetext{b}{Does not warrant a subdwarf classification on the new \cite{lepine2007} system.}
\tablenotetext{c}{Established as the esdM7.5 standard in \cite{lepine2007}.}
\tablenotetext{d}{Established as the usdM7.5 standard in \cite{lepine2007}.}
\tablenotetext{e}{Established as the sdM8 standard in \cite{lepine2007}.}
\tablenotetext{f}{See \cite{cushing-schizo} for more information on the classification and interpretation of LSR J1610$-$0040.}
\end{deluxetable}

\clearpage

\begin{deluxetable}{lccccccc}
\tabletypesize{\scriptsize}
\rotate
\tablewidth{8in}
\tablenum{4}
\tablecaption{New M and L Dwarfs Spectral Standards in the Near-infrared\label{nir_stds}}
\tablehead{
\colhead{Designation} & 
\colhead{Other Name} &
\colhead{Adopted Near-IR} &
\colhead{2MASS $J$} &
\colhead{2MASS $J-H$} &
\colhead{2MASS $J-K_s$} &
\colhead{Published Opt.} &
\colhead{Optical}  \\
\colhead{J2000 coords} &
\colhead{}&
\colhead{Spec.\ Type}&
\colhead{(mag)} &
\colhead{(mag)} &
\colhead{(mag)} &
\colhead{Spec.\ Ty.} &
\colhead{Ref.} \\
\colhead{(1)} &
\colhead{(2)} &
\colhead{(3)} &
\colhead{(4)} &
\colhead{(5)} &
\colhead{(6)} &
\colhead{(7)} &
\colhead{(8)}
}
\startdata
2MASS J07193126+3249482  & Gl 270   & M0 &   7.184$\pm$0.020 & 0.643$\pm$0.039 & 0.808$\pm$0.028 & M0 & 3 \\
2MASS J11200526+6550470  & Gl 424   & M1 &   6.306$\pm$0.018 & 0.576$\pm$0.027 & 0.772$\pm$0.025 & M1 & 8 \\
2MASS J02135359$-$3202282& Gl 91    & M2 &   6.961$\pm$0.020 & 0.634$\pm$0.033 & 0.868$\pm$0.026 & M2 & 8 \\
2MASS J19165526+0510086  & Gl 752A  & M3 &   5.583$\pm$0.025 & 0.654$\pm$0.040 & 0.910$\pm$0.028 & M3 & 3 \\
2MASS J05420897+1229252  & Gl 213   & M4 &   7.124$\pm$0.013 & 0.497$\pm$0.028 & 0.735$\pm$0.015 & M4 & 3 \\
2MASS J01031971+6221557  & Gl 51    & M5 &   8.611$\pm$0.021 & 0.597$\pm$0.035 & 0.893$\pm$0.025 & M5 & 3 \\
2MASS J02162977+1335136  & LHS 1375 & M6 &   9.871$\pm$0.021 & 0.557$\pm$0.030 & 0.890$\pm$0.021 & M6 & 8 \\
2MASS J16553529$-$0823401& vB 8     & M7 &   9.776$\pm$0.026 & 0.575$\pm$0.038 & 0.960$\pm$0.033 & M7 & 3 \\
\enddata
\tablecomments{References: (1) \cite{kirkpatrick2000}, (2) \cite{cruz2003}, (3) \cite{kirkpatrick1991}, 
(4) \cite{kirkpatrick1995}, (5) \cite{kirkpatrick1999}, (6) \cite{kirkpatrick2008}, (7)
\cite{gizis2000}, (8) unpublished from Kirkpatrick, as listed in DwarfArchives.org.}
\end{deluxetable}

\clearpage

\begin{deluxetable}{lllccc}
\tabletypesize{\scriptsize}
\rotate
\tablewidth{7in}
\tablenum{5}
\tablecaption{New Low-gravity Objects\label{lowg_table}}
\tablehead{
\colhead{2MASS Designation} & 
\colhead{Opt.\ Sp.\ Ty.} &
\colhead{NIR Sp.\ Ty.} &
\colhead{J-H} &
\colhead{H-K$_s$} &
\colhead{J-K$_s$} \\
\colhead{(1)} &
\colhead{(2)}&
\colhead{(3)} &
\colhead{(4)} &
\colhead{(5)} &
\colhead{(6)} 
}
\startdata
2MASS J00040288$-$6410358 & L1 $\gamma$ & --- &     0.955$\pm$0.103 & 0.822$\pm$0.087 & 1.777$\pm$0.113 \\
2MASS J02340093$-$6442068 & L0 $\gamma$ & --- &     0.883$\pm$0.089 & 0.590$\pm$0.098 & 1.473$\pm$0.093 \\
2MASS J03032042$-$7312300 & L2 $\gamma$ & --- &     1.041$\pm$0.137 & 0.775$\pm$0.120 & 1.816$\pm$0.137 \\
2MASS J04062677$-$3812102 & L0 $\gamma$ & L1 pec &  1.057$\pm$0.162 & 0.598$\pm$0.154 & 1.655$\pm$0.172 \\
2MASS J05341594$-$0631397 & M8 $\gamma$ & M8 pec &  0.685$\pm$0.124 & 0.433$\pm$0.137 & 1.118$\pm$0.124 \\ 
2MASS J15575011$-$2952431 & M9 $\delta$ & M9 pec &  0.866$\pm$0.159 & 0.605$\pm$0.153 & 1.471$\pm$0.163 \\
\enddata
\tablecomments{Optical spectral types use the Greek suffix designations of \cite{cruz2009} in 
which $\gamma$ is used to denote low gravity (as opposed to normal [$\alpha$, usually dropped for convenience] 
or intermediate [$\beta$] gravity). 
Objects of extremely low gravity are further denoted by $\delta$, as originally outlined in 
\cite{kirkpatrick2005,kirkpatrick2006}.}
\end{deluxetable}

\begin{deluxetable}{lllccc}
\tabletypesize{\scriptsize}
\rotate
\tablewidth{7in}
\tablenum{6}
\tablecaption{New Red L Dwarfs\label{redL_table}}
\tablehead{
\colhead{2MASS Designation} & 
\colhead{Opt.\ Sp.\ Ty.} &
\colhead{NIR Sp.\ Ty.} &
\colhead{J-H} &
\colhead{H-K$_s$} &
\colhead{J-K$_s$} \\
\colhead{(1)} &
\colhead{(2)}&
\colhead{(3)} &
\colhead{(4)} &
\colhead{(5)} &
\colhead{(6)} 
}
\startdata
2MASS J13313310+3407583   & L0     & L1 pec (red)&   0.932$\pm$0.041 & 0.516$\pm$0.038 & 1.448$\pm$0.035 \\
2MASS J18212815+1414010\tablenotemark{a}   & L4.5   & L5 pec (red)&   1.035$\pm$0.031 & 0.746$\pm$0.028 & 1.781$\pm$0.032 \\
2MASS J21481633+4003594\tablenotemark{a}   & L6     & L6.5 pec (red)& 1.364$\pm$0.041 & 1.018$\pm$0.038 & 2.382$\pm$0.037 \\
2MASS J23174712$-$4838501 & L4 pec & L6.5 pec (red)& 1.225$\pm$0.044 & 0.744$\pm$0.036 & 1.969$\pm$0.048 \\
2MASS J23512200+3010540   & L5.5   & L5 pec (red)&   1.278$\pm$0.123 & 0.553$\pm$0.093 & 1.831$\pm$0.118 \\
\enddata
\tablenotetext{a}{See \cite{looper2008} for in-depth discussion.}
\end{deluxetable}

\clearpage

\begin{deluxetable}{lllccc}
\tabletypesize{\scriptsize}
\rotate
\tablewidth{7in}
\tablenum{7}
\tablecaption{New M and L Subdwarfs\label{sd_table}}
\tablehead{
\colhead{2MASS Designation} & 
\colhead{Opt.\ Sp.\ Ty.} &
\colhead{NIR Sp.\ Ty.} &
\colhead{J-H} &
\colhead{H-K$_s$} &
\colhead{J-K$_s$} \\
\colhead{(1)} &
\colhead{(2)}&
\colhead{(3)} &
\colhead{(4)} &
\colhead{(5)} &
\colhead{(6)} 
}
\startdata
2MASS J00054517+0723423   & M4     & sdM6 &    0.467$\pm$0.072 & 0.314$\pm$0.089 & 0.781$\pm$0.072 \\
2MASS J00524886+1712434   & ---    & esdM5 &   0.471$\pm$0.124 & 0.180$\pm$0.156 & 0.651$\pm$0.134 \\
2MASS J00554279+1301043   & M6     & d/sdM7 &  0.650$\pm$0.112 & 0.332$\pm$0.124 & 0.982$\pm$0.114 \\
2MASS J03073402$-$1736375 & usdM2  & --- &     0.313$\pm$0.112 & 0.390$\pm$0.148 & 0.703$\pm$0.132 \\
2MASS J03440399+2730504   & sdM3.5 & --- &     0.549$\pm$0.078 & 0.227$\pm$0.100 & 0.776$\pm$0.089 \\
2MASS J04470652$-$1946392 & sdM7.5 & sdM7.5 &  0.230$\pm$0.084 & 0.153$\pm$0.134 & 0.383$\pm$0.123 \\
2MASS J06211484+6558164   & esdM0  & $<$esdM5& 0.472$\pm$0.042 & 0.199$\pm$0.040 & 0.671$\pm$0.034 \\
2MASS J06453153$-$6646120 & sdL8   & sdL8 &    0.874$\pm$0.044 & 0.385$\pm$0.051 & 1.259$\pm$0.050 \\
2MASS J10462067+2354307   & ---    & sdM6 &    0.440$\pm$0.106 & 0.166$\pm$0.156 & 0.606$\pm$0.145 \\
2MASS J11582077+0435014   & sdL7   & sdL7 &    0.927$\pm$0.085 & 0.245$\pm$0.091 & 1.172$\pm$0.084 \\
2MASS J14270666+4808164   & ---    & $\le$sdM5&0.455$\pm$0.064 & 0.076$\pm$0.095 & 0.531$\pm$0.090 \\
2MASS J16130315+6502051   & ---    & sdM6 &    0.535$\pm$0.101 & 0.256$\pm$0.113 & 0.791$\pm$0.107 \\
2MASS J17561080+2815238   & sdL1   & L1 pec (blue)&    0.577$\pm$0.053 & 0.322$\pm$0.058 & 0.899$\pm$0.053 \\
2MASS J18355309$-$3217129 & M6.5   & d/sdM7 &  0.368$\pm$0.101 & 0.358$\pm$0.129 & 0.726$\pm$0.105 \\
2MASS J23470713+0219127   & ---    & d/sdM7 &  0.489$\pm$0.040 & 0.304$\pm$0.041 & 0.793$\pm$0.036 \\
\enddata
\end{deluxetable}

\begin{deluxetable}{lcccc}
\tabletypesize{\scriptsize}
\rotate
\tablewidth{5in}
\tablenum{8}
\tablecaption{Known L Subdwarfs\tablenotemark{a}\label{sdLs}}
\tablehead{
\colhead{2MASS Designation} & 
\colhead{Ref.} &
\colhead{Opt.\ Type} &
\colhead{NIR Type} &
\colhead{2MASS J-K$_s$} \\
\colhead{J2000 coords} &
\colhead{} &
\colhead{} &
\colhead{} &
\colhead{} \\
\colhead{(1)} &
\colhead{(2)} &
\colhead{(3)} &
\colhead{(4)} &
\colhead{(5)}
}

\startdata
2MASS J05325346+8246465   &  2 & (e?)sdL7   & (e?)sdL7      & 0.26$\pm$0.16 \\
2MASS J06164006$-$6407194 &  3 & sdL5       & sdL:          & --- \\
2MASS J06453153$-$6646120 &  1 & sdL8       & sdL8          & 1.26$\pm$0.05 \\
2MASS J11582077+0435014   &  1 & sdL7       & sdL7          & 1.17$\pm$0.08 \\
SDSS J125637.13$-$022452.4&  4 & sdL3.5     & sdL3.5        & --- \\
SDSS J141624.08+134826.7\tablenotemark{b}& 6& sdL7& sdL7    & 1.03$\pm$0.03 \\
2MASS J16262034+3925190   &  5 & sdL4       & sdL4          & $-$0.03$\pm$0.08 \\
2MASS J17561080+2815238   &  1 & sdL1       & L1 pec (blue) & 0.90$\pm$0.05 \\
\enddata
\tablenotetext{a}{Another purported L subdwarf, ULAS J135058.86+081506.8, has been published by \cite{lodieu2010},
but the limited spectroscopic evidence cannot distinguish between an L subdwarf and a blue L dwarf.}
\tablenotetext{b}{Both \cite{schmidt2009} and \cite{bowler2009} prefer to call this object a blue L dwarf rather 
than an L subdwarf. Although
low metallicity signatures are more subtle than in 2MASS J0532+8246, we believe that they still point to a
subdwarf nature for this object.}
\tablecomments{References: (1) this paper, (2) \cite{burgasser2003}, (3) \cite{cushing2009}, (4) Sivarani et al. via \cite{burgasser2007},
(5) \cite{burgasser2004}, (6) \cite{schmidt2009} and \cite{bowler2009}.}
\end{deluxetable}

\clearpage

\begin{deluxetable}{lllccc}
\tabletypesize{\scriptsize}
\rotate
\tablewidth{7in}
\tablenum{9}
\tablecaption{New Blue L Dwarfs\label{blueL_table}}
\tablehead{
\colhead{2MASS Designation} & 
\colhead{Opt.\ Sp.\ Ty.} &
\colhead{NIR Sp.\ Ty.} &
\colhead{J-H} &
\colhead{H-K$_s$} &
\colhead{J-K$_s$} \\
\colhead{(1)} &
\colhead{(2)}&
\colhead{(3)} &
\colhead{(4)} &
\colhead{(5)} &
\colhead{(6)} 
}
\startdata
2MASS J00150206+2959323   & L7     & L7.5 pec (blue)&         0.932$\pm$0.117 & 0.744$\pm$0.111 & 1.676$\pm$0.109 \\
2MASS J03001631+2130205   & ---    & L6 pec (sl.\ blue)&      1.146$\pm$0.103 & 0.420$\pm$0.091 & 1.566$\pm$0.098 \\
2MASS J10461875+4441149   & ---    & L5 pec (sl.\ blue)&      0.966$\pm$0.082 & 0.526$\pm$0.079 & 1.492$\pm$0.078 \\
2MASS J11181292$-$0856106 & L6     & L6 pec (blue) &          0.970$\pm$0.107 & 0.590$\pm$0.103 & 1.560$\pm$0.102 \\
2MASS J13023811+5650212   & L2     & L3: pec (sl.\ blue) &    0.891$\pm$0.160 & 0.497$\pm$0.162 & 1.388$\pm$0.173 \\
2MASS J14403186$-$1303263 & L1     & L1 pec (sl.\ blue)&      0.660$\pm$0.096 & 0.468$\pm$0.105 & 1.128$\pm$0.082 \\
2MASS J19495702+6222440   & L2 pec (blue)    & L2 pec (sl.\ blue)&      0.889$\pm$0.181 & 0.720$\pm$0.173 & 1.609$\pm$0.167 \\
2MASS J21513979+3402444   & ---    & L7 pec (sl.\ blue)&      0.908$\pm$0.224 & 0.813$\pm$0.202 & 1.721$\pm$0.187 \\
\enddata
\end{deluxetable}

\clearpage

\begin{deluxetable}{lcccccclll}
\tabletypesize{\scriptsize}
\rotate
\tablewidth{9in}
\tablenum{10}
\tablecaption{Known Blue L Dwarfs\label{blue_Ls}}
\tablehead{
\colhead{2MASS Designation} & 
\colhead{Discovery} &
\colhead{${\mu}$} &
\colhead{Dist.} &
\colhead{V$_{tan}$} &
\colhead{Kinematic} &
\colhead{Opt.} &
\colhead{Near-IR} &
\colhead{Sp.\ Type} &
\colhead{2MASS J-K$_s$} \\
\colhead{J2000 coords} &
\colhead{Ref.} &
\colhead{($\arcsec$/yr)} &
\colhead{(pc)} &
\colhead{(km/s)} &
\colhead{Ref.} &
\colhead{Sp.\ Type} &
\colhead{Sp.\ Type} &
\colhead{Ref.} &
\colhead{(mag)} \\
\colhead{(1)} &
\colhead{(2)} &
\colhead{(3)} &
\colhead{(4)} &
\colhead{(5)} &
\colhead{(6)} &
\colhead{(7)} &
\colhead{(8)} &
\colhead{(9)} &
\colhead{(10)} 
}

\startdata
2MASS J00150206+2959323    & 1 & 0.50$\pm$0.09 & 21:         & 50$\pm$9   & 1 & L7     & L7.5 pec (blue)      & 1      & 1.676$\pm$0.109 \\
2MASS J03001631+2130205    & 1 & 0.27$\pm$0.19 & 22:         & 28$\pm$20  & 1 & ---    & L6 pec (sl.\ blue)   & 1      & 1.566$\pm$0.098 \\
SDSS J080531.84+481233.0\tablenotemark{c} & 2 & 0.46$\pm$0.02 & 15$\pm$3    & 32$\pm$6   & 8 & L4     & L9$\pm$1.5           & 14,2   & 1.290$\pm$0.055 \\ 
SIPS J0921$-$2104\tablenotemark{a} & 6 & 0.94$\pm$0.02 & 12$\pm$1    & 56$\pm$4   & 8 & L1.5   & L4$\pm$1.5           & 11,10  & 1.089$\pm$0.033 \\
SDSS J093109.56+032732.5   & 2 & 0.63$\pm$0.03 & 36$\pm$8    & 108$\pm$23 & 8 & ---    & L7.5$\pm$1.5         & 2      & --- \\ 
SDSS J103321.92+400549.5   & 5 & 0.24$\pm$0.02 & 45$\pm$8    & 53$\pm$10  & 8 & ---    & L6                   & 5      & --- \\
2MASS J10461875+4441149    & 1 & 0.27$\pm$0.07 & 24:         & 31$\pm$8   & 1 & ---    & L5 pec (sl.\ blue)   & 1      & 1.492$\pm$0.078 \\
2MASS J11181292$-$0856106  & 1 & 0.59$\pm$0.14 & 21:         & 59$\pm$14  & 1 & L6     & L6 pec (blue)        & 1      & 1.560$\pm$0.101 \\
\enddata
\tablecomments{References: (1) this paper, (2) \cite{knapp2004}, (3) \cite{cruz2003}, (4) \cite{folkes2007}, (5) \cite{chiu2006}, 
(6) \cite{deacon2005}, (7) \cite{radigan2008}, (8) \cite{faherty2009}, (9) \cite{sheppard2008}, (10) \cite{burgasser2008}, 
(11) \cite{reid2008}, (12) \cite{zhang2009} (13) \cite{gizis2000}, (14) \cite{hawley2002}.}
\tablenotetext{a}{SIPS J0921$-$2104: Also known as 2MASS J09211411$-$2104446.}
\tablenotetext{b}{2MASS J1711+4028: Also known as G 203-50B. Optical spectral type is determined here for the first time using 
a spectrum acquired on 2009 Oct 11 (UT) at Keck-I with LRIS (1200s integration).}
\tablenotetext{c}{SDSS J0805+4812: Suspected binary comprised of mid-L and mid-T components; \cite{burgasser2007}.}
\end{deluxetable}

\clearpage

\begin{deluxetable}{lllccc}
\tabletypesize{\scriptsize}
\rotate
\tablewidth{7in}
\tablenum{11}
\tablecaption{New T Dwarfs\label{T_table}}
\tablehead{
\colhead{2MASS Designation} & 
\colhead{Opt.\ Sp.\ Ty.} &
\colhead{NIR Sp.\ Ty.} &
\colhead{J-H} &
\colhead{H-K$_s$} &
\colhead{J-K$_s$} \\
\colhead{(1)} &
\colhead{(2)}&
\colhead{(3)} &
\colhead{(4)} &
\colhead{(5)} &
\colhead{(6)} 
}
\startdata
2MASS J11061191+2754215   & ---    & T2.5 &    0.674$\pm$0.069 & 0.349$\pm$0.074 & 1.023$\pm$0.066 \\
2MASS J13243553+6358281   & ---    & T2.5 pec &1.020$\pm$0.088 & 0.518$\pm$0.082 & 1.538$\pm$0.090\\
2MASS J14044948$-$3159330 & ear-T  & T3   &    0.622$\pm$0.092 & 0.417$\pm$0.116 & 1.039$\pm$0.113 \\
2MASS J15111466+0607431\tablenotemark{a}   & ---    & T2 pec &  1.061$\pm$0.109 & 0.411$\pm$0.125 & 1.472$\pm$0.127 \\
2MASS J21265916+7617440   & L7     & T0 pec &  0.750$\pm$0.050 & 0.426$\pm$0.054 & 1.176$\pm$0.050 \\
\enddata
\tablenotetext{a}{Discovered independently by \cite{chiu2006}.}
\end{deluxetable}

\clearpage

\begin{deluxetable}{lllccc}
\tabletypesize{\scriptsize}
\rotate
\tablewidth{7in}
\tablenum{12}
\tablecaption{Other Interesting Discoveries\label{other_table}}
\tablehead{
\colhead{2MASS Designation} & 
\colhead{Opt.\ Sp.\ Ty.} &
\colhead{NIR Sp.\ Ty.} &
\colhead{J-H} &
\colhead{H-K$_s$} &
\colhead{J-K$_s$} \\
\colhead{(1)} &
\colhead{(2)}&
\colhead{(3)} &
\colhead{(4)} &
\colhead{(5)} &
\colhead{(6)} 
}
\startdata
2MASS J04205430$-$7327392 & ---       & ---         & 1.354$\pm$0.034 & 1.602$\pm$0.032 & 2.956$\pm$0.032 \\
2MASS J09073735$-$1457036 & ---       & ---         & 0.132$\pm$0.115 & 0.343$\pm$0.183 & 0.475$\pm$0.156 \\
2MASS J16002647$-$2456424 & ---       & M7.5 pec    & 0.600$\pm$0.084 & 0.301$\pm$0.103 & 0.901$\pm$0.084 \\
2MASS J18284076+1229207   & M8 sl pec & M7.5 pec    & 0.471$\pm$0.066 & 0.566$\pm$0.082 & 1.037$\pm$0.078 \\
2MASS J21403907+3655563   & ---       & M8 pec      & 0.539$\pm$0.103 & 0.396$\pm$0.132 & 0.935$\pm$0.127 \\
\enddata
\end{deluxetable}

\clearpage

\begin{deluxetable}{clllcccc}
\tabletypesize{\scriptsize}
\tablewidth{7in}
\tablenum{13}
\tablecaption{Measured Velocities and Distance Estimates\label{RV_table}}
\tablehead{
\colhead{2MASS Designation} & 
\colhead{Source\tablenotemark{a}} &
\colhead{Sp.\ Ty.} &
\colhead{Obs.\ Date} &
\colhead{$v_{rad}$} &
\colhead{Dist.\ Est.} &
\colhead{$v_{tan}$} &
\colhead{Tot.\ Space Vel.} \\
\colhead{} & 
\colhead{} &
\colhead{} &
\colhead{(UT)} &
\colhead{(km/s)} &
\colhead{(pc)} &
\colhead{(km/s)} &
\colhead{(km/s)} \\
\colhead{(1)} &
\colhead{(2)}&
\colhead{(3)} &
\colhead{(4)} &
\colhead{(5)} &
\colhead{(6)} &
\colhead{(7)} &
\colhead{(8)} 
}
\startdata
\\
\multicolumn{5}{l}{Low-gravity Objects}\\
\\
2MASS J00040288$-$6410358 & optical & L1 $\gamma$ & 2006 Nov 21 &  $-$57$\pm$50 &  57 &  92$\pm$54 & 108$\pm$74 \\
2MASS J02340093$-$6442068 & optical & L0 $\gamma$ & 2005 Dec 04 &  $-$46$\pm$50 &  45 & 133$\pm$109& 140$\pm$120\\
2MASS J03032042$-$7312300 & optical & L2 $\gamma$ & 2006 Nov 21 &  $-$21$\pm$50 &  59 & 382$\pm$256& 383$\pm$261\\ 
2MASS J04062677$-$3812102 & optical & L0 $\gamma$ & 2009 Oct 11 &  $-$76$\pm$50 &  90 & 103$\pm$120& 128$\pm$130\\
(ditto)                   & optical & L0 $\gamma$ & 2009 Oct 12 &  $-$67$\pm$50 &  90 & 103$\pm$120& 123$\pm$130\\
\enddata
\tablenotetext{a}{Type of spectrum, as given in Table \ref{spec_followup}, from which the listed spectral type and radial velocity were measured.}
\tablecomments{See \S\ref{lowg-section} and \ref{subdwarf-section} for an object-by-object discussion of the
distance estimates for the low-gravity objects and subdwarfs. Distances to the red and blue L dwarfs have been 
estimated with the \cite{looper2008} relations and use the
listed spectral type and measured 2MASS J magniutde.}
\end{deluxetable}

\clearpage

\clearpage

\begin{figure}
\epsscale{0.9}
\plotone{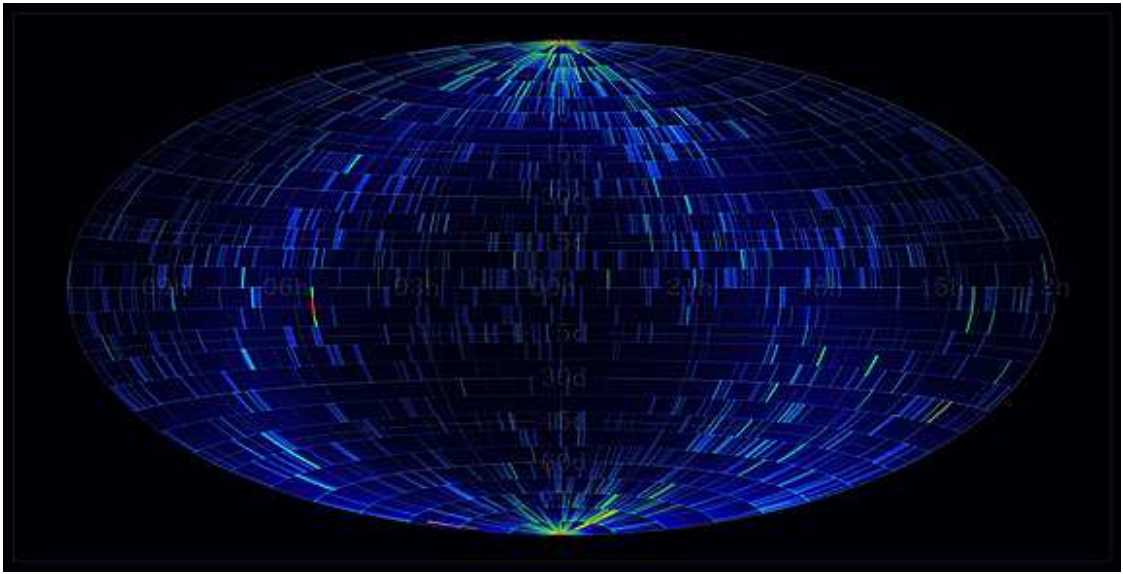}
\caption{Equatorial projection showing multi-epoch coverage for the 2MASS 1X survey. The ``thermal'' color stretch shows the number of coverages corresponding to each reobserved tile, where blue is the fewest reobservations and red the most. Multi-epoch coverage of Orion can be seen as the red and green strip at center left.  \label{1X_coverage}}
\end{figure}

\clearpage

\begin{figure}
\epsscale{0.9}
\plotone{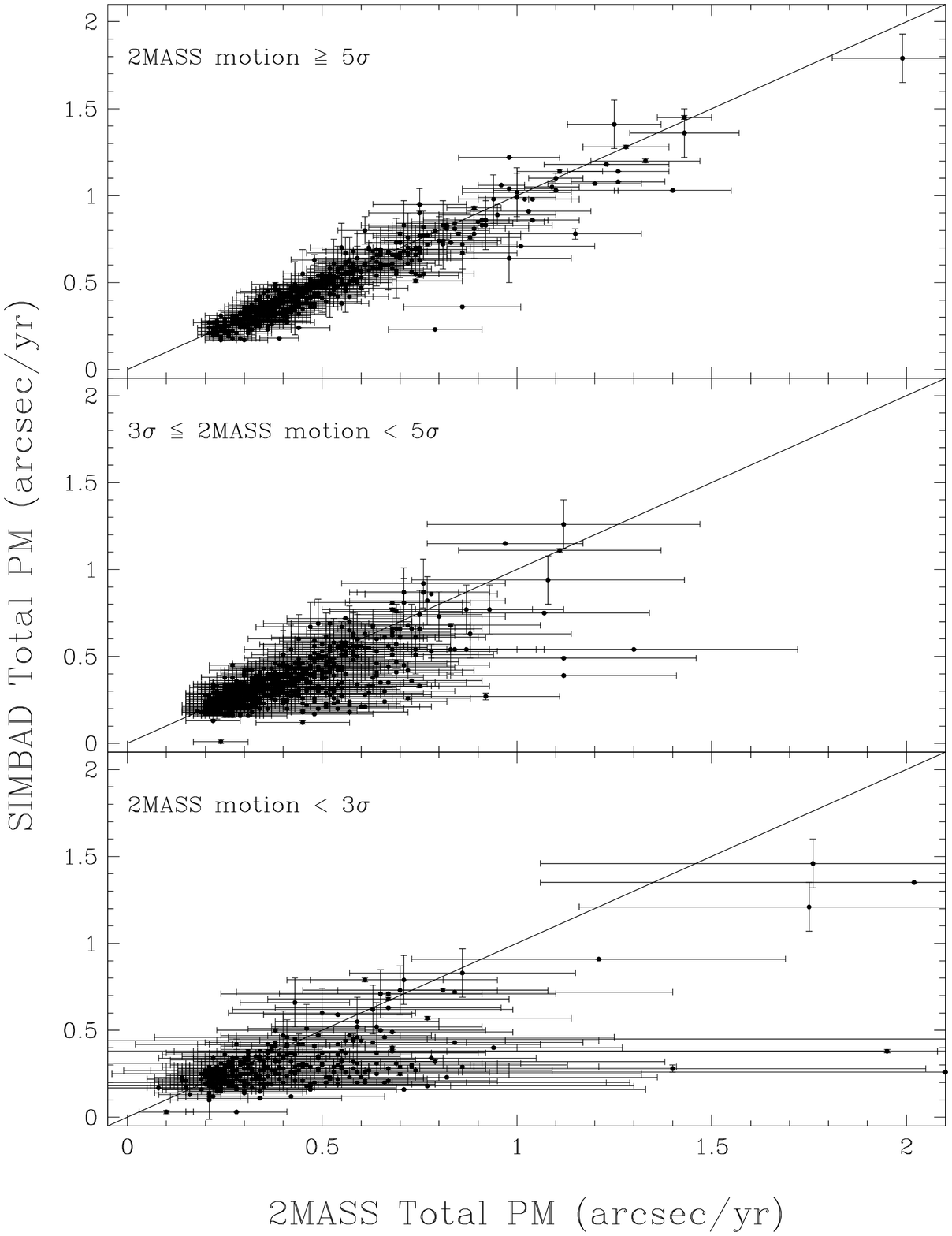}
\caption{Comparison of 2MASS derived proper motions to those listed at SIMBAD. \label{SIMBADpm_vs_2MASSpm}}
\end{figure}

\clearpage

\begin{figure}
\epsscale{0.9}
\plotone{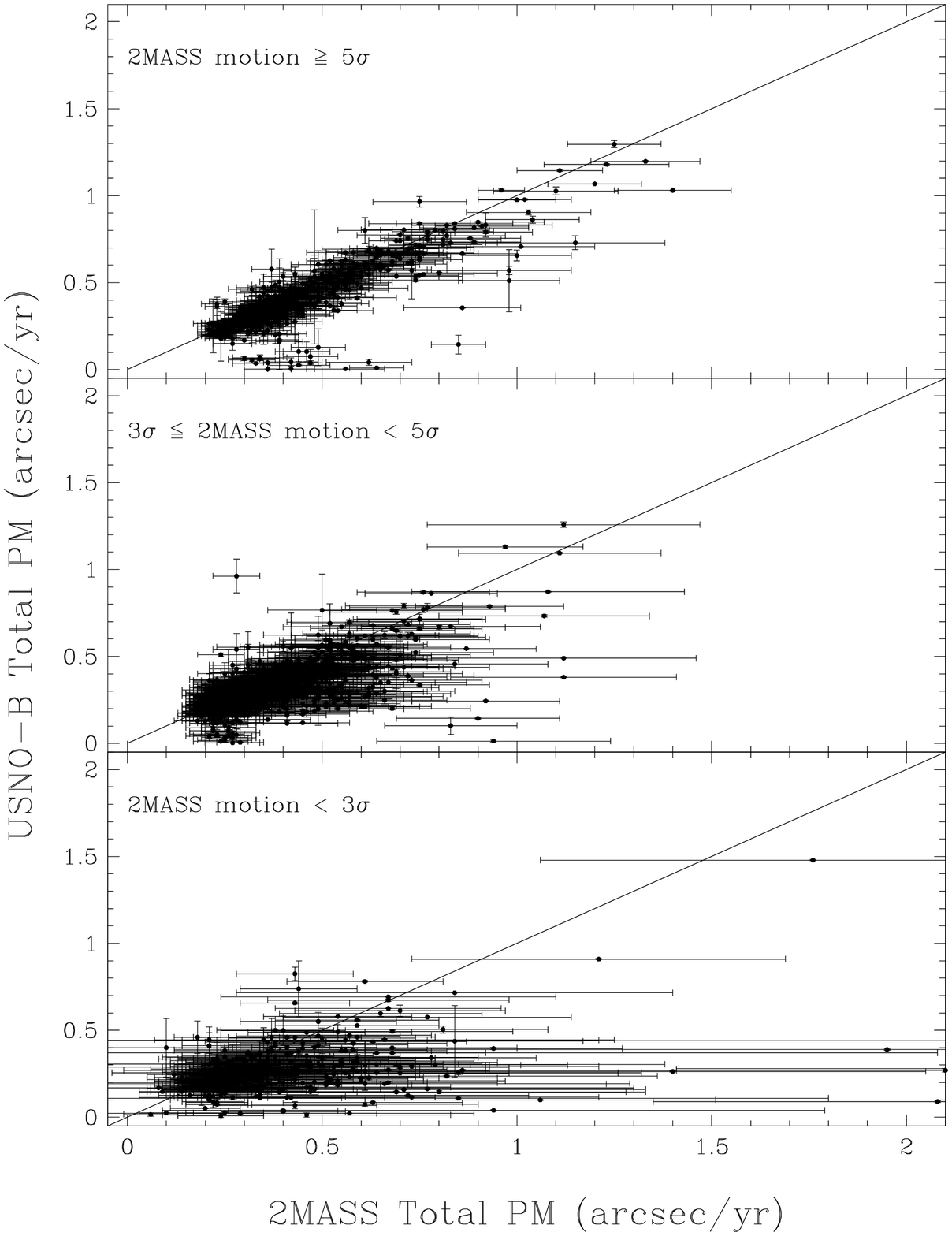}
\caption{Comparison of 2MASS derived proper motions to those listed in USNO-B Catalog. \label{USNOBpm_vs_2MASSpm}}
\end{figure}

\clearpage

\begin{figure}
\epsscale{0.9}
\plotone{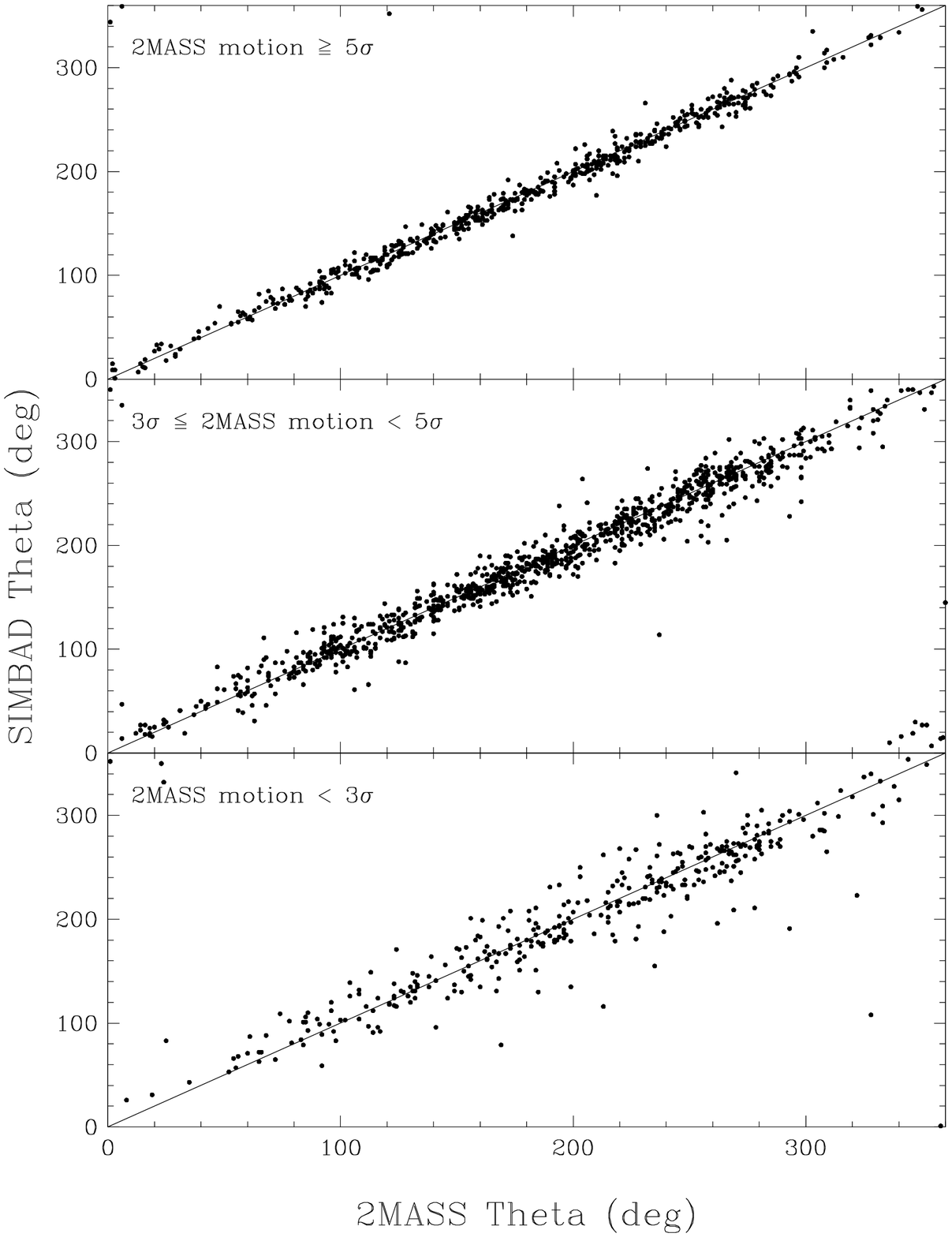}
\caption{Comparison of 2MASS derived proper motion thetas to those listed at SIMBAD. \label{SIMBADpm_vs_2MASSpm_angles}}
\end{figure}

\clearpage

\begin{figure}
\epsscale{0.9}
\plotone{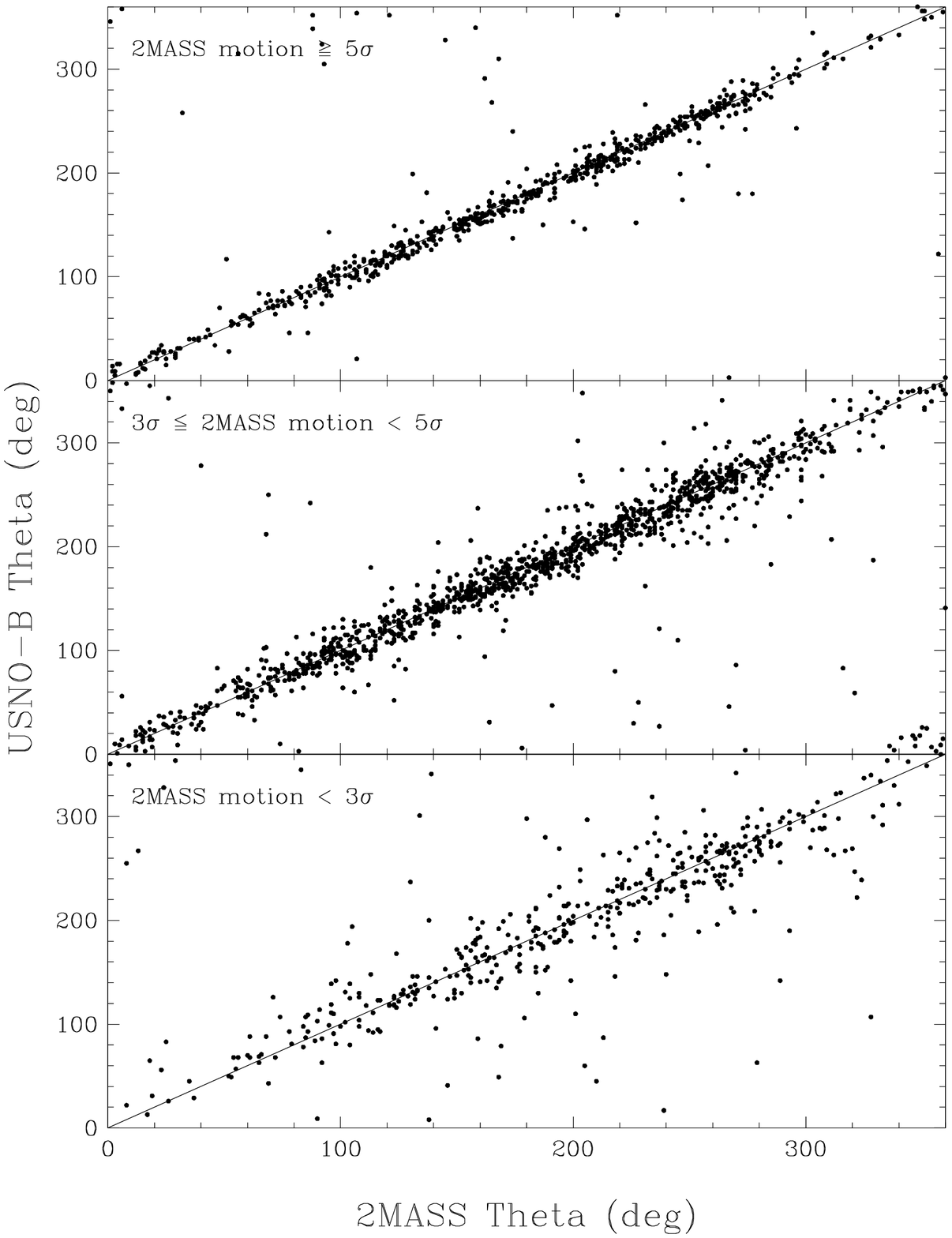}
\caption{Comparison of 2MASS derived proper motion thetas to those listed in USNO-B Catalog. \label{USNOBpm_vs_2MASSpm_angles}}
\end{figure}

\clearpage

\begin{figure}
\epsscale{0.7}
\plotone{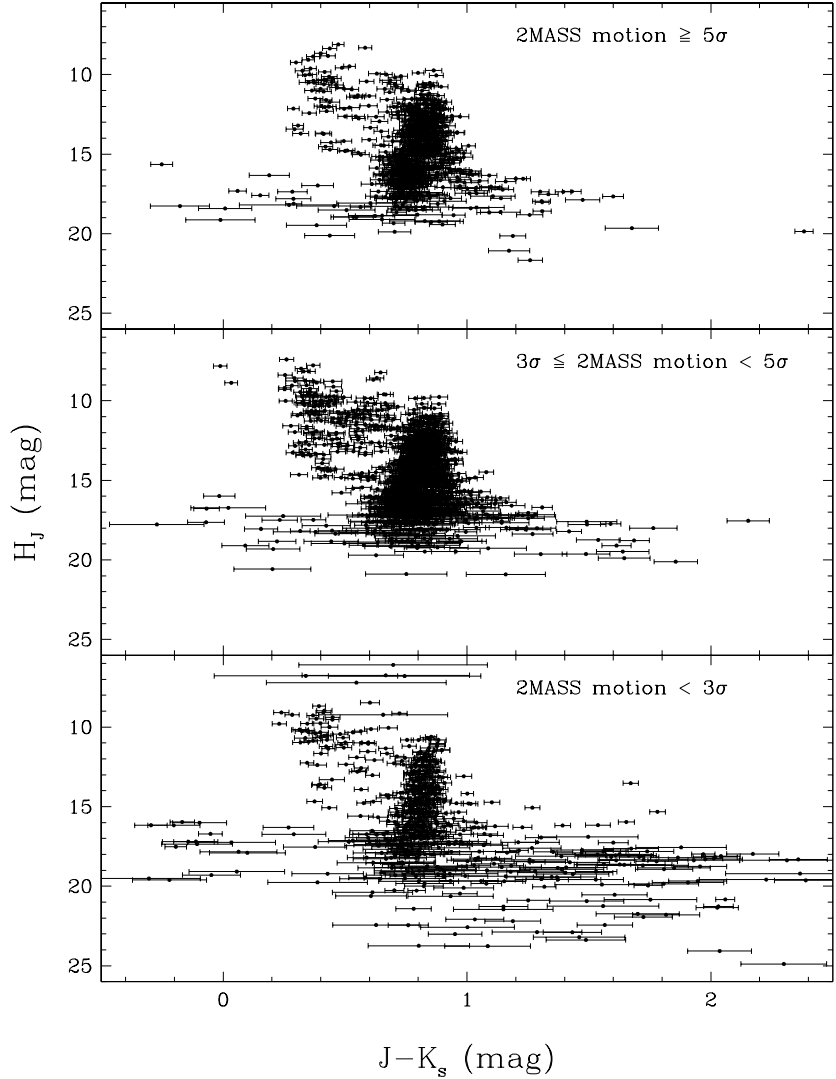}
\caption{Reduced proper motion at J-band (aka $H_J$) vs $J-K_s$ color for 2MASS PM discoveries at various confidence-of-motion values.
As described further in the text, the spectral sequence from pre-K to late-L begins in the
upper left of the diagram, extends down the center of the figure at $0.8 < J-K_s < 0.9$, and continues toward the lower right.
T dwarfs, which have a wide range
of near-infrared color ($-1 < J-K_s < 2$) but are intrinsically faint (large $H_J$), span the 
entire lower portion of the diagram. White dwarfs fall at lower left because they tend to have blue near-infrared colors
but are intrinsically faint for their motion (large $H_J$). 
Old, low-metalicity objects (the subdwarfs) are pushed down and to the left relative to normal field objects because 
of their bluer colors and higher
relative motions. Many of the objects with $0.6 < J-K_s < 0.9$ and $H_J > 15$ are M subdwarfs
falling below and to the left of the main popluation of field M dwarfs.
\label{red_motion_vs_JK}}
\end{figure}

\clearpage

\begin{figure}
\epsscale{0.9}
\plotone{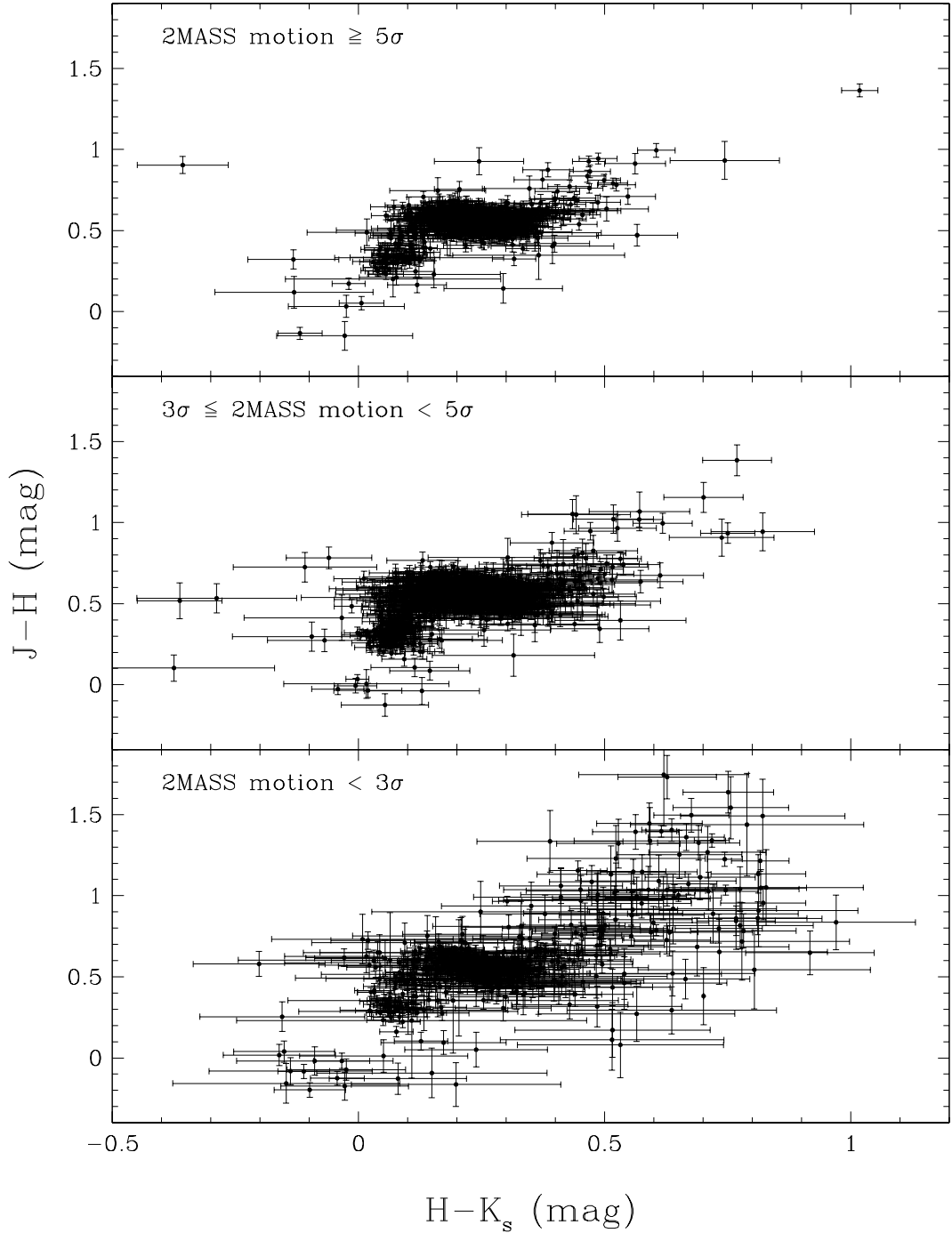}
\caption{The $J-H$ vs. $H-K_s$ color-color plot of all 2MASS proper motion sources. This figure can be compared to Figure 
\ref{IR_color_color_colored}, which shows the location of field dwarfs of spectral type B to L.
\label{JH_vs_HK}}
\end{figure}

\clearpage

%

\begin{figure}
\epsscale{0.9}
\plotone{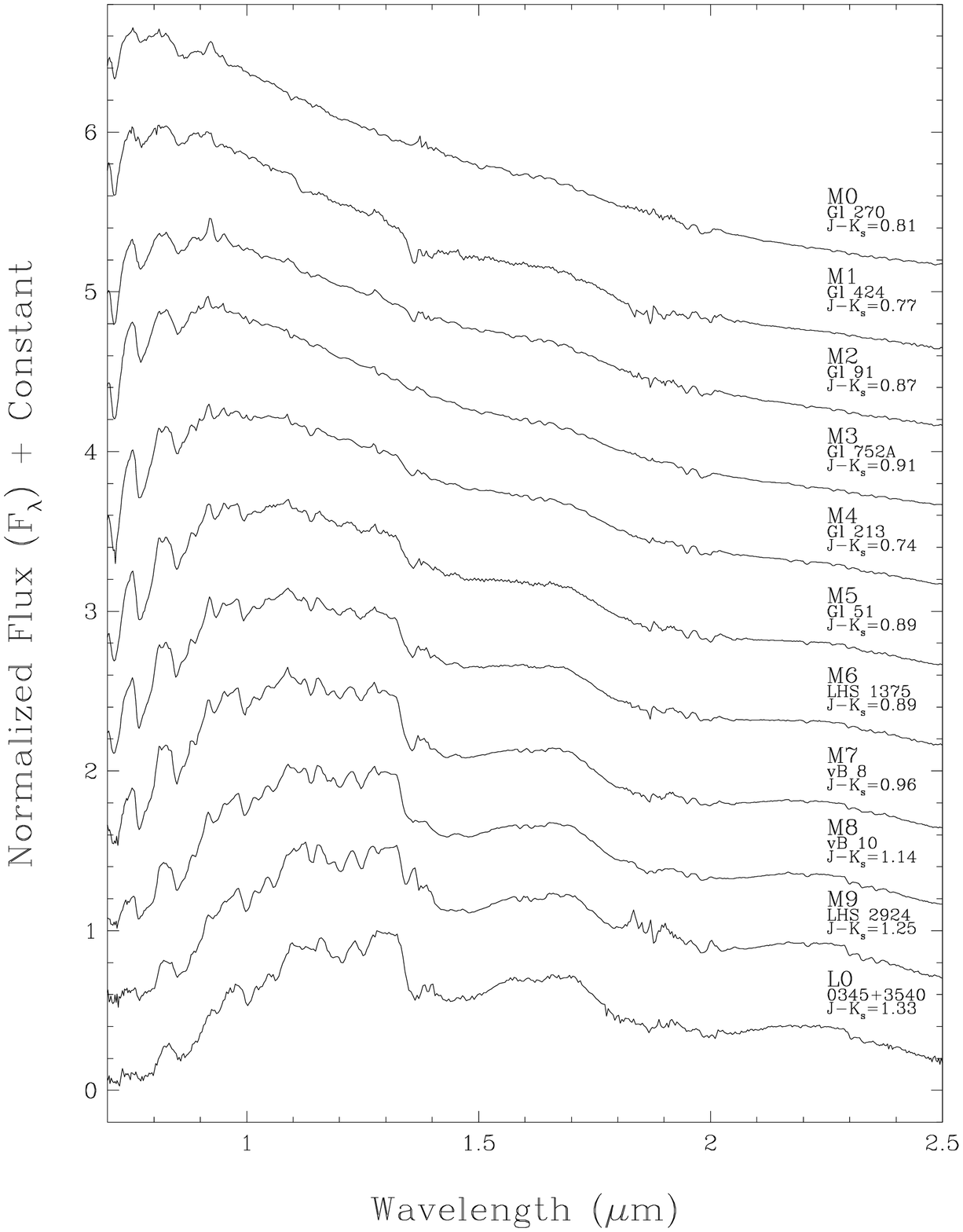}
\caption{A standard dwarf spectral sequence covering 0.7 to 2.5 $\mu$m for M0 through L0. All spectra were taken
with SpeX in prism mode. Spectra have been normalized to one at 1.28 $\mu$m and integer offsets added to separate the
spectra vertically. \label{NIR_M_stds}}
\end{figure}

\clearpage

\begin{figure}
\epsscale{0.9}
\plotone{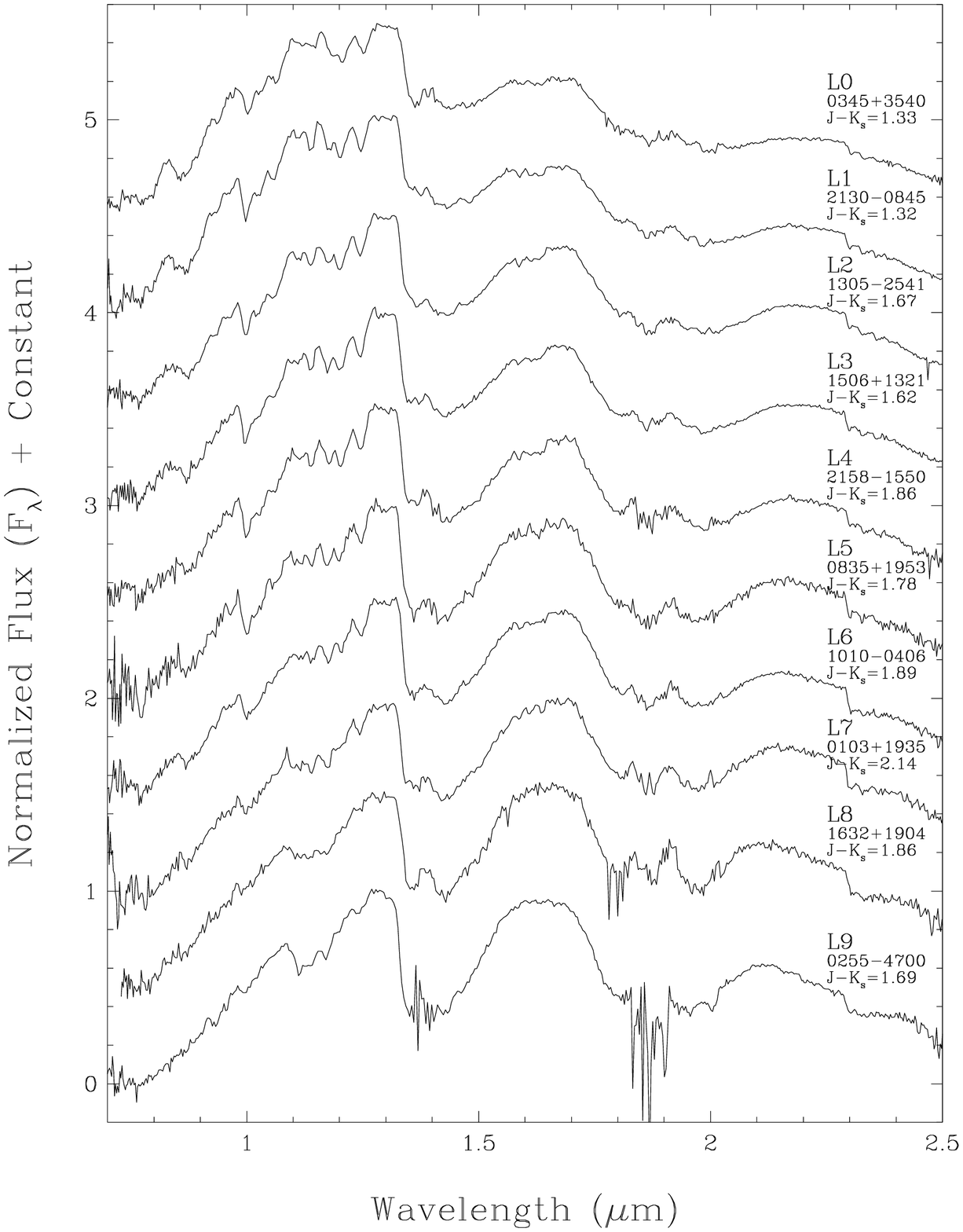}
\caption{A standard dwarf spectral sequence covering 0.7 to 2.5 $\mu$m for L0 through L9. All spectra were taken
with SpeX in prism mode. Spectra have been normalized to one at 1.28 $\mu$m and integer offsets added to separate the
spectra vertically. \label{NIR_L_stds}}
\end{figure}

\clearpage

\begin{figure}
\epsscale{0.9}
\plotone{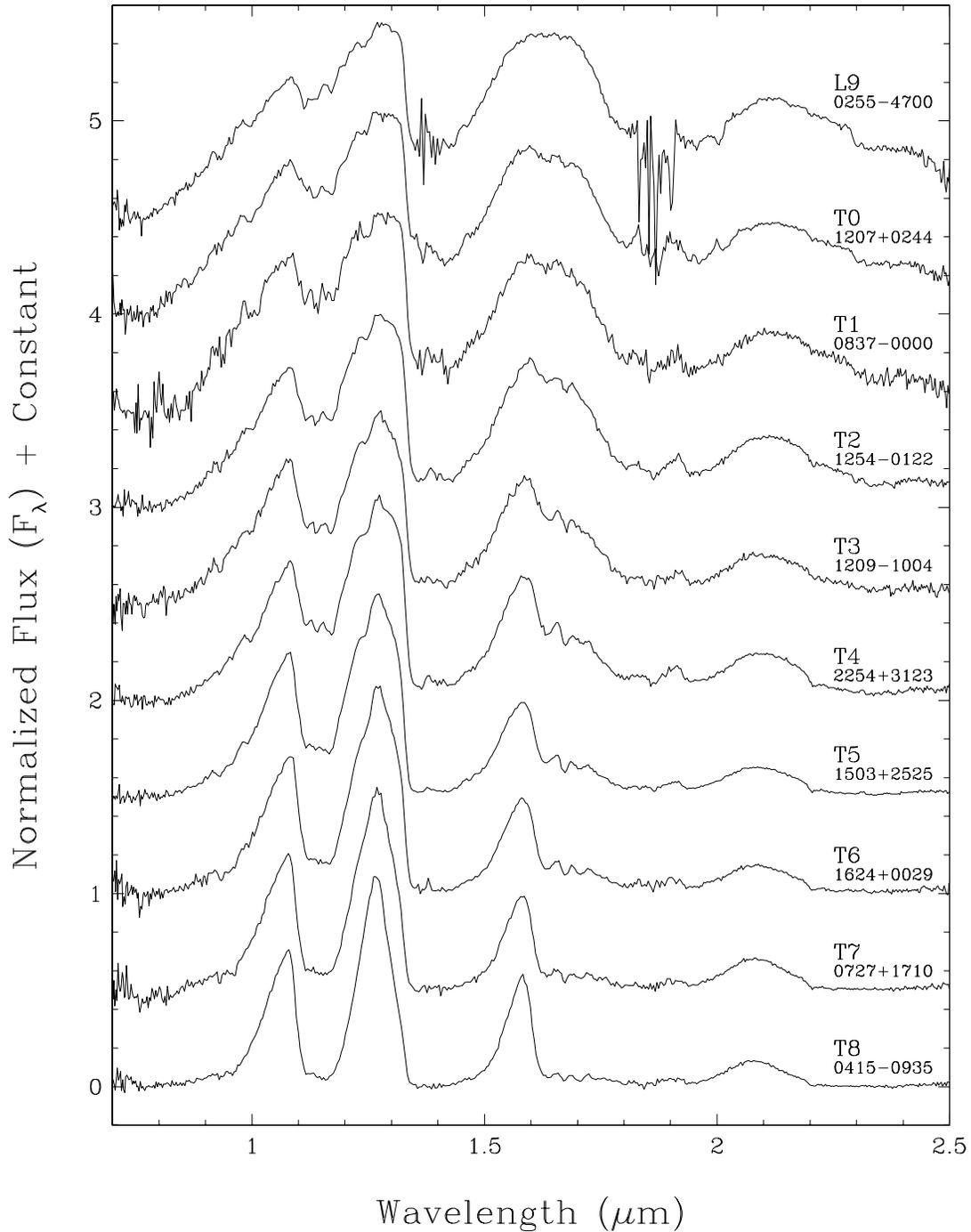}
\caption{A standard dwarf spectral sequence covering 0.7 to 2.5 $\mu$m from L9 through T8. All spectra were taken
with SpeX in prism mode. Spectra have been normalized to one at 1.28 $\mu$m and integer offsets added to separate the
spectra vertically. This figure is adapted from Figure 2 of  \cite{burgasser2006} and is shown here to provide continuity with
the M and L dwarf sequences in the previous two figures. \label{NIR_T_stds}}
\end{figure}

\clearpage

\begin{figure}
\epsscale{0.9}
\plotone{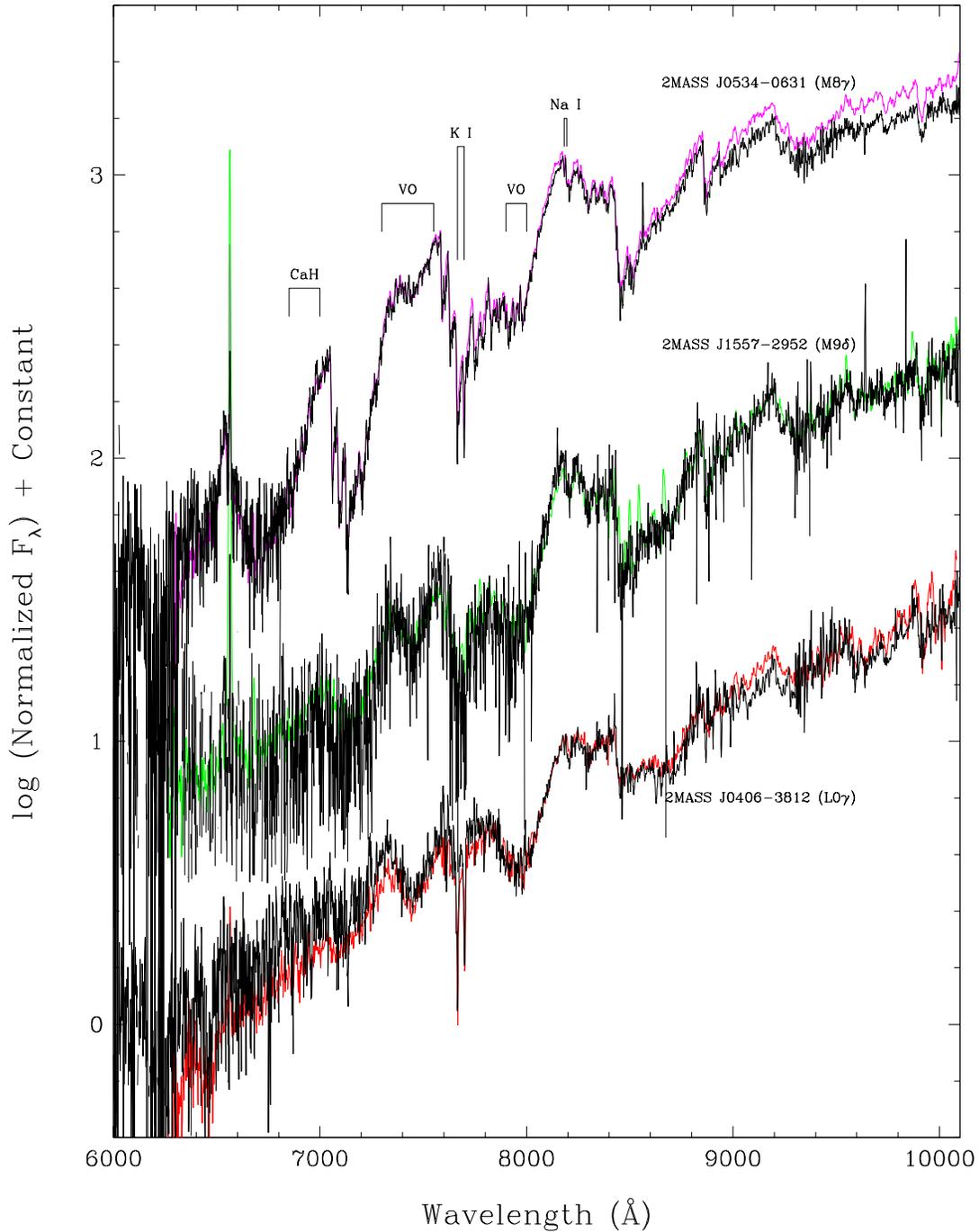}
\caption{Optical spectra of low-gravity dwarf discoveries (black lines) with late-M to early-L types. Overplotted for comparison
are the low-gravity M8 dwarf 2MASS J1207$-$3932 (magenta), the low-gravity M9 dwarf KPNO-Tau 12 (green), and the L0$\gamma$ dwarf 
2MASS J0141$-$4633 (red). Spectra are normalized to one at 8250 \AA\ and integer offsets added when needed to separate the 
spectra vertically. \label{oddones_lowg_opt1}}
\end{figure}

\clearpage

\begin{figure}
\epsscale{0.9}
\plotone{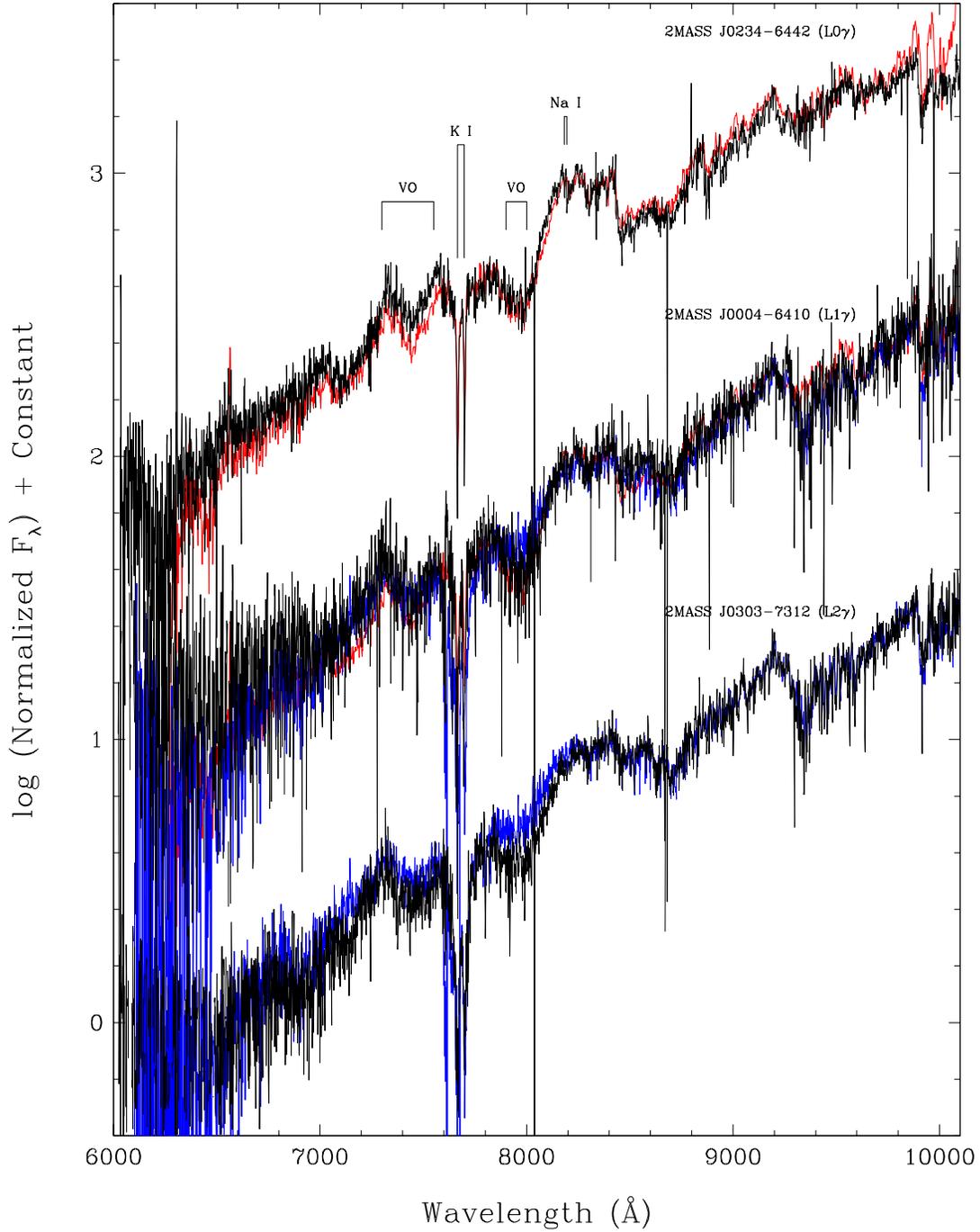}
\caption{Optical spectra of low-gravity dwarf discoveries (black lines) with early-L types. Overplotted for comparison
are the L0$\gamma$ dwarf 2MASS J0141$-$4633
(red) and the L2$\gamma$ dwarf 2MASS J2322$-$6151 (blue).  Spectra are normalized to one at 8250 \AA\ and integer 
offsets added when needed to separate the 
spectra vertically.\label{oddones_lowg_opt2}}
\end{figure}

\clearpage

\begin{figure}
\epsscale{0.9}
\plotone{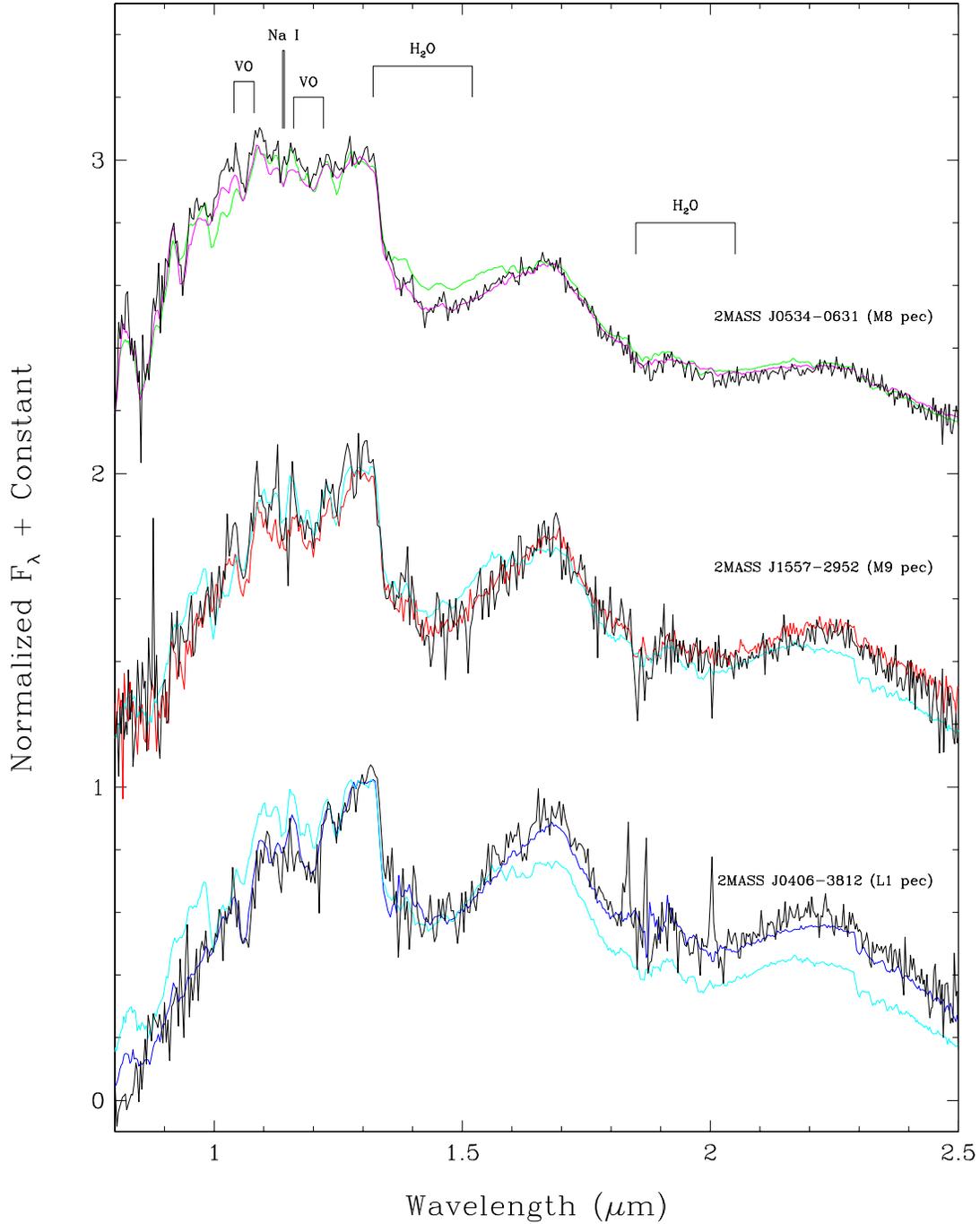}
\caption{Near-infrared spectra of low-gravity discoveries (black lines). Overplotted for comparison 
are the standard M8 dwarf vB 10 (green), the low-gravity M8 dwarf 2MASS J1207$-$3932 (magenta), the low-gravity M9
dwarf KPNO-Tau 12 (red), the standard
L1 dwarf 2MASS J2130$-$0845 (cyan), and the low-gravity L1 dwarf 2MASS J0141$-$4633 (blue).  Spectra are normalized to one at 1.28 $\mu$m 
and integer offsets added when needed to separate the 
spectra vertically. \label{oddones_lowg_IR}}
\end{figure}

\clearpage

\begin{figure}
\epsscale{0.9}
\plotone{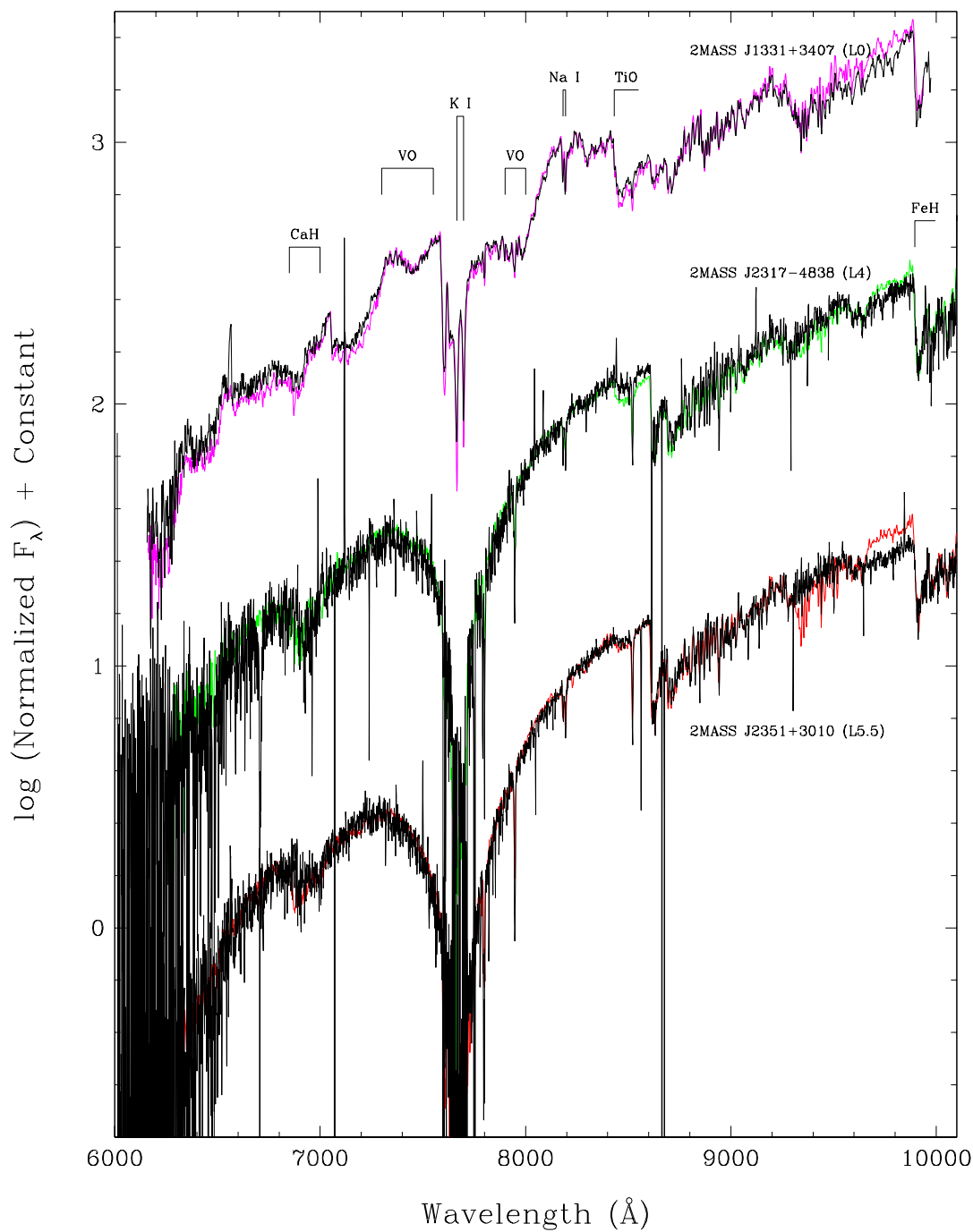}
\caption{Optical spectra of three red L dwarf discoveries (black lines). Overplotted for comparison
are optical standards: the L0 dwarf 2MASS J0345+2540 (magenta), L4 dwarf 2MASS J1155+2307 (green), and 
L5 dwarf DENIS J1228-1547 (red).  Spectra are normalized to one at 8250 \AA\ and integer offsets added when needed to separate the 
spectra vertically.\label{oddones_redL_opt}}
\end{figure}

\clearpage

\begin{figure}
\epsscale{0.9}
\plotone{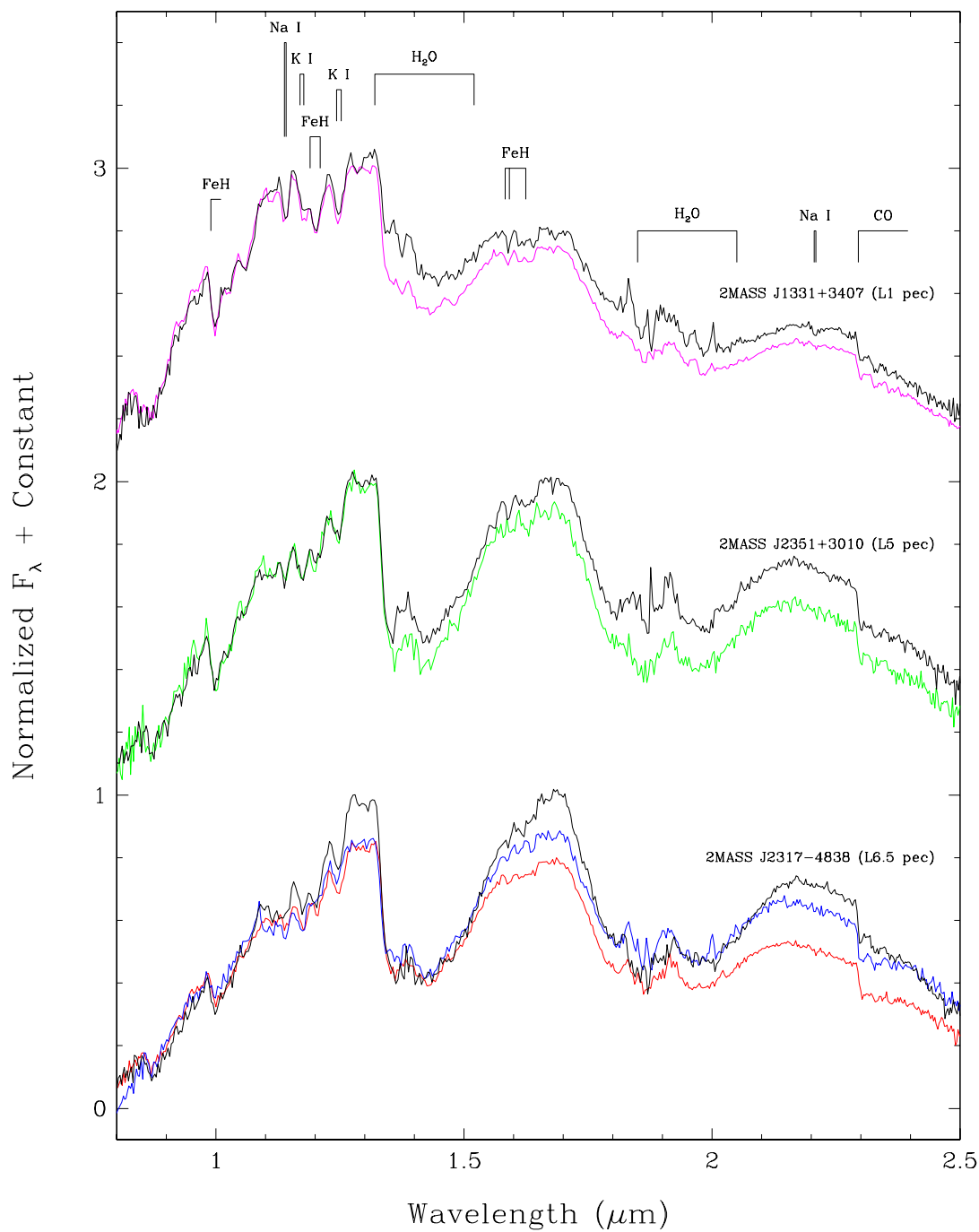}
\caption{Near-infrared spectra of three red L dwarf discoveries (black lines). Overplotted for comparison
are near-infrared standards: the L1 dwarf 2MASS J2130$-$0845 (magenta), L5 dwarf 2MASS J0835+1953 (green), L6 dwarf 2MASS J1010$-$0406 (red)
and L7 dwarf 2MASS J0103+1935 (blue).  Spectra are generally normalized to one at 1.28 $\mu$m and integer offsets added when needed to separate the 
spectra vertically. The spectra of 2MASS J1010$-$0406 and 2MASS J0103+1935 have been scaled at 1.0 $\mu$m to match the flux level of
2MASS J2317$-$4838 to maximize the agreement among all three spectra at the shortest wavelengths shown.\label{oddones_redL_IR}}
\end{figure}

\clearpage

\begin{figure}
\epsscale{0.9}
\plotone{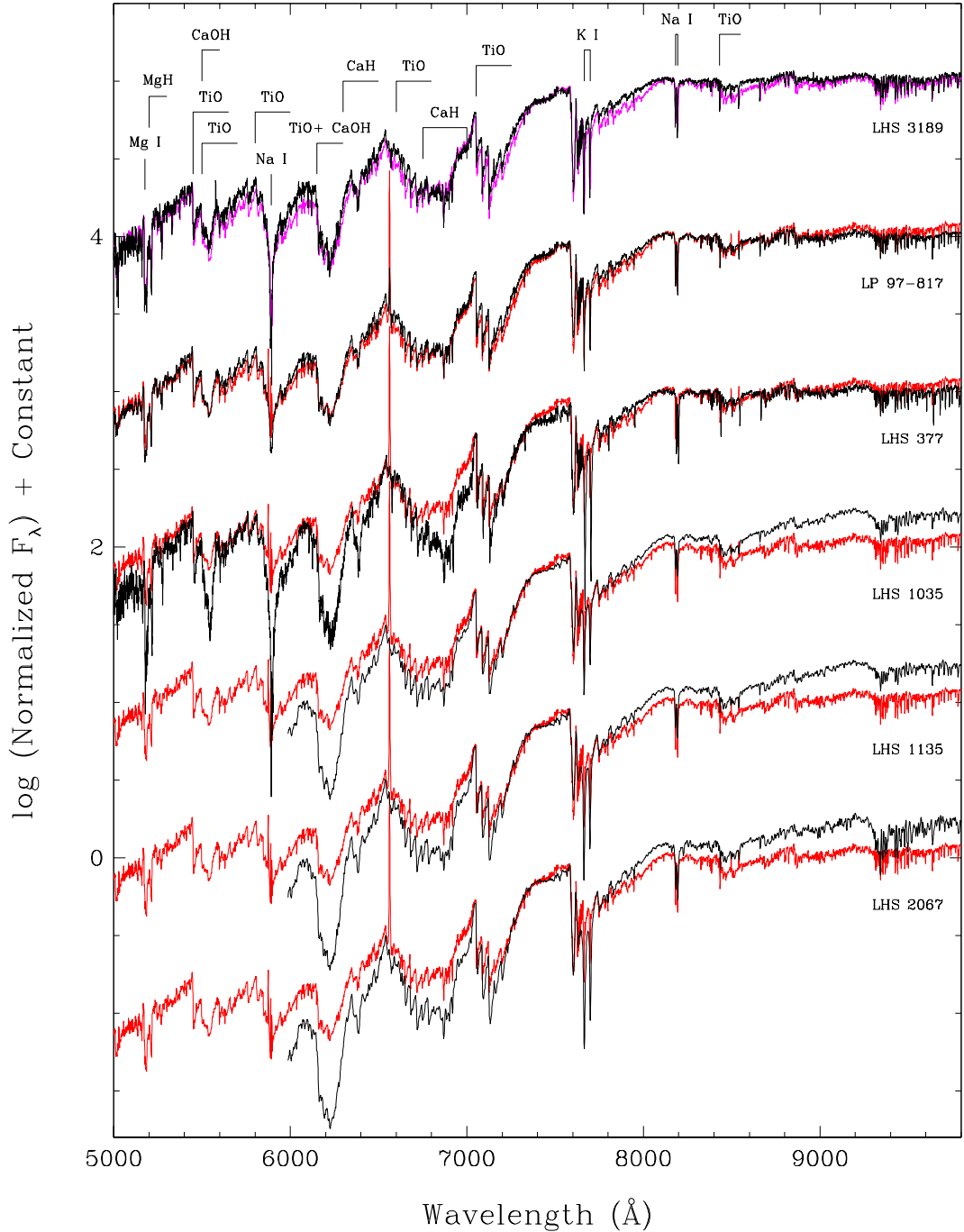}
\caption{Optical spectra from Keck/DEIMOS (top three objects, in black) or Gemini-North/GMOS (bottom three objects, in black) compared
to normal dwarfs of type M4.5 (LHS 3001, magenta) or M5 (Gliese 51, red) taken with Keck/DEIMOS. Spectra are renormalized so that the
target and comparison spectra overlap either at 7500 or 8250 \AA. Integer offsets have been added when needed to separate the 
spectra vertically.
\label{sd_or_not}}
\end{figure}

\clearpage

\begin{figure}
\epsscale{0.9}
\plotone{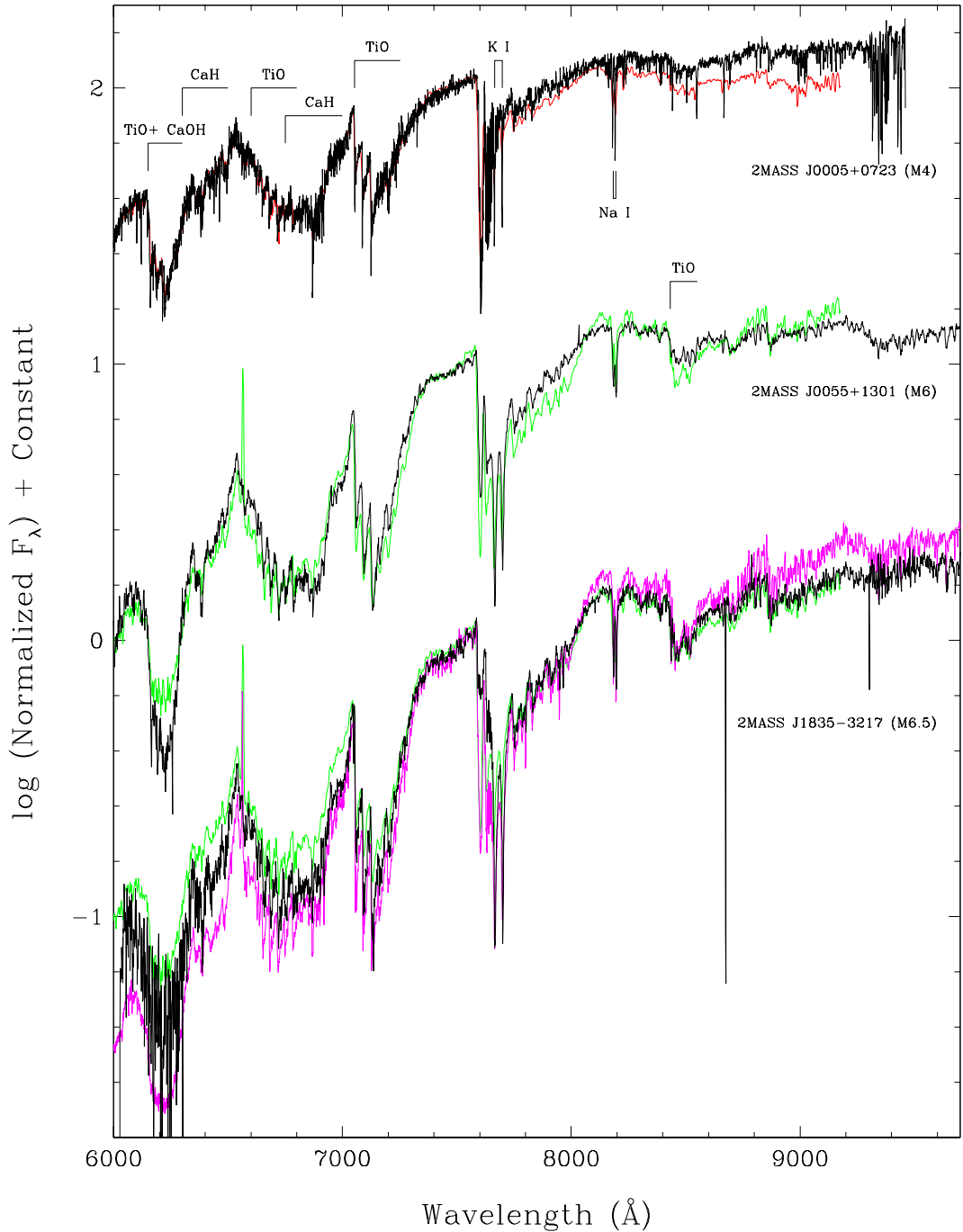}
\caption{Optical spectra for three objects (black lines) showing intermediate dwarf/subdwarf characteristics in the near-infrared.
Overplotted for comparison are optical standards: the M4 dwarf Gl 402 (red), the M6 dwarf Wolf 359 (green), and the M7 dwarf vB 8
 (magenta). Spectra are normalized 
to one at 7500 \AA. Integer offsets have been added when needed to separate the 
spectra vertically.
\label{oddones_sd_opt3}}
\end{figure}

\clearpage

\begin{figure}
\epsscale{0.9}
\plotone{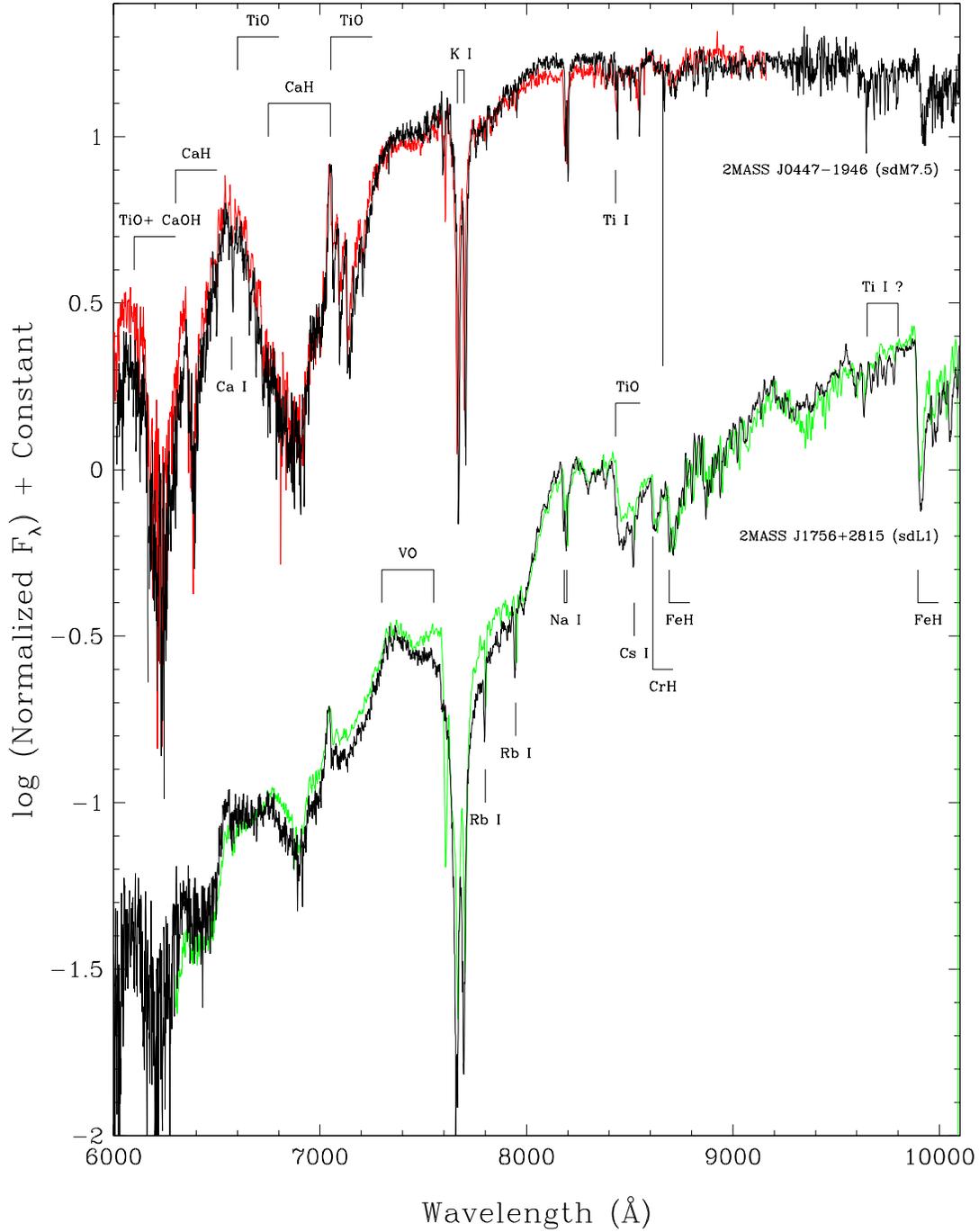}
\caption{Optical spectra of two late-type subdwarf discoveries (black lines). Overplotted for comparison
are optical standards: the sdM8 LSPM J1425+7102 (red) and the L1 dwarf 2MASS J2130$-$0845 (green).
Spectra are normalized to one at 8250 \AA\ and integer offsets added when needed to separate the 
spectra vertically.
\label{oddones_sd_opt1}}
\end{figure}

\clearpage

\begin{figure}
\epsscale{0.9}
\plotone{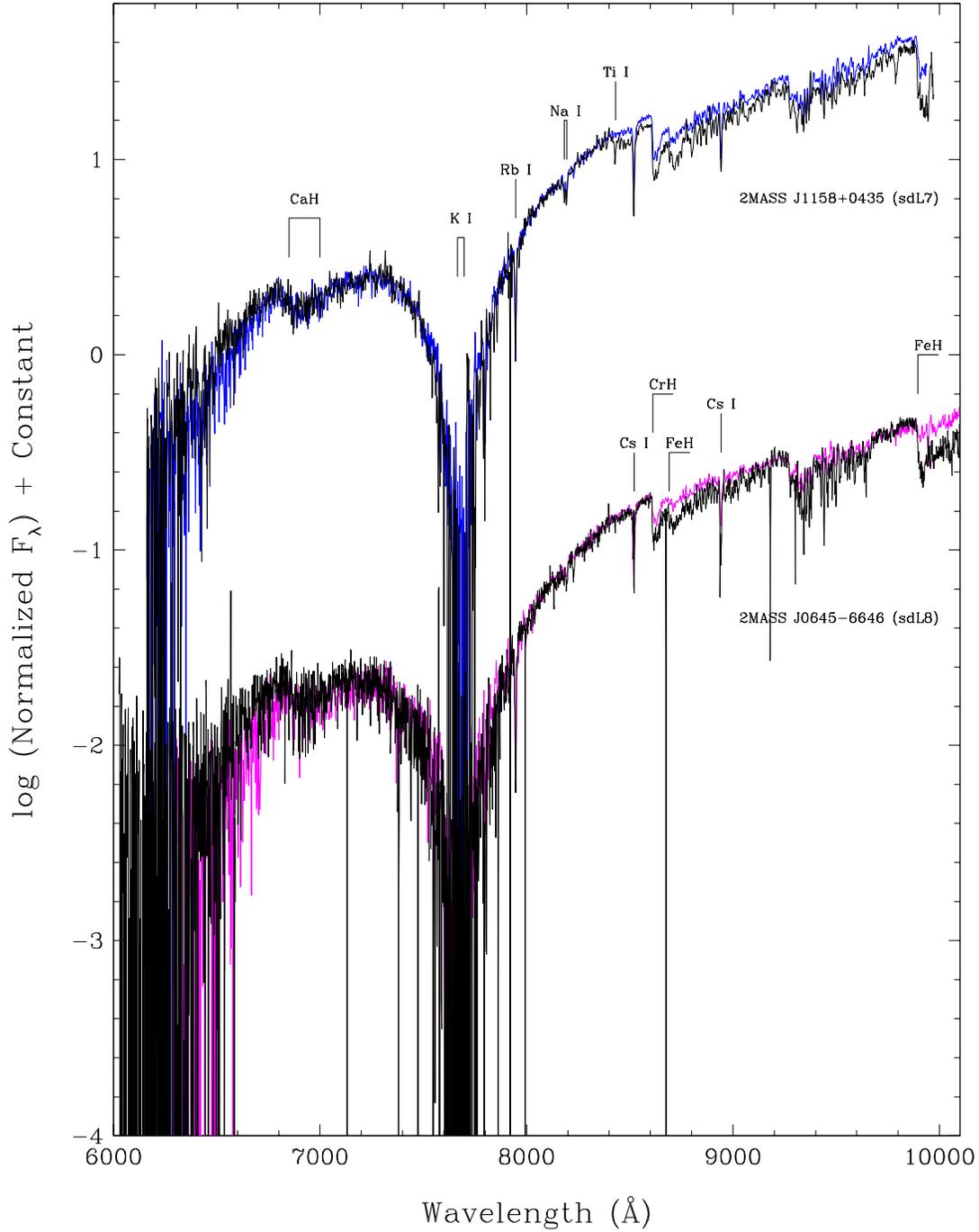}
\caption{Optical spectra of two late-type L subdwarf discoveries (black lines). Overplotted for comparison
are optical standards: the L7 dwarf DENIS J0205$-$1159 (blue) and the L8 dwarf 2MASS J1632+1904 (magenta). 
Spectra are normalized to one at 8250 \AA\ and integer offsets added when needed to separate the 
spectra vertically.
\label{oddones_sd_opt2}}
\end{figure}

\clearpage

\begin{figure}
\epsscale{0.9}
\plotone{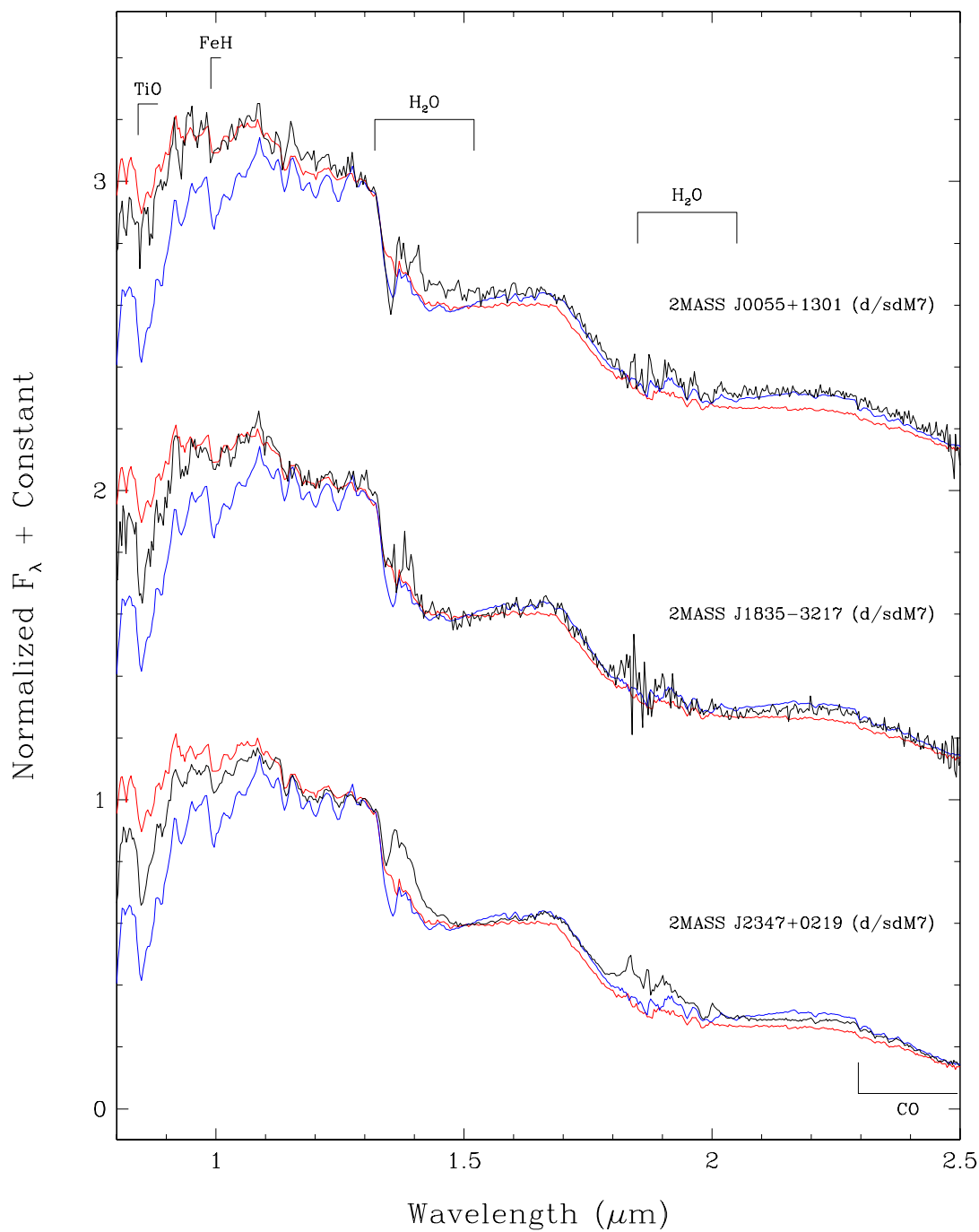}
\caption{Near-infrared spectra of three discoveries having near-IR classifications intermediate between M7
and sdM7 (black lines). Overplotted for comparison are the optically classified sdM7 LHS 377 (red) and the
near-infrared M7 standard vB8 (blue).  Spectra are normalized to one at 1.28 $\mu$m and integer offsets added when needed to separate the 
spectra vertically.
\label{oddones_sd_IR2}}
\end{figure}

\clearpage

\begin{figure}
\epsscale{0.9}
\plotone{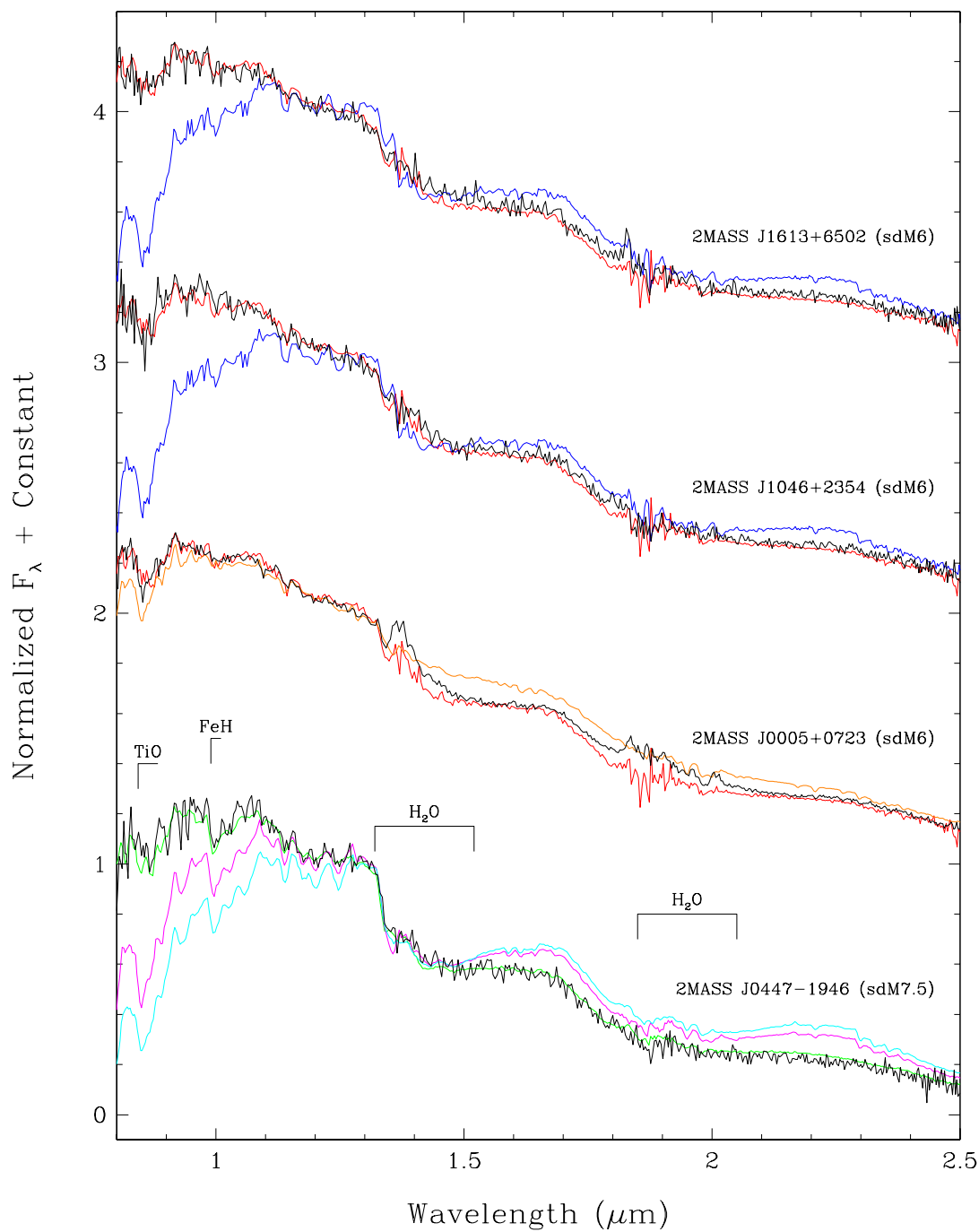}
\caption{Near-infrared spectra of four late-M subdwarf discoveries (black lines). Overplotted for comparison
are near-infrared spectra of the optically classified sdM6 LHS 1074 (red) and the optically classified sdM7.5
LSR J2036+5100 (green). Also plotted for comparison are near-infrared M dwarf standards: the M4 Gl 213 (orange),
the M6 LHS 3933 (blue), the M7 vB 8 (magenta), and the M8 vB 10 (cyan).
Spectra are normalized to one at 1.28 $\mu$m and integer offsets added when needed to separate the 
spectra vertically.
\label{oddones_sd_IR1}}
\end{figure}

\clearpage

\begin{figure}
\epsscale{0.9}
\plotone{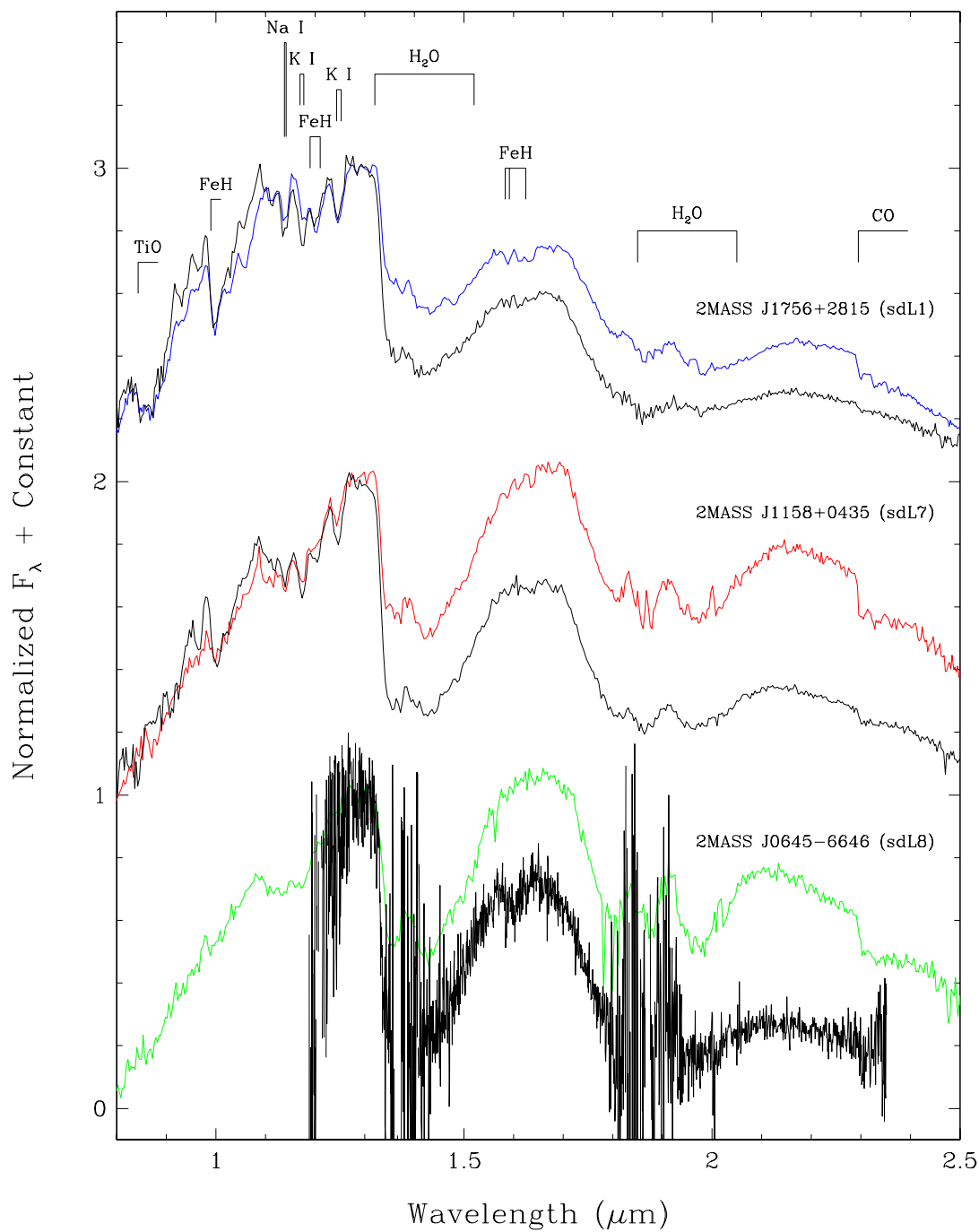}
\caption{Near-infrared spectra of three subdwarf L discoveries (black lines). Overplotted for comparison
are near-infrared standards: the L1 dwarf 2MASS J2130$-$0845 (blue), the L7 dwarf 2MASS J0103+1935 (red), and the L8
dwarf 2MASS J1632+1904 (green). Spectra are normalized to one at 1.28 $\mu$m and integer offsets added when needed to separate the 
spectra vertically.
\label{oddones_sd_IR3}}
\end{figure}

\clearpage

\begin{figure}
\includegraphics[scale=0.65,angle=270]{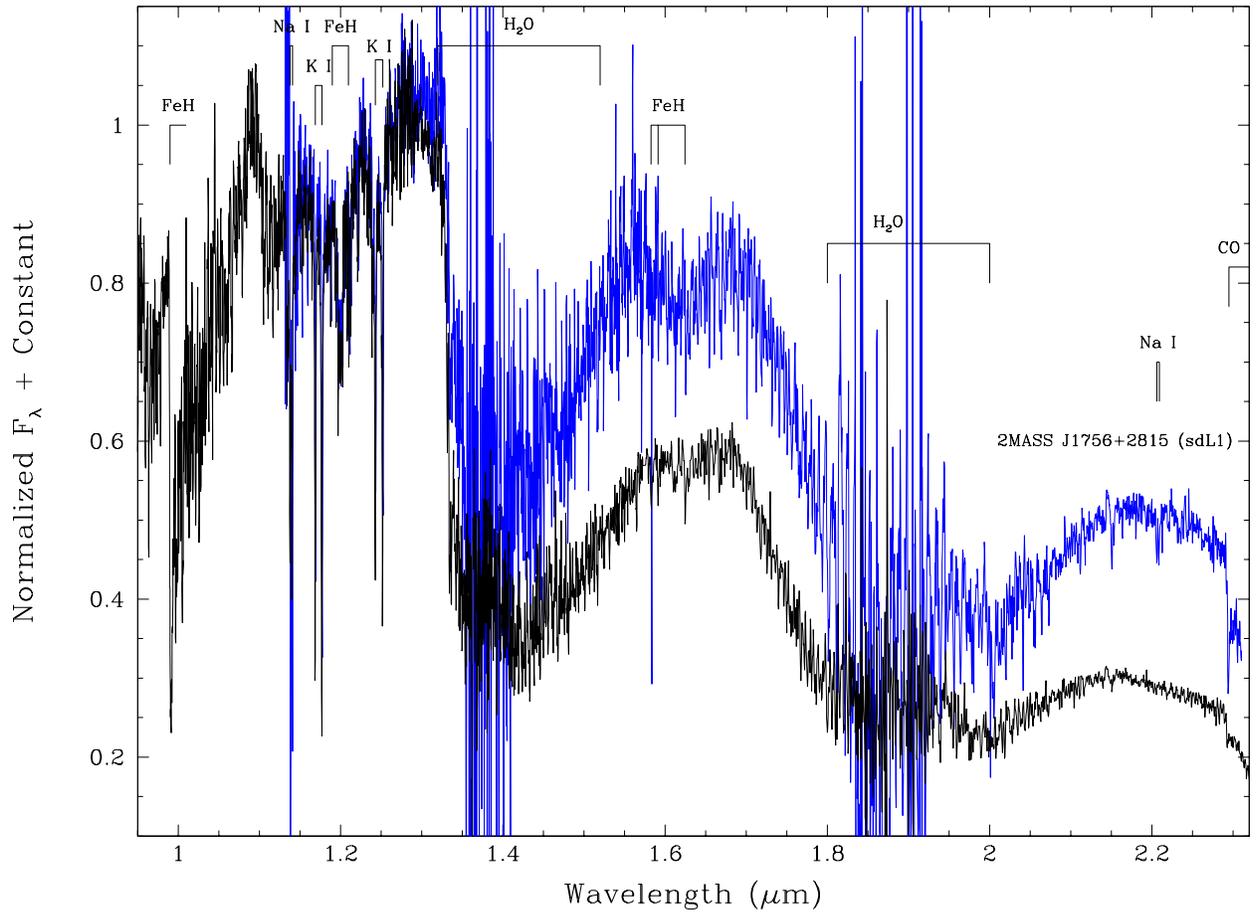}
\caption{NIRSPEC data for the sdL1 2MASS J1756+2815 (black line) compared to the normal L1 dwarf
2MASS J1658+7027 (blue).  Spectra are normalized to one at 1.28 $\mu$m.
\label{oddones_sd_IR4}}
\end{figure}

\clearpage

\begin{figure}
\includegraphics[scale=0.65,angle=270]{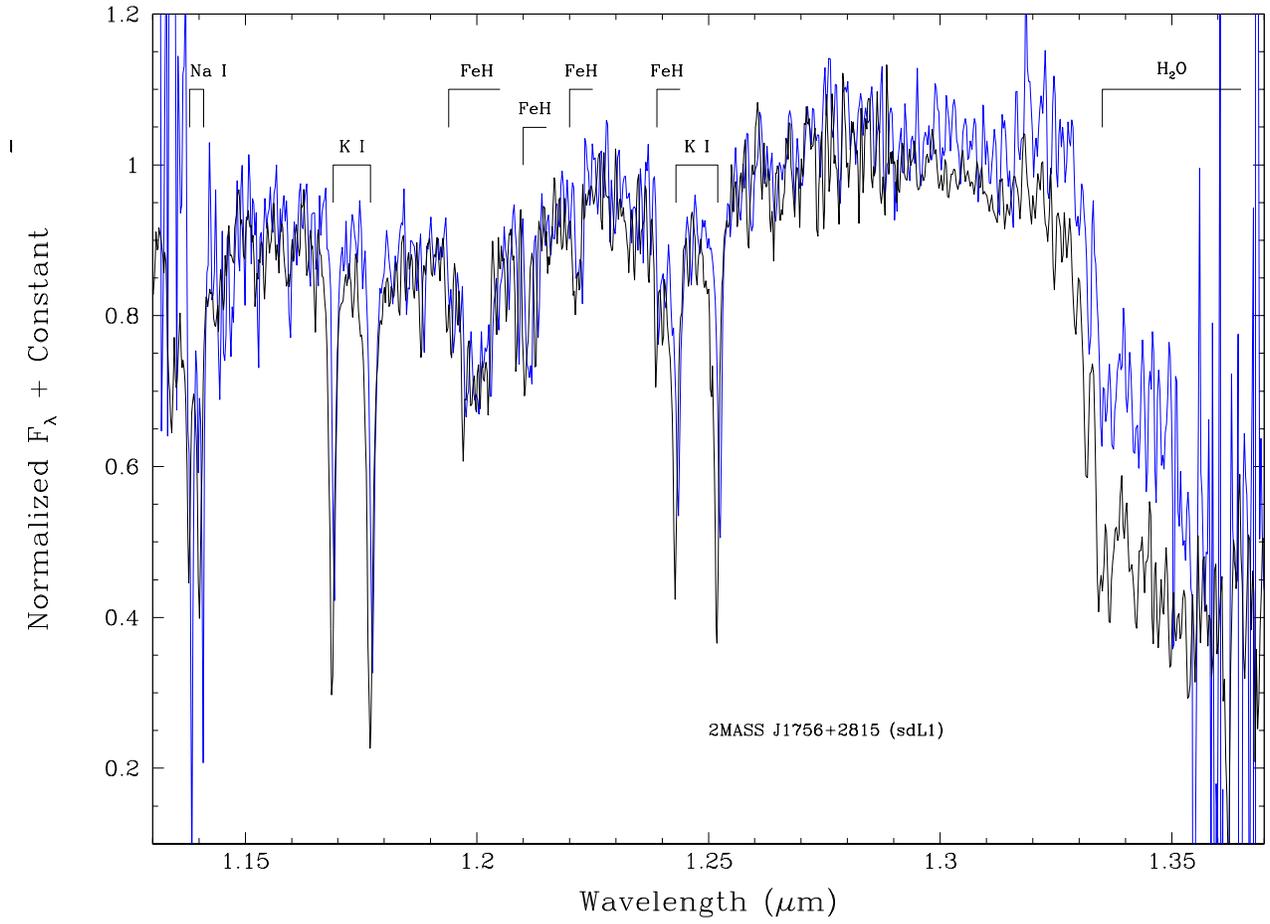}
\caption{Blow-up of Figure \ref{oddones_sd_IR4} showing the J-band portion from 1.13 to 1.37 $\mu$m. 
Spectra are normalized to one at 1.28 $\mu$m.
\label{oddones_sd_IR4b}}
\end{figure}

\clearpage

\begin{figure}
\includegraphics[scale=0.65,angle=270]{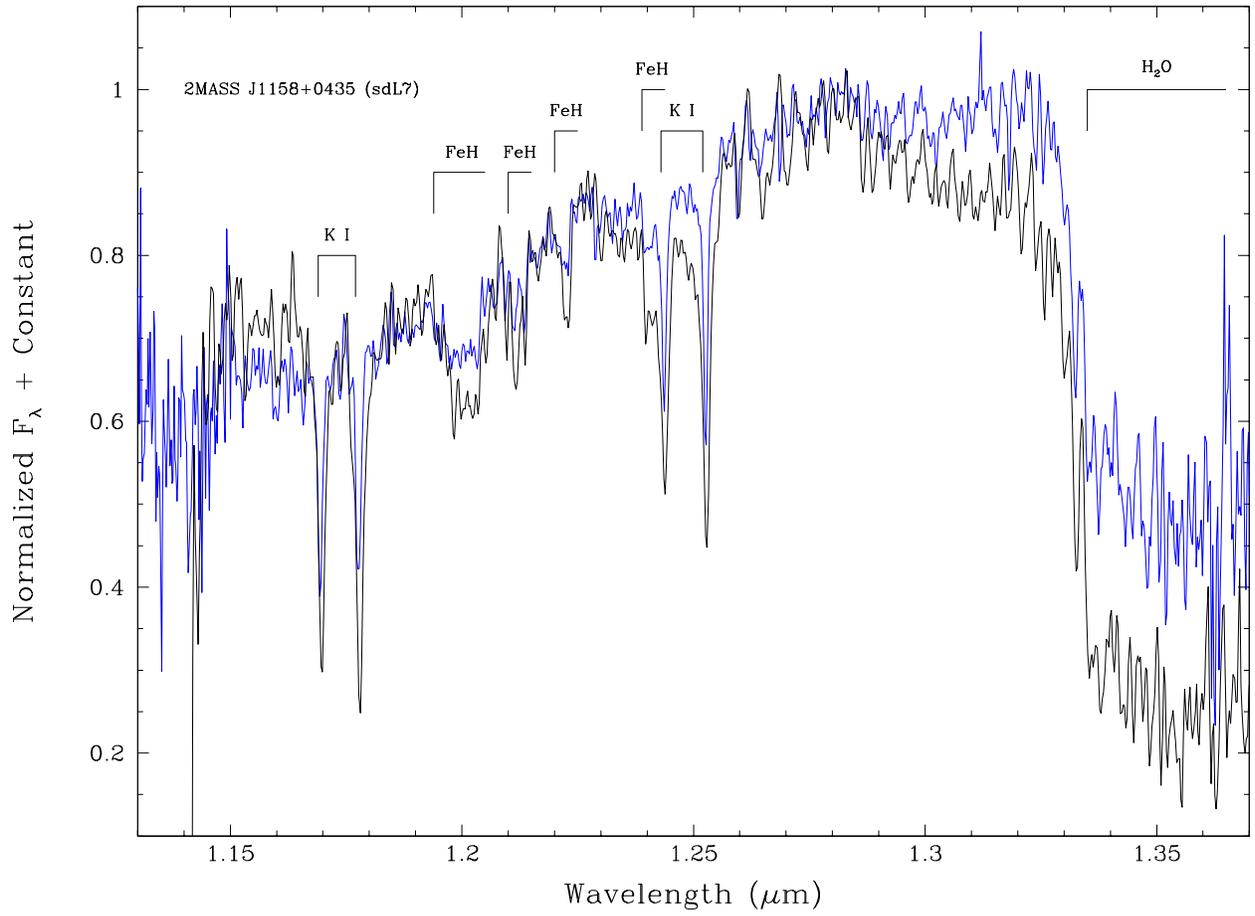}
\caption{NIRSPEC data for the sdL7 2MASS J1158+0435 (black line) compared to the standard L7 dwarf
2MASS J0103+1935 (blue). Spectra are normalized to one at 1.28 $\mu$m.
\label{oddones_sd_IR5}}
\end{figure}

\clearpage

\begin{figure}
\includegraphics[scale=0.65,angle=270]{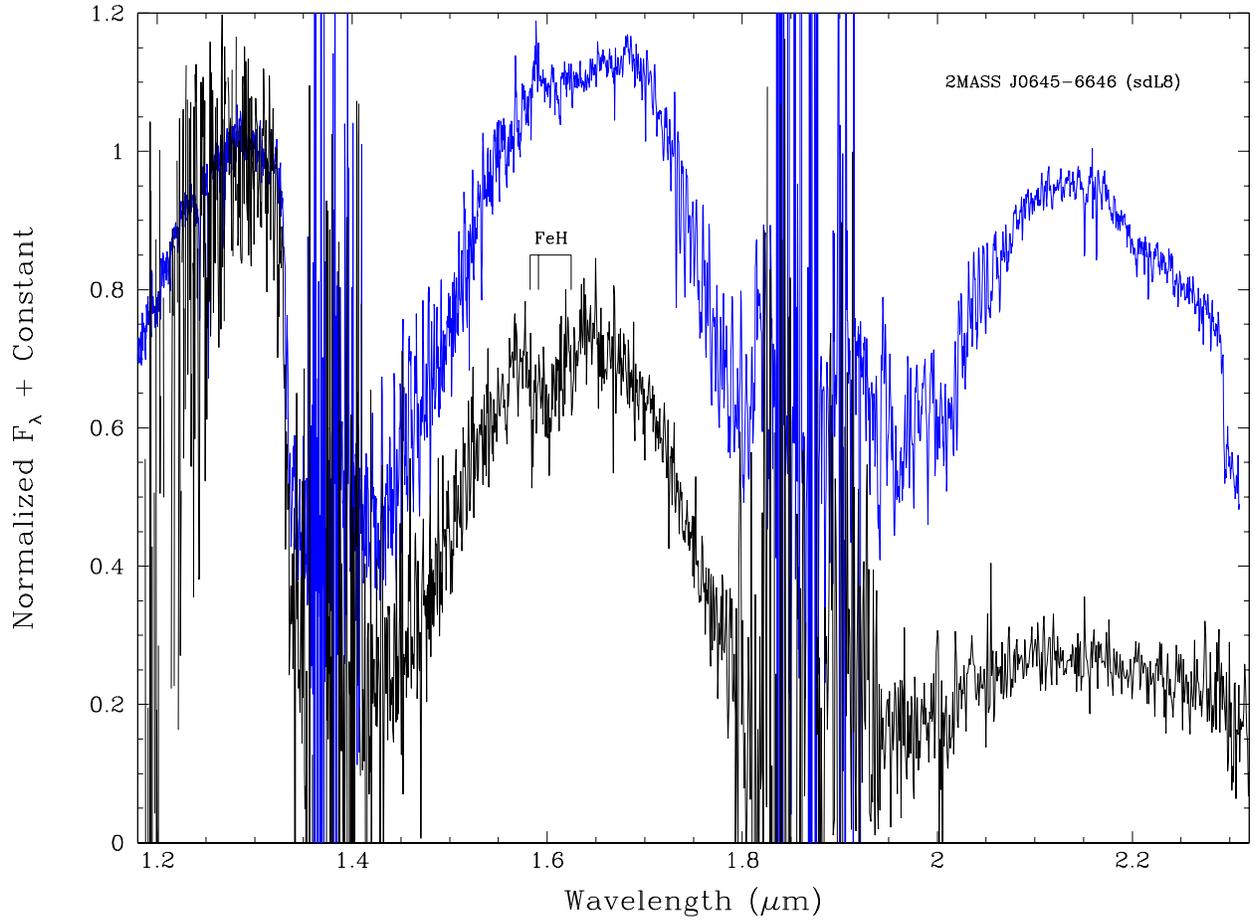}
\caption{Blow-up of the OSIRIS spectrum of the sdL8 discovery 2MASS J0645$-$6646 (black line)
compared to the NIRSPEC spectrum of the near-infrared L8 standard 2MASS J1632+1904 (blue). Both
spectra have been normalized to one at 1.28 $\mu$m.
\label{oddones_sd_IR6}}
\end{figure}

\clearpage

\begin{figure}
\includegraphics[scale=0.65,angle=270]{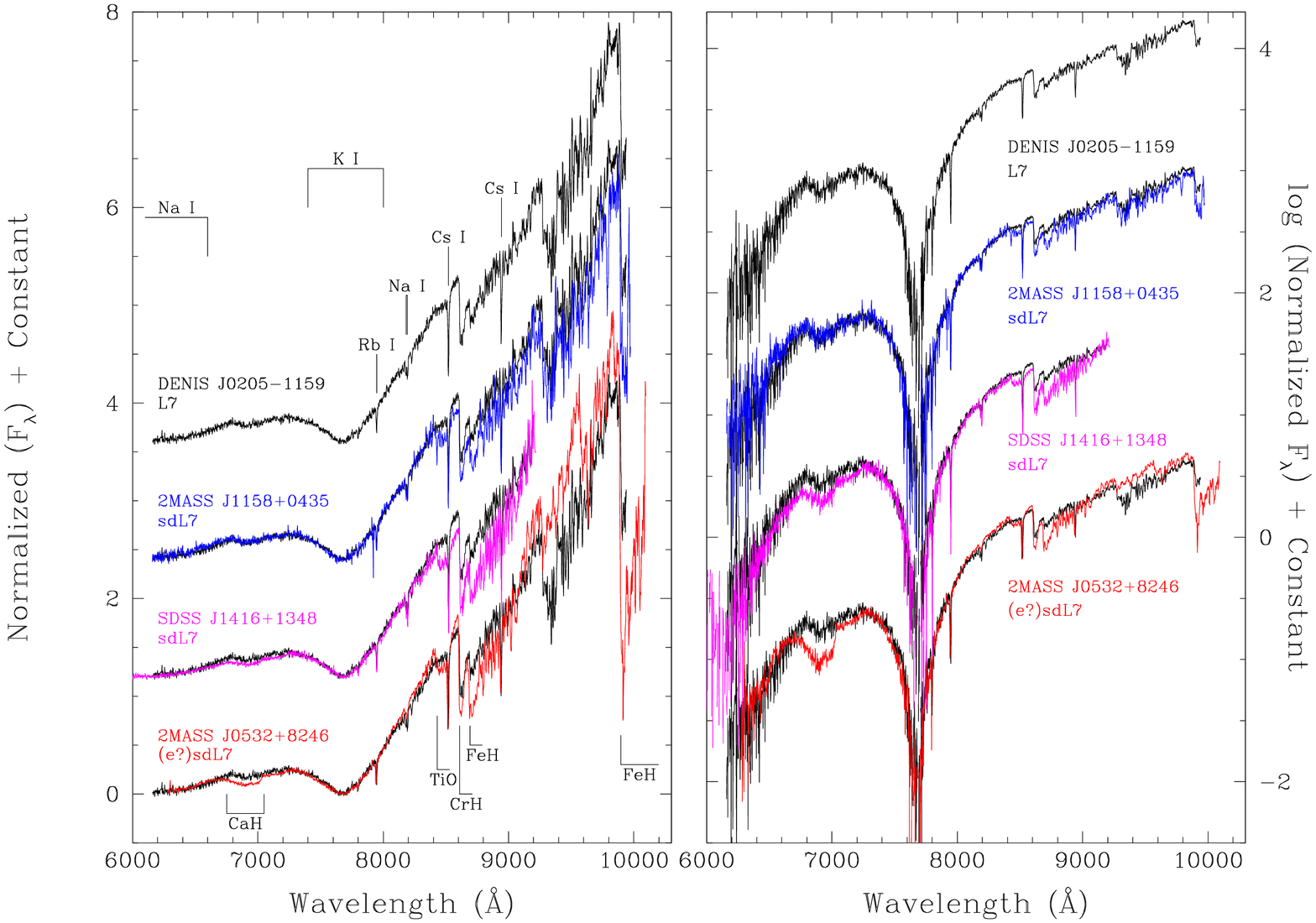}
\caption{The L7 metallicity sequence at optical wavelengths. Shown are the optical L7 dwarf
standard DENIS J0205$-$1159 (black line) from \cite{kirkpatrick1999}, the new sdL7 2MASS J1158+0435 (blue) from this paper,
the bright sdL7 SDSS J1416+1348 (magenta) from Schmidt et al.\ (submitted), and the more extreme L7 subdwarf
2MASS J0532+8246 (red) from \cite{burgasser2003}.
Spectra have been normalized to one at
8250 \AA\ and offsets in multiple units of 1.2 have been added to space the spectra vertically. The left panel shows the
flux in linear units and the right panel shows the flux in logarithmic units.
\label{sdL7s_opt}}
\end{figure}

\clearpage

\begin{figure}
\epsscale{0.9}
\plotone{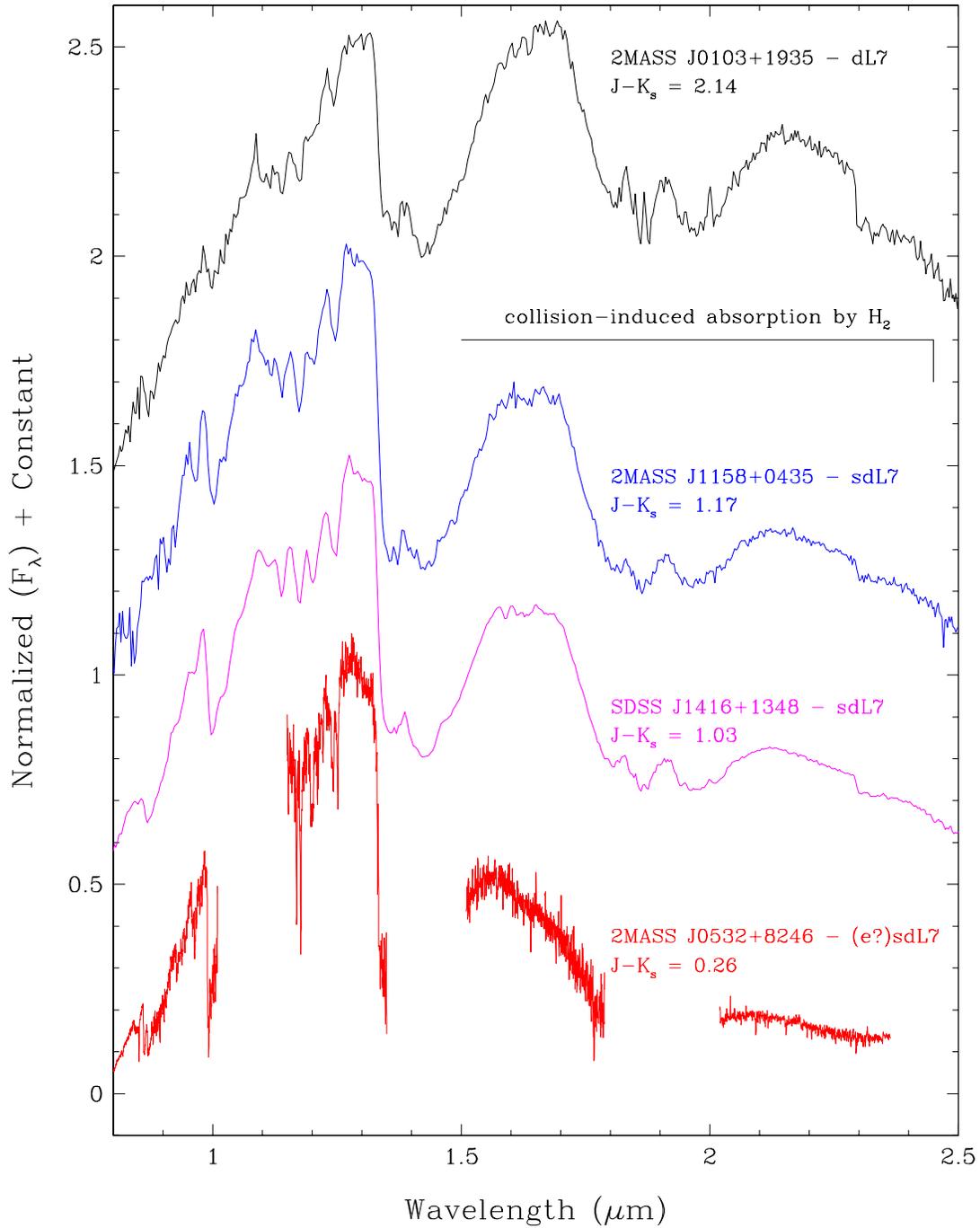}
\caption{The L7 metallicity sequence at near-infrared wavelengths. Color coding is the same as in 
Figure \ref{sdL7s_opt}. Spectra have been normalized to 1.28 $\mu$m, and vertical offsets are in multiple units of 0.5.
\label{sdL7s_nir}}
\end{figure}

\clearpage

\begin{figure}
\epsscale{0.9}
\plotone{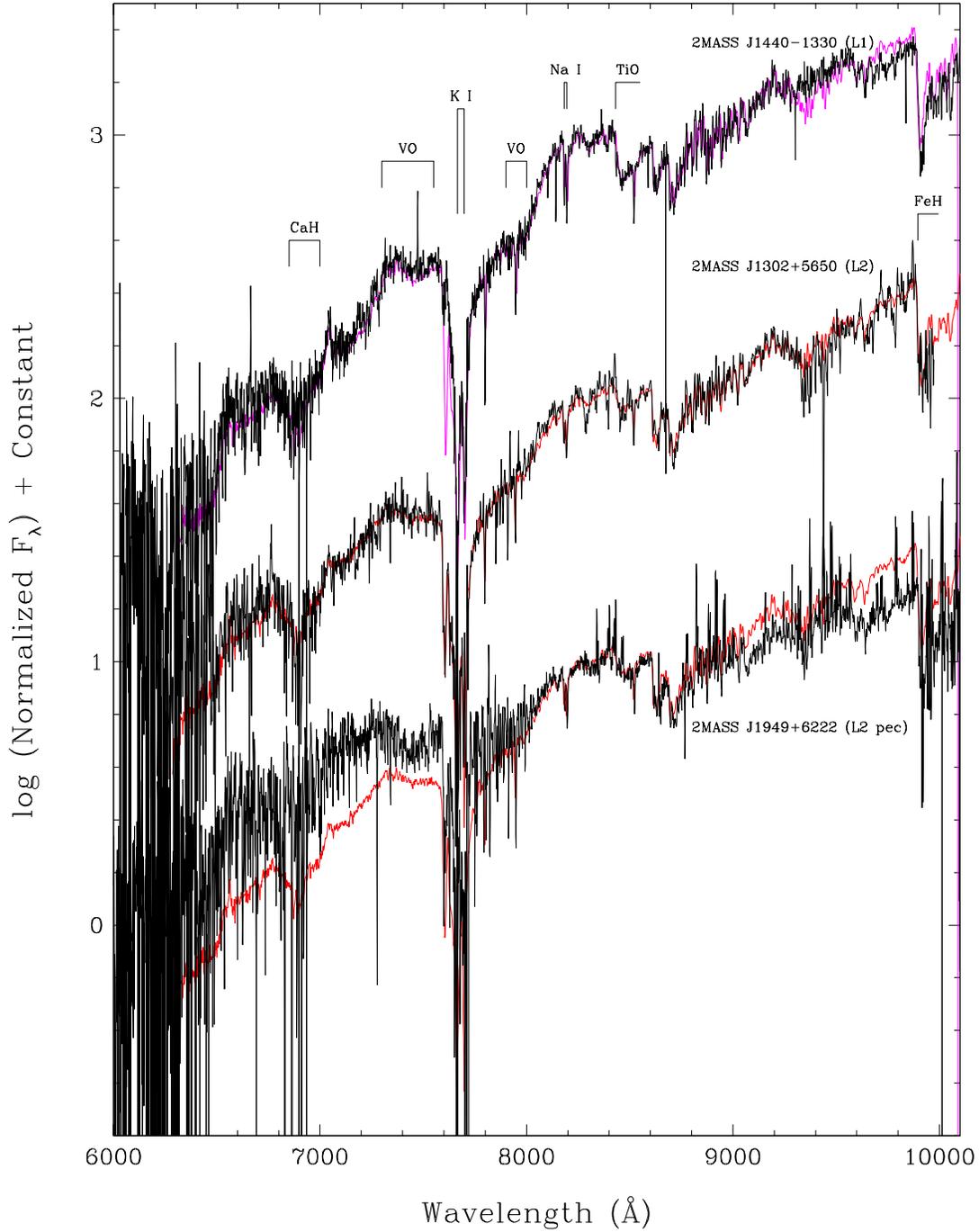}
\caption{Optical spectra of three blue L dwarfs with early-L spectral types (black lines). Overplotted for comparison
are optical standards: the L1 dwarf 2MASS J1439+1929 (magenta) and the L2 dwarf Kelu-1 (red). 
Spectra are normalized to one at 8250 \AA\ and integer offsets added when needed to separate the 
spectra vertically.\label{oddones_blueL_opt1}}
\end{figure}

\clearpage

\begin{figure}
\epsscale{0.9}
\plotone{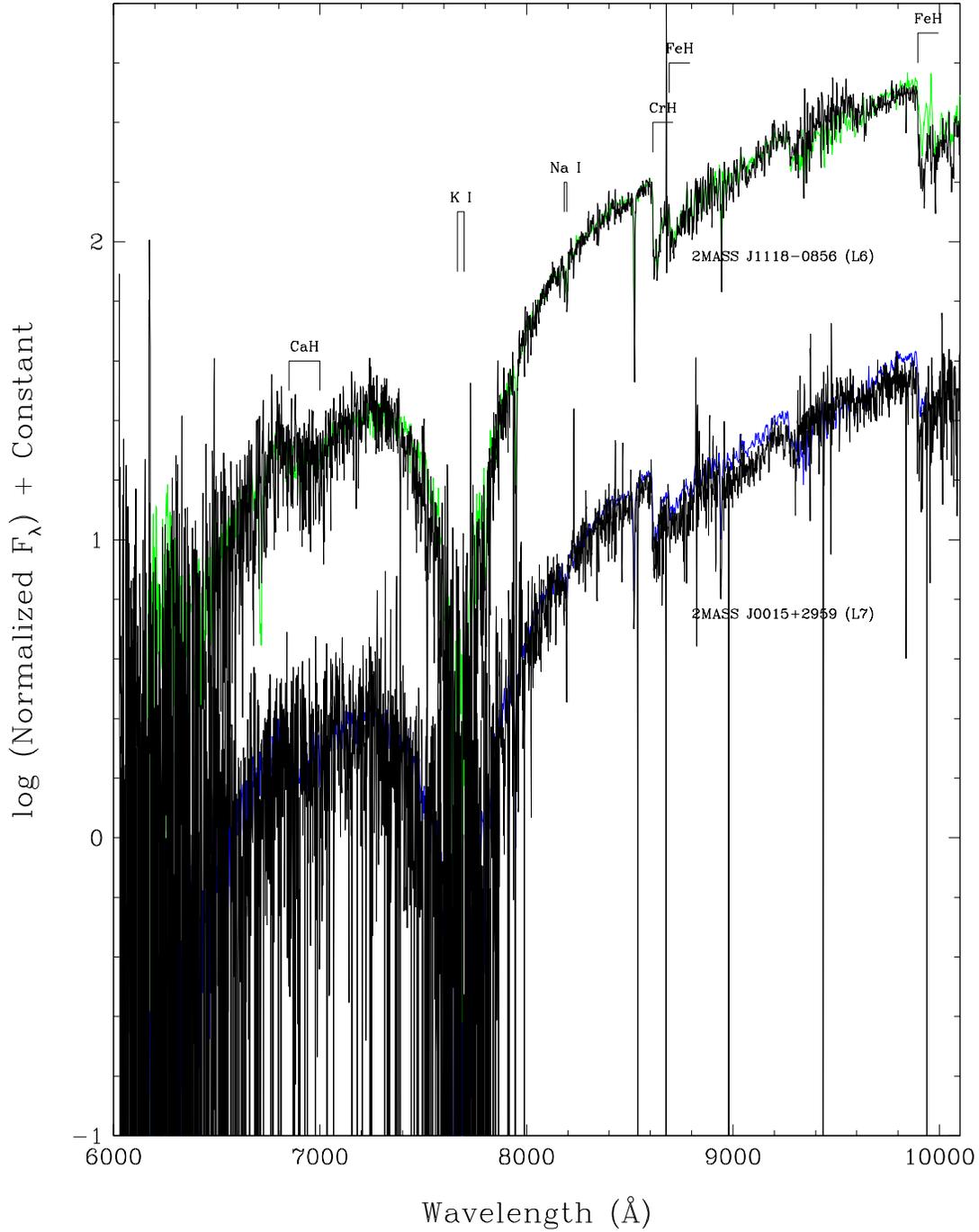}
\caption{Optical spectra of two blue L dwarfs with mid- to late-L spectral types (black lines). Overplotted for comparison
are optical standards: the L6 dwarf 2MASS J0850+1057 (green) and the L7 dwarf DENIS 0205$-$1159 (blue). 
Spectra are normalized to one at 8250 \AA\ and integer offsets added when needed to separate the 
spectra vertically.\label{oddones_blueL_opt2}}
\end{figure}

\clearpage

\begin{figure}
\epsscale{0.9}
\plotone{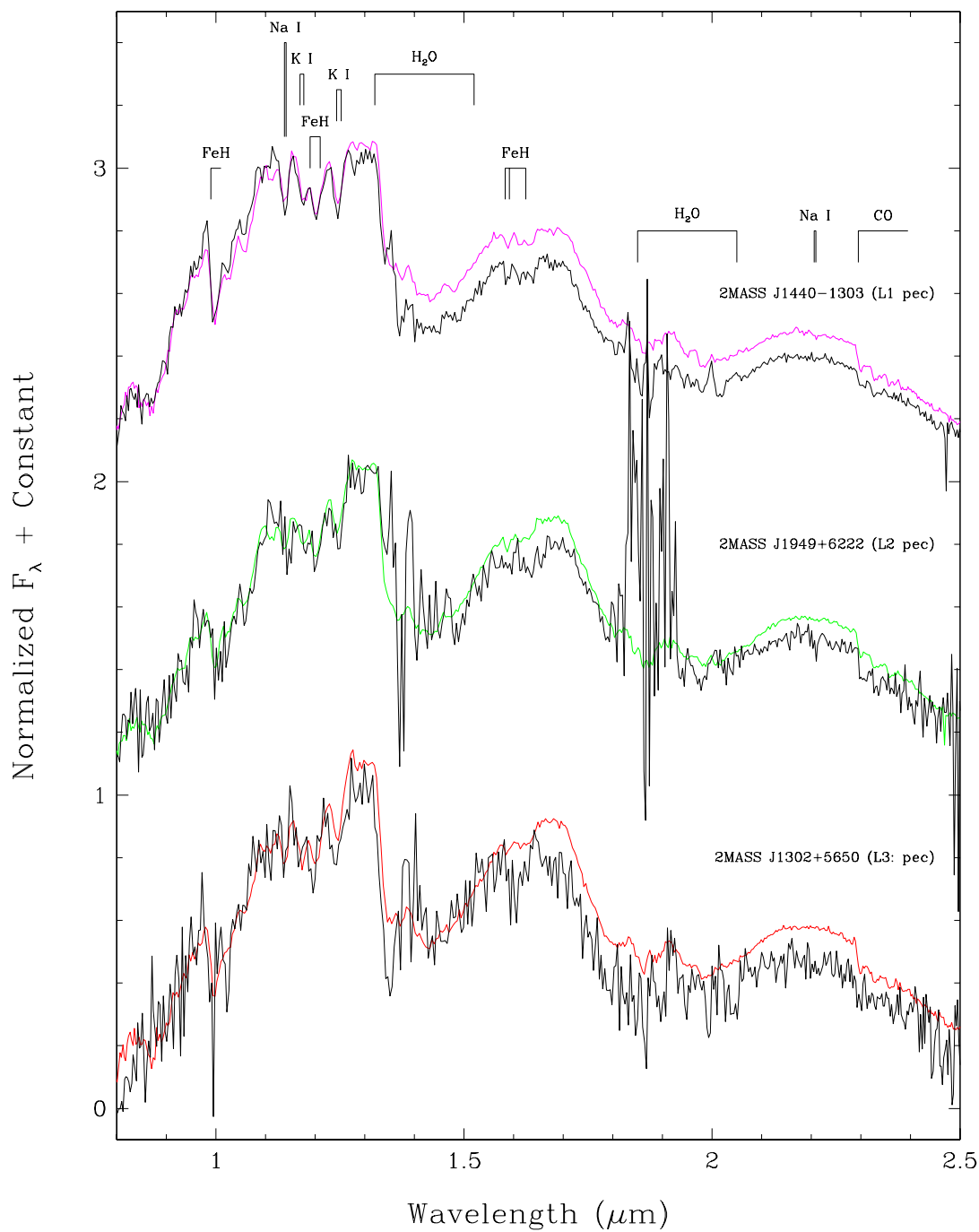}
\caption{Near-infrared spectra of three blue L dwarf discoveries at early-L types (black lines). Overplotted for comparison
are near-infrared standards: the L1 dwarf 2MASS J2130$-$0845 (magenta), L2 dwarf Kelu-1 (green), and L3 dwarf 
2MASS J1506+1321 (red). Spectra are normalized to one at 1.28 $\mu$m and integer offsets added when needed to separate the 
spectra vertically.\label{oddones_blueL_IR1}}
\end{figure}

\clearpage

\begin{figure}
\epsscale{0.9}
\plotone{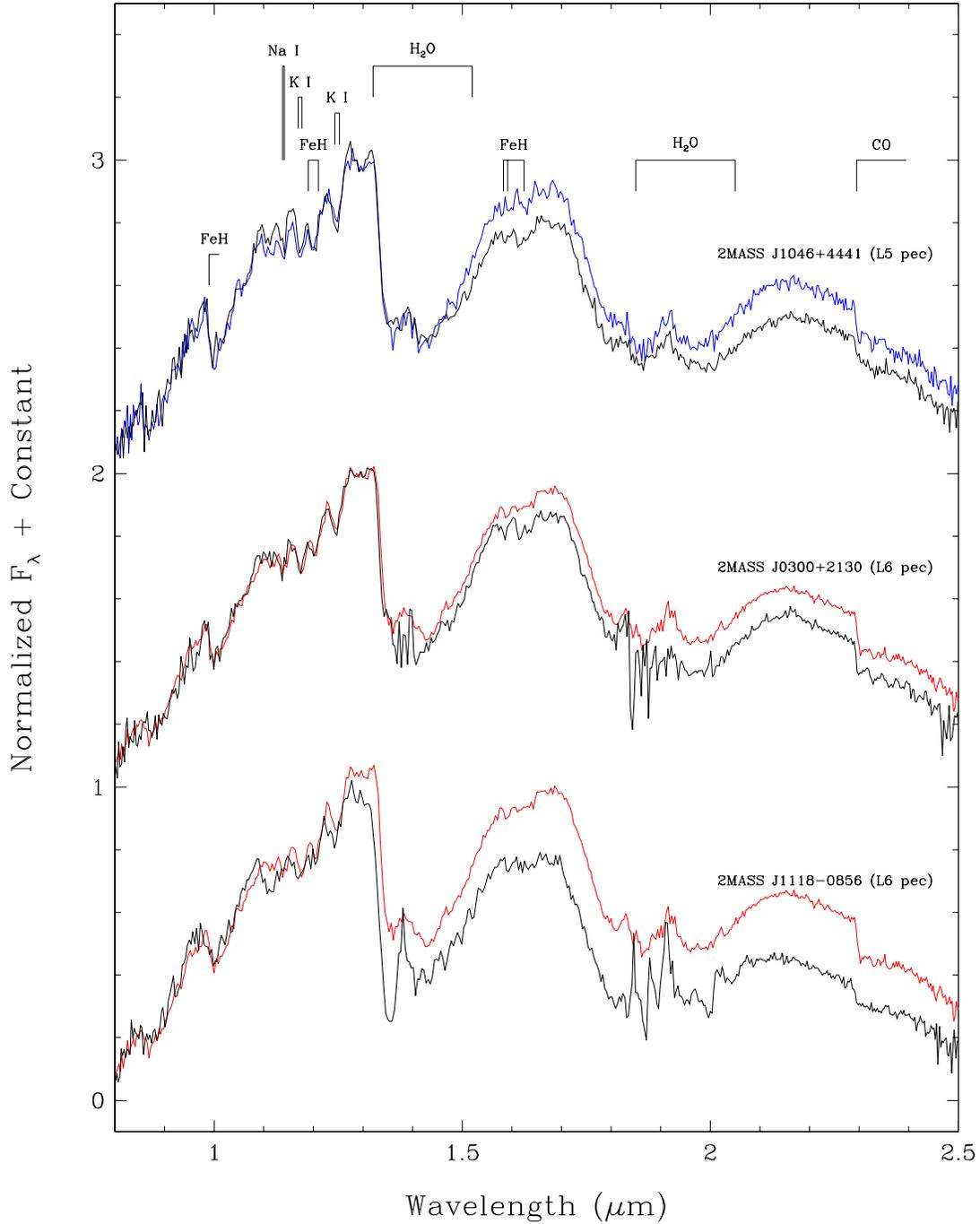}
\caption{Near-infrared spectra of three blue L dwarf discoveries at mid-L types (black lines). Overplotted for comparison
are near-infrared standards: the L5 dwarf 2MASS J0835+1953 (blue) and L6 dwarf 2MASS J1010$-$0406 (red). 
Spectra are normalized to one at 1.28 $\mu$m and integer offsets added when needed to separate the 
spectra vertically.
\label{oddones_blueL_IR2}}
\end{figure}

\clearpage

\begin{figure}
\epsscale{0.9}
\plotone{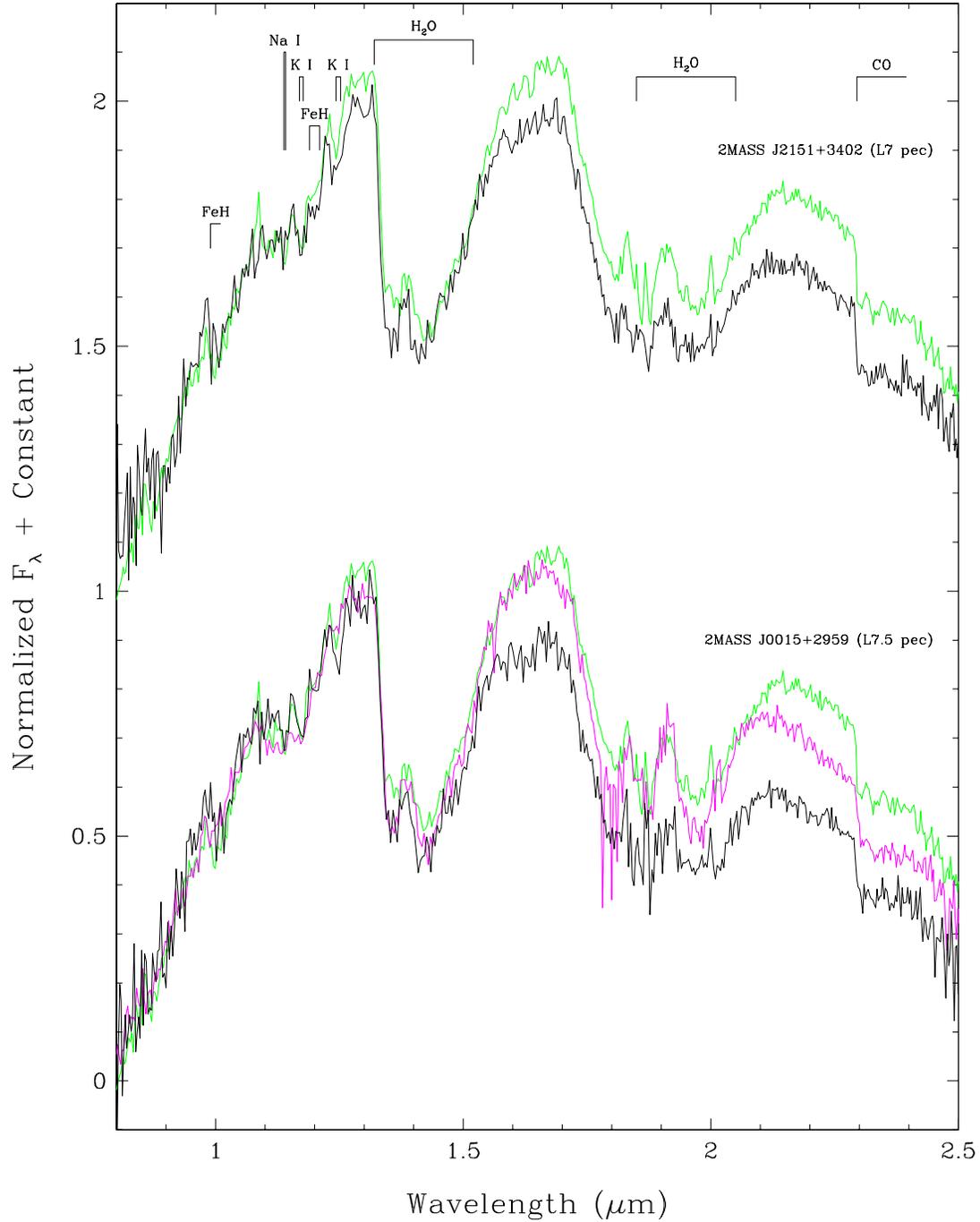}
\caption{Near-infrared spectra of two blue L dwarf discoveries at late-L types (black lines). 
Overplotted for comparison are near-infrared standards: the L7 dwarf 2MASS J0103+1935 (green) 
and L8 dwarf 2MASS J1632+1904 (magenta). Spectra are normalized to one at 1.28 $\mu$m and integer offsets added when needed to separate the 
spectra vertically.\label{oddones_blueL_IR3}}
\end{figure}

\clearpage

\begin{figure}
\includegraphics[scale=0.65,angle=270]{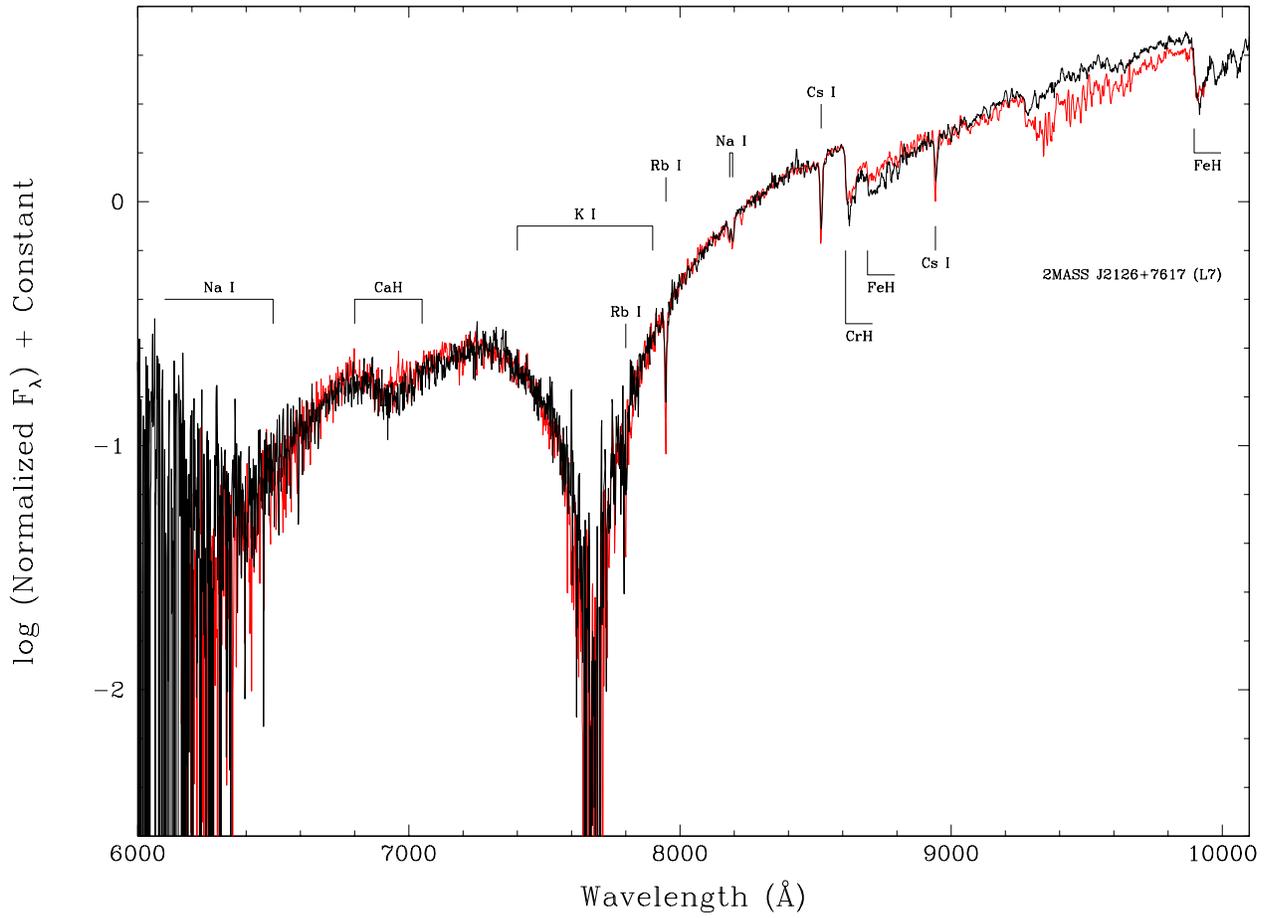}
\caption{Subaru-FOCAS data for 2MASS J2126+7617 (black line) overplotted with the optical L7 standard
DENIS J0205$-$1159 (red). Spectra are normalized to one at 8250 \AA. \label{oddones_T_opt}}
\end{figure}

\clearpage

\begin{figure}
\includegraphics[scale=0.65,angle=270]{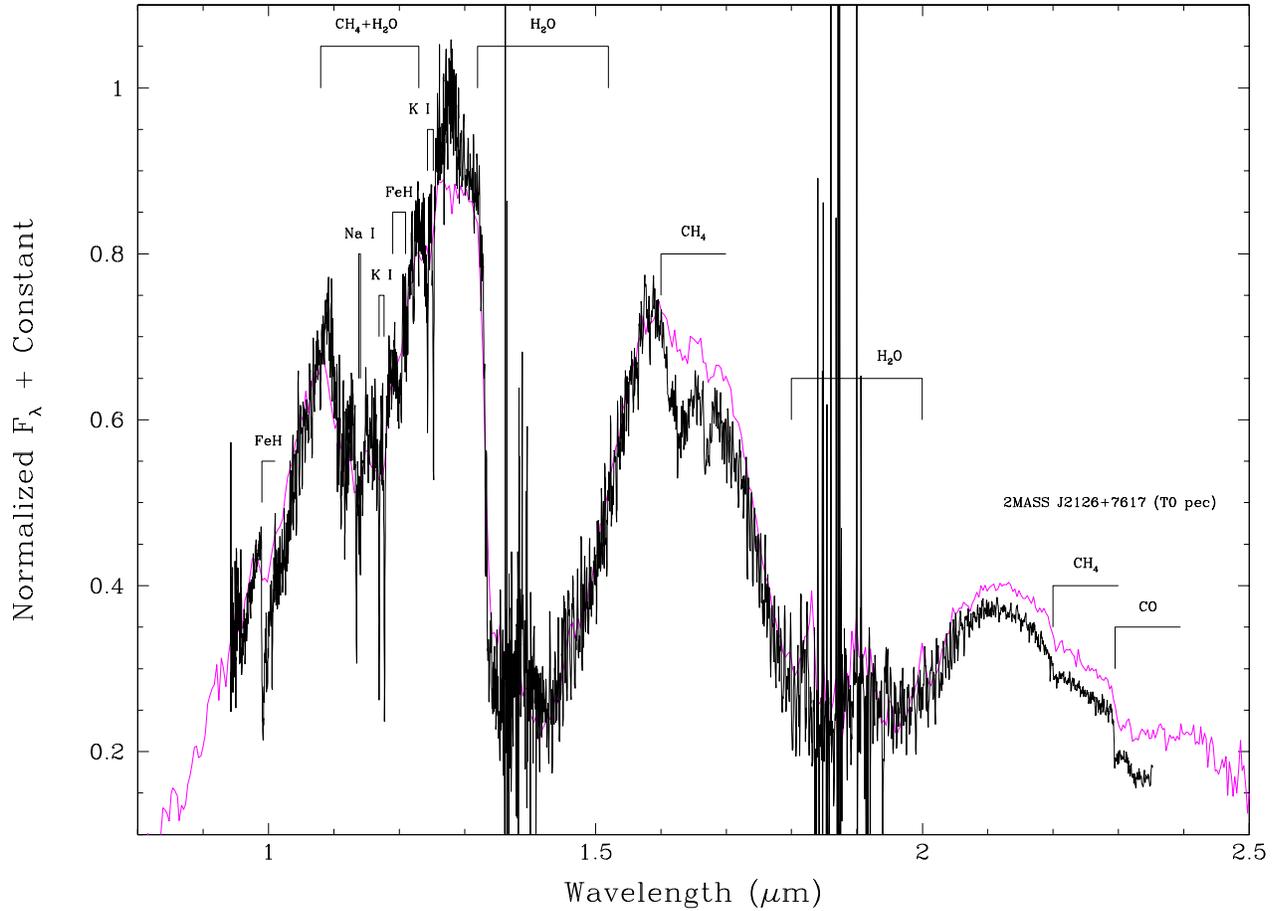}
\caption{The Keck-NIRSPEC spectrum of 2MASS J2126+7617, normalized to one
at 1.28 $\mu$m (black). Overploted for comparison is the spectrum of the near-infrared T0 standard 
SDSS J1207+0244 (magenta), normalized to provide 
the best fit to the water bands of 2MASS J2126+7617. \label{oddones_T_IR}}
\end{figure}

\clearpage

\begin{figure}
\epsscale{0.9}
\plotone{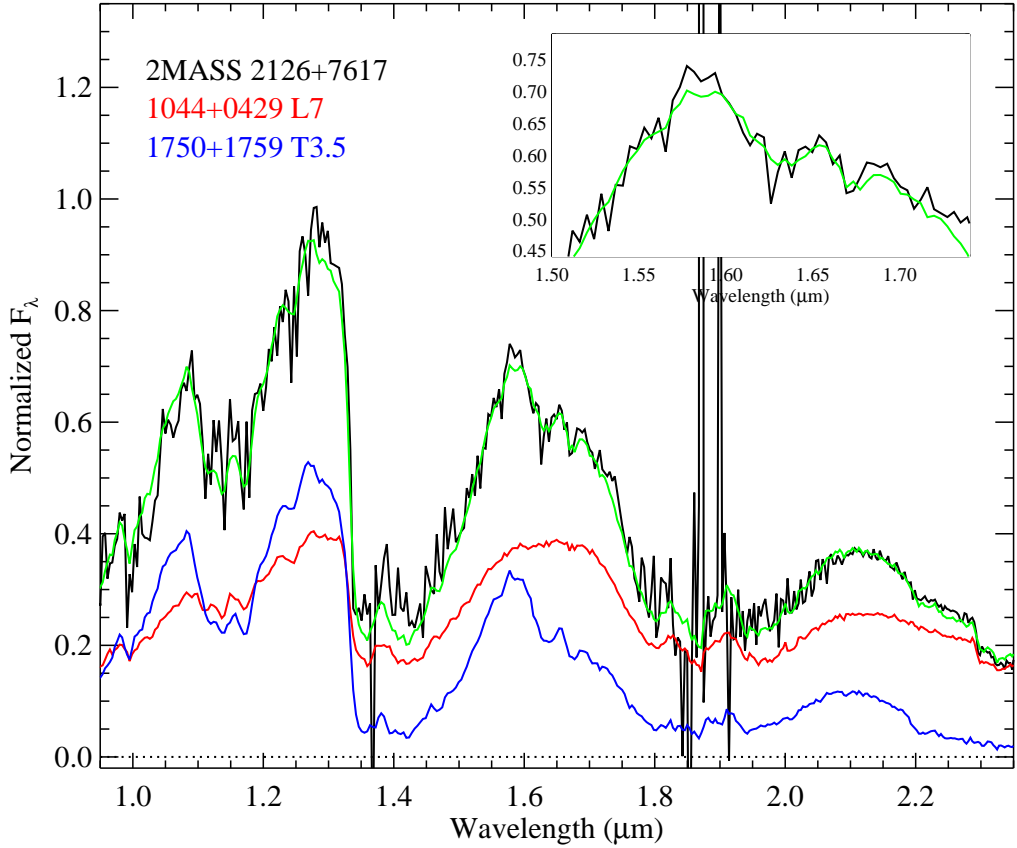}
\caption{Spectral template of the best-fit binary (green line) compared to 
the near-infrared spectrum of 2MASS J2126+7617 (black line).  
Data for this source have been smoothed with a Gaussian kernel to match the resolution of the SpeX 
templates ($\lambda/{\Delta}\lambda$ $\approx$ 120), and normalized in the 1.2--1.3~$\mu$m window.  
The binary template is scaled to minimize its $\chi^2$ deviations.  
The primary (the L7 2MASS J1044+0429, red line) and secondary components (the T3.5 SDSS J1750+1759, 
blue line) of the best-fitting binary template are scaled according to their contribution to the 
combined-light template, based on the \cite{looper2008} $M_{K_s}$/spectral type relation. 
Inset box shows a close-up of the 1.5--1.75~$\mu$m region. \label{2MASS2126_fit}}
\end{figure}

\clearpage

\begin{figure}
\epsscale{0.9}
\plotone{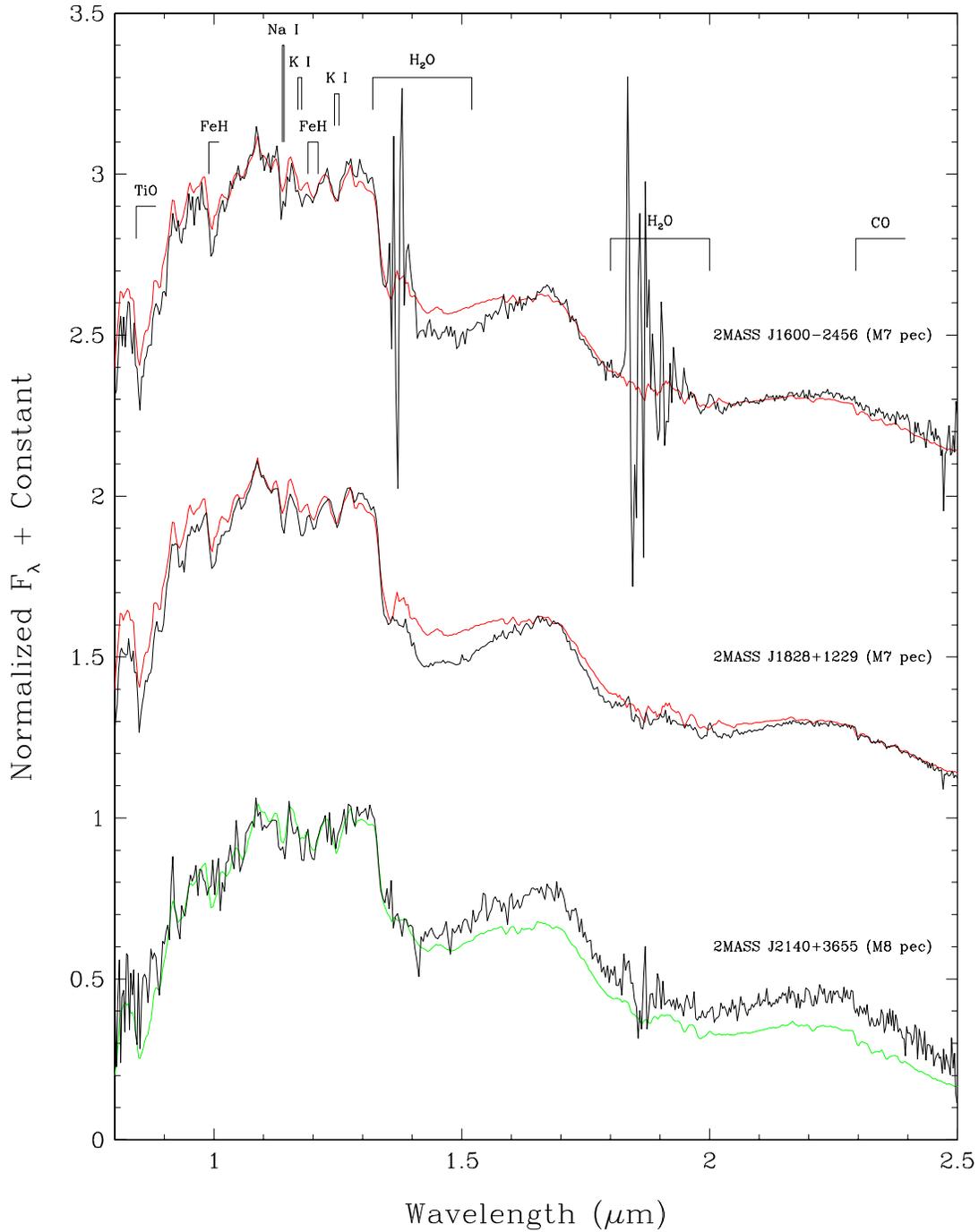}
\caption{Near-infrared spectra of three slightly unusual spectra (black lines) that do not fit 
any of the other categories. Overploted for comparison are spectra of the near-infrared M7 standard 
vB 8 (red) and M8 standard vB 10 (green). Spectra are normalized to one at 1.28 $\mu$m and integer 
offsets added when needed to separate the spectra vertically.
\label{oddones_other_IR}}
\end{figure}

\clearpage

\begin{figure}
\includegraphics[scale=0.65,angle=270]{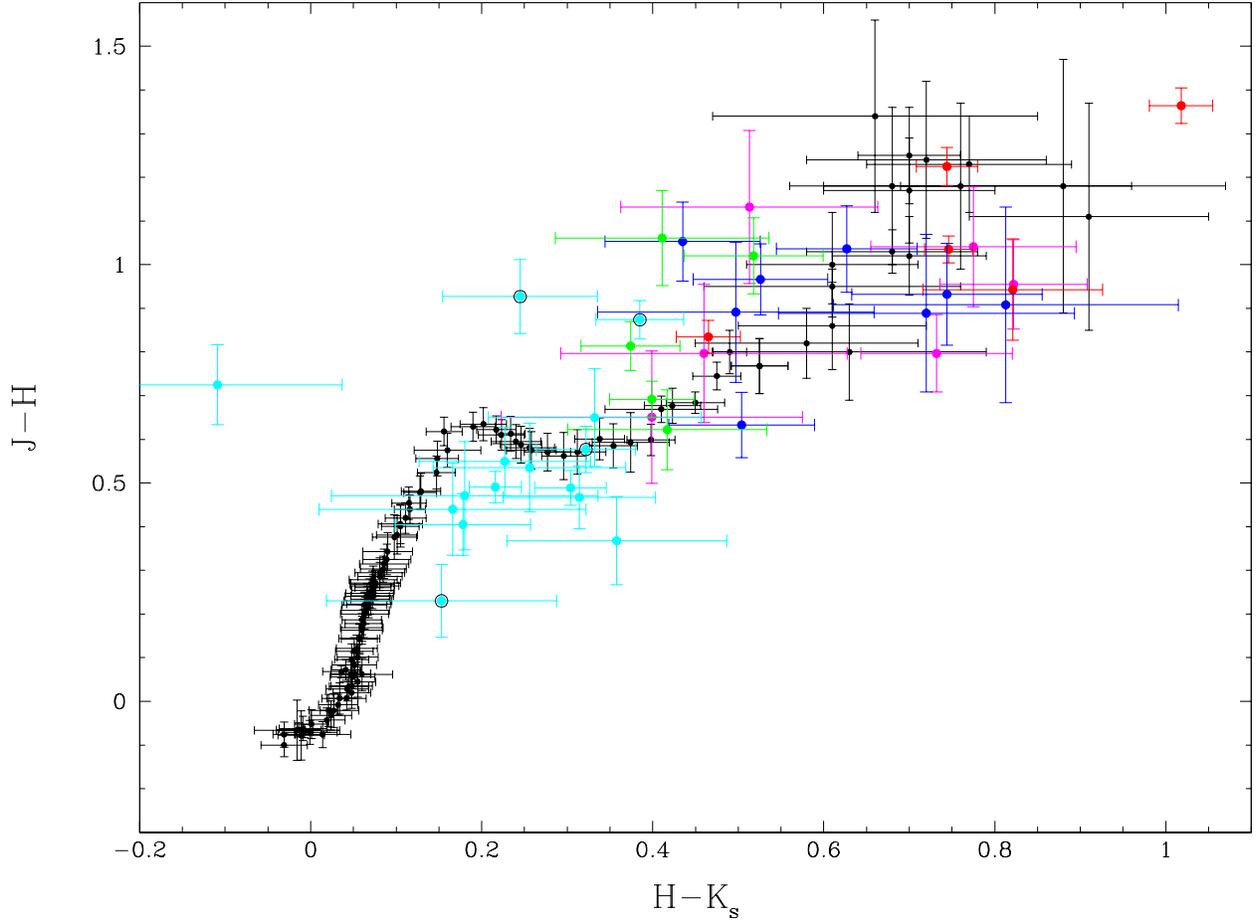}
\caption{2MASS J-H vs.\ H-K$\_s$ color-color diagram. 
Black points represent averaged colors from the 2MASS
All-Sky Point Source Catalog for dwarfs ranging in type from B2 (lower left) through L8
(upper right), sampled at every half spectral subclass.
Objects from Tables
\ref{lowg_table}, \ref{redL_table}, \ref{sd_table}, \ref{blueL_table}, and \ref{T_table} are shown with colored points.
Low-gravity (young) M and L dwarfs are shown in magenta, red L dwarf discoveries are in red,
M and L subdwarfs are in cyan (with the sdM9.5, sdL1, sdL7, and sdL8 further highlighted with black
open circles), blue L dwarf discoveries in blue, and T dwarf discoveries in green.
\label{IR_color_color_colored}}
\end{figure}

\clearpage

\begin{figure}
\includegraphics[scale=0.65,angle=270]{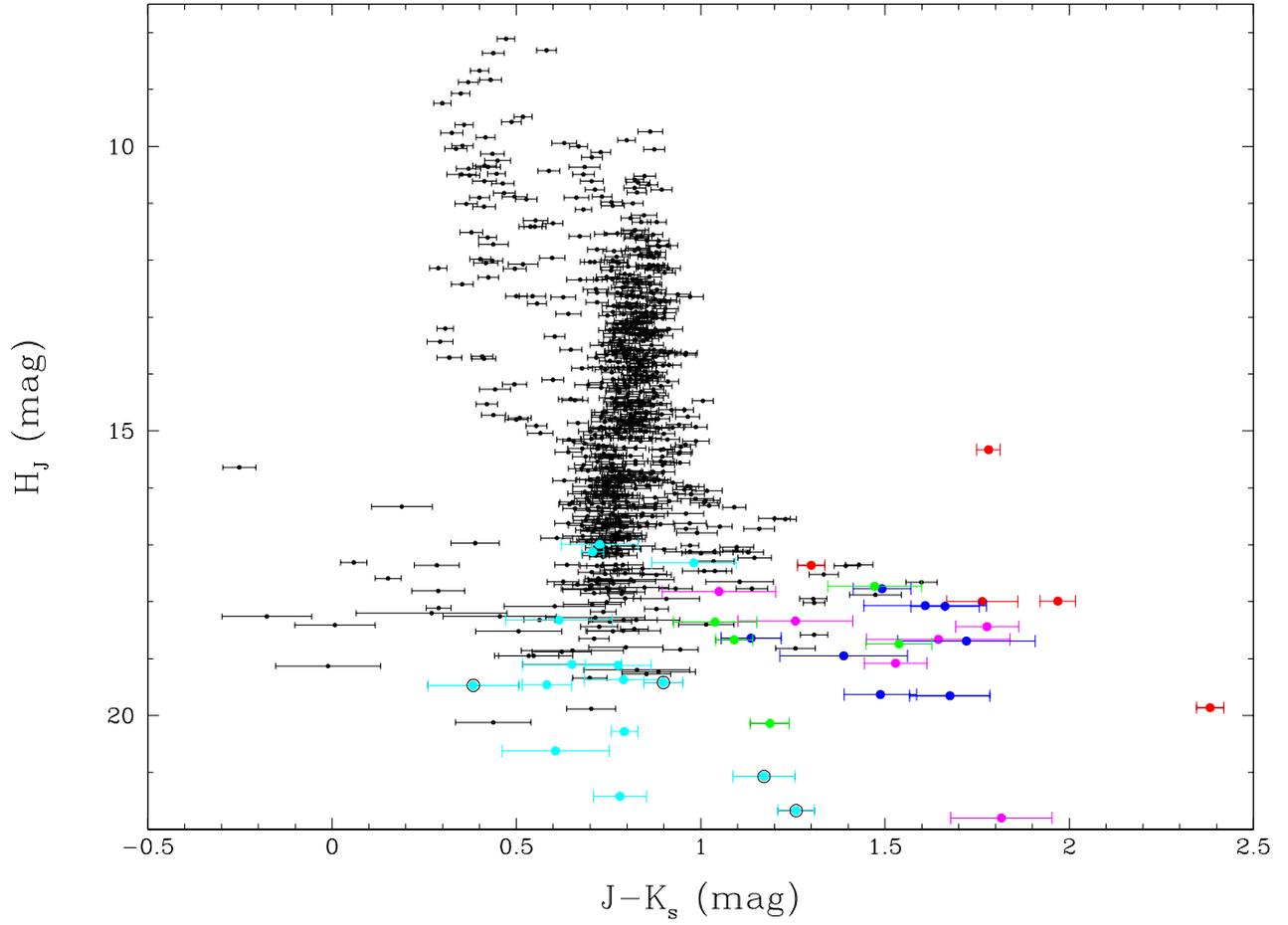}
\caption{Blow-up of the top panel of Figure \ref{red_motion_vs_JK}. Color coding is the same
as that in Figure \ref{IR_color_color_colored}
\label{red_motion_JK_colored}}
\end{figure}

\clearpage

\begin{figure}
\includegraphics[scale=0.65,angle=270]{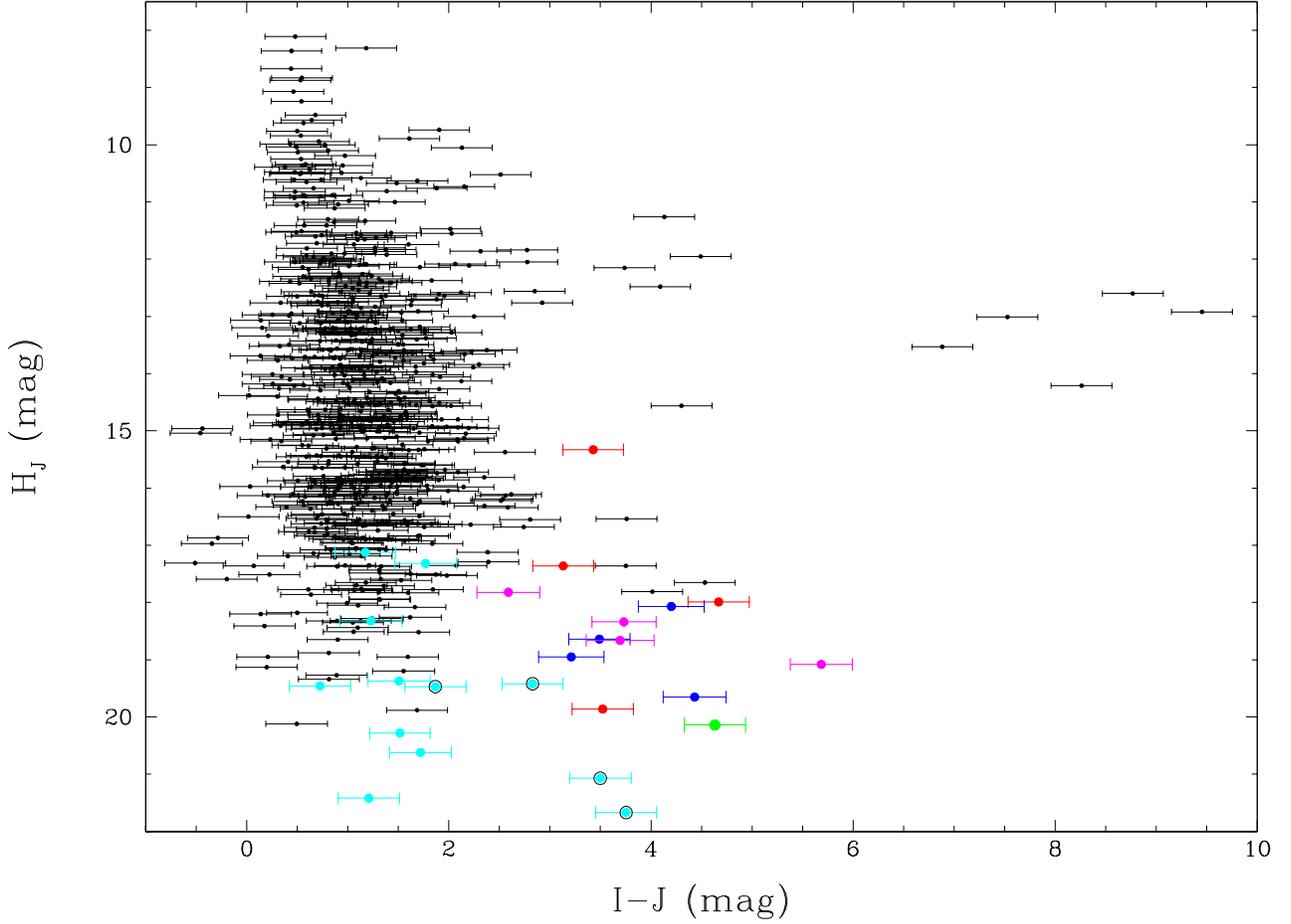}
\caption{Reduced proper motion at J-band plotted against I-J color. Black points represent
those 2MASS proper motion objects whose motions are at least five times the astrometric error.
Objects from Tables \ref{redL_table}, \ref{sd_table}, and \ref{T_table} are shown with colored points.
Red L dwarf discoveries are shown in red,
M and L subdwarfs are in cyan, and T dwarf discoveries in green. (None of the low-gravity or
blue L dwarfs were detected in the USNO-B I band and hence cannot be displayed.) For a few
of the brighter red L dwarfs, L subdwarfs, and T dwarfs, I-J colors limits are shown as
triangles. In these cases, an I-band limit of 18.5 mag is assumed. (Note that objects with
$I-J > 6$ have erroneous $I$-band magnitudes from the USNO-B.)
\label{red_motion_IJ_colored}}
\end{figure}

\end{document}